\documentclass[a4paper,12pt]{article}
\pdfoutput=1
\usepackage{amssymb,amsmath,bm}
\usepackage{color}
\usepackage{slashed}
\usepackage{cite}
\usepackage{verbatim}
\usepackage[latin1]{inputenc}
\usepackage{amsfonts}
\usepackage{lscape}
\usepackage{amsthm}
\usepackage{booktabs}
\usepackage{array}
\usepackage{rotating}
\usepackage{float}
\usepackage{multirow}
\usepackage{graphicx,rotating,amsmath,multirow,xcolor,colortbl,xcolor}
\usepackage{hyperref,pdflscape,slashed}
\definecolor{rosso}{cmyk}{0,1,1,0.4}
\definecolor{rossos}{cmyk}{0,1,1,0.55}
\definecolor{rossoc}{cmyk}{0,1,1,0.2}
\definecolor{blu}{cmyk}{1,1,0,0.3}
\definecolor{blus}{cmyk}{1,1,0,0.6}
\definecolor{bluc}{cmyk}{1,1,0,0.1}
\definecolor{verde}{cmyk}{0.92,0,0.59,0.25}
\definecolor{verdec}{cmyk}{0.92,0,0.59,0.15}
\definecolor{verdes}{cmyk}{0.92,0,0.59,0.4}
\definecolor{Gray}{gray}{0.95}
\usepackage{slashed}
\usepackage{cancel}
\usepackage{soul}

\font\tenrsfs=rsfs10 at 12pt
\font\sevenrsfs=rsfs7
\font\fiversfs=rsfs5
\newfam\rsfsfam
\textfont\rsfsfam=\tenrsfs
\scriptfont\rsfsfam=\sevenrsfs
\scriptscriptfont\rsfsfam=\fiversfs
\def\mathscr#1{{\fam\rsfsfam\relax#1}}

\usepackage[small,bf]{caption}
\hypersetup{colorlinks,bookmarksopen,bookmarksnumbered,
linkcolor=blus,pdfstartview=FitH,urlcolor=rossos,citecolor=verde}

\newcommand{\bea}{\begin{eqnarray}}
\newcommand{\eea}{\end{eqnarray}}
\newcommand{\beq}{\begin{equation}}
\newcommand{\eeq}{\end{equation}}
\newcommand{\beqa}{\begin{eqnarray}}
\newcommand{\eeqa}{\end{eqnarray}}

\newcommand{\SU}{\text{SU}}
\newcommand{\SO}{\text{SO}}

\newcommand{\U}{\text{U}}

\def\eq#1{eq.~(\ref{#1})}
\def\Fig#1{fig.~\ref{#1}}

\def\Tab#1{table~\ref{#1}}

\def\Eq#1{eq.~(\ref{#1})}
\def\Sec#1{sect.~\ref{#1}}
\newcommand{\MM}[1]{\textit{\textcolor{magenta}{(MC: #1)}}}
\newcommand{\newc}{\newcommand}
\newc{\gsim}{\lower.7ex\hbox{$\;\stackrel{\textstyle>}{\sim}\;$}}
\newc{\lsim}{\lower.7ex\hbox{$\;\stackrel{\textstyle<}{\sim}\;$}}
\newc{\gev}{\,{\rm GeV}}
\newc{\mev}{\,{\rm MeV}}
\newc{\ev}{\,{\rm eV}}
\newc{\kev}{\,{\rm keV}}
\newc{\tev}{\,{\rm TeV}}

\newc{\mz}{M_Z}
\newc{\mpl}{M_*}
\newc{\mw}{m_{\rm weak}}
\newc{\nr}[1]{N^c_R{}_{#1}}

\def\beq{\begin{equation}}
\def\eeq{\end{equation}}
\def\bea{\begin{eqnarray}}
\def\eea{\end{eqnarray}}
\def\bitem{\begin{itemize}}
\def\eitem{\end{itemize}}

\newcommand{\be}{\begin{equation}}
\newcommand{\ee}{\end{equation}}

\newc{\ie}{{\it i.e.}}          \newc{\etal}{{\it et al.}}
\newc{\eg}{{\it e.g.}}          \newc{\etc}{{\it etc.}}
\newc{\cf}{{\it c.f.}}

\def\inv{^{\raise.15ex\hbox{${\scriptscriptstyle -}$}\kern-.05em 1}}
\def\lbar{{\lower.35ex\hbox{$\mathchar'26$}\mkern-10mu\lambda}} %lambda bar

\def\eq#1{Eq.(\ref{#1})}

\def\to{\rightarrow}

\let\<=\langle
\let\>=\rangle

\let\+=\uparrow
\newcommand{\mb}[1]{\boldsymbol{#1}}

\newcommand{\kahler}{K\"{a}hler }

\newcommand{\rep}[1]{\mathbf{#1}}
\newcommand{\conjrep}[1]{\overline{\mathbf{#1}}}

\newcommand{\IZ}{\mathbb{Z}}

\newcommand{\GeV}{\,\text{GeV}}

\begin{document}

%{\hfill CERN-TH-2016-223}

\vspace{2cm}

\begin{center}
\boldmath

{\textbf{\LARGE In Pursuit of New Paradigms
}}
\unboldmath

\bigskip

\vspace{0.4 truecm}

{\bf Matthew McCullough}
 \\[5mm]

{\it CERN, Theoretical Physics Department, Geneva, Switzerland}\\[2mm]

\vspace{2cm}

{\bf TASI 2024}
\end{center}

%\begin{quote}
%BUSSTEPP 2017.
%\end{quote}

\thispagestyle{empty}
\vfill

%\newpage

\tableofcontents
%\newpage
%%%%%%%%%%%%%%%%%%%%%%%%%%%%%%%%%%%%%%%%%

\section{Paradigm = Effective Field Theory?}
What is a paradigm?  \href{https://en.wikipedia.org/wiki/Paradigm}{Wikipedia} says the origins of the word are Greek, from the word `paradeigma'.  Wikipedia goes on, clarifying that this means `an isolated example by which a general rule illustrated', however the simplest translation of this `general rule' is just `pattern'.  Through the eyes of a physicist or mathematician, an example of a pattern we might observe is characterised by its symmetry, exact or approximate, and by the objects, however abstract, it is composed of.  However, the symmetry itself transcends, conceptually, the objects or manner in which it may be manifested.  We may come across the same symmetry in wildly different settings.

What, then, is a paradigm in the context theoretical physics?  The notion of a pattern in the more general sense maps neatly into our modern conception of symmetries.  The `isolated example' would then correspond to the collection of objects which manifest the symmetry.  For us, these would be the quantum fields which transform under the symmetry.  For any specific symmetry there are an infinite number of possibilities for the fields, classified by the way in which they transform under the symmetry transformations, and one could consider any number of them in combination to construct an `isolated example'.  This is much like music and the keys on a piano.  They symmetry would be the specific choice of key, the fields (representations) would be the specific notes of that key that will be played, and the EFT would be a specific piece of music.

By this stage of TASI you should be reasonably familiar with effective field theories (EFTs).  An EFT is composed of its symmetries (including approximate ones) and the fields in fixed representations of that symmetry.  With these ingredients one allows all possible interactions between the fields, consistent with the symmetries, with the strength of those interactions determined by the underlying short-distance physics which, itself, is not required for an accurate description of some physical process on some length scale for which the EFT is employed.

So an EFT is the combination of a symmetry and the objects which manifest it; the fields.  So I think I can fairly argue that for all intents and purposes Paradigm = EFT.  The pursuit of new paradigms sounds far more exciting, and less prosaic, than the pursuit of new EFTs, but it's really the same thing in my book.  Almost any paradigm shift I can think of in fundamental physics was, on reflection, the discovery of a new EFT.  One exception for which a defendable argument could be formed was the discovery of quantum mechanics, but maybe let's discuss that over coffee.\footnote{During these lectures it was also pointed out that in some sense classical mechanics is an EFT for quantum mechanics.}

These lectures will thus be concerned with the present EFT (paradigm) we have, the Standard Model (SM) of particle physics and the EFTs (paradigms) that may replace it.  I understand you have not had a dedicated introduction to EFTs during TASI, nor is there time to provide one here, so I'll give you a quick example to illustrate the basics and then provide some supplementary material in these notes.

\subsubsection*{Intro to the ideas of EFT, by example.}
\begin{figure}[t]
\centering
\includegraphics[height=2.0in]{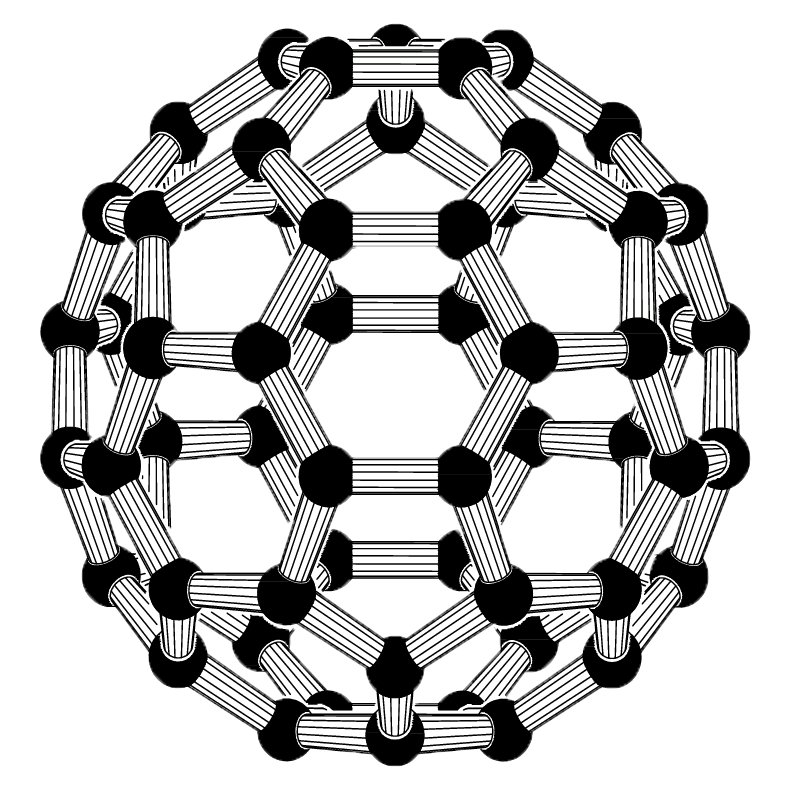}  
\caption{Buckyball.  Thanks Wikipedia.}
\label{fig:buckyball}
\end{figure}

Consider a soccer ball, or, at least, its shape.  This shape is a truncated icosahedron.  Interestingly, objects with this shape arise naturally, for instance Buckminsterfullerene (Buckyball) depicted in \Fig{fig:buckyball}.  Buckyballs are molecules of 60 carbon atoms, with a roughly spherical shape.  Suppose you hold a point charge a long way from the Buckyball, at a distance far greater than its physical size.  You move the charge around and map out the electrostatic potential you measure.  What form does it take?

One could go ahead and calculate this in terms of the nuclei and electrons, determining the orbitals of the electrons for this large molecule and so on.  However, in my view there is a smarter way to proceed, guided by symmetries.  If the Buckyball were exactly rotationally invariant then the electrostatic potential would be too.  However, it carries no overall charge, so it can't have the usual electrostatic potential we are familiar with for point charges.  We'll have to think a bit harder.

Let's recall the electrostatic potential for a dipole.  A dipole carries no overall electric charge, but it does have an electrostatic potential.  Let's align it with the $z$-direction and assume it is composed of unit charges separated by a distance $d$.  I'll set all electromagnetic constants to $1$, since it is the scaling we are interested in.  The potential in cylindrical coordinates, chosen because one rotational symmetry is preserved, is
\bea
V\left(z,r=\sqrt{x^2+y^2} \right) & \propto & \frac{1}{\sqrt{r^2 + (z-d/2)^2}} -\frac{1}{\sqrt{r^2 + (z+d/2)^2}} \\
& \approx & \frac{d z}{(r^2 + z^2)^{3/2}} +...~~,
\eea
where the higher order terms are sub-leading for small $d$.  This is the famous dipole potential.  We could have foreseen this by thinking about symmetries.  The dipole breaks the full $\SO(3)$ rotational symmetry down to $\SO(2)$.  If one is a long way from the dipole this breaking is small.  The smallest representation of $\SO(3)$ which breaks the symmetry in this way is simply a vector $\bold{d}$.  Us particle theorists would call $\bold{d}$ a `spurion', which is an object (representation) we think of as transforming under a global symmetry in a fixed way, which also takes on a background non-zero value.  The global symmetry here being spatial rotations.  Importantly, the size of the dipole is characteristic of the microscopic length scale of the object and if we take its value to zero a symmetry is restored.

In this view we can write everything in a manifestly $\SO(3)$-symmetric way and determine the effects of this background value by setting the spurion to its non-zero background value. Thus, to leading order in $\bold{d}$, we already know the only form the electrostatic potential could have taken is
\be
V(\bold{r}) \approx \frac{\bold{d}\cdot \bold{r}}{r^3} ~~,
\ee
where the form of the denominator follows from the fact that, dimensionally, the potential has units of inverse length since it is, microscopically, comprised of the superposition of the electrostatic potentials of individual point charges.  On to the Buckyball.

Let's outline a more general strategy.  Imagine there is a scale separation (this is a \emph{very} important component of the EFT approach) such that we are measuring the electrostatic potential of an object from a distance far greater that its size.  Consider the symmetries preserved ($\mathcal{P}$) by the existence of that object in space.  Determine the smallest dimension `spurion' representation $\mathcal{R}$ of the spatial rotations $\mathcal{S}$ for which a background value of $\mathcal{R}$ preserves $\mathcal{P}$.  Write down the leading term in an electrostatic potential in an expansion in $\mathcal{R}$.   Thus
\be
V(\bold{r}) \approx r_0^\mathcal{N} \frac{\mathcal{R}^{a1,...,a_\mathcal{N}} r^{a_1} ... r^{a_\mathcal{N}}}{r^{1+2 \mathcal{N}}} +...~~,
\ee
where $r_0$ is a characteristic microscopic length scale of the object and $\mathcal{N}$ is the number of indices of the spurion representation.  A Buckyball respects the icosahedral point group of discrete rotations $I_h$.  The lowest order multipole (representation) which preserves this symmetry is the six-index 64-pole!\footnote{See `Multipoles and Symmetry', Gelessus, Thiel and Weber (1995).} Thus the electrostatic potential of a Buckyball is
\bea
V(\bold{r}) &  \approx & r_0^6 \frac{\mathcal{R}_6^{a1,...,a_6} r^{a_1} ... r^{a_6}}{r^{13}} ~~, \\
&  \approx & \frac{r_0^6 (x^2-y^2) (y^2-z^2) (z^2-x^2)}{r^{13}} ~~,
\eea
where $\mathcal{R}_6$ is 6-index symmetric traceless irrep of $\SO(3)$ whose individual non-vanishing components take a form preserving $I_h$ and $r_0$ is a characteristic microscopic length scale of the Buckyball.

Let's consider terms at $\mathcal{O}(R^2)$.  Here there are a variety of possibilities for the way in which they can be arranged, amongst them
\beq
\frac{\left(\mathcal{R}_6^{a1,...,a_6} r^{a_1} ... r^{a_6}\right)^2}{r^{25}} , ... , \frac{\mathcal{R}_6^{a1,...,a_6} \mathcal{R}_6^{a1,...,a_6} }{r^{13}} ~~.
\eeq
Thus we see that at leading order we expect an electrostatic potential which scales as $V(r) \propto r^{-7}$, however terms scaling as $r^{-13}$ arise at higher orders in $r_0/r$ and so-on.

There are a few key lessons here that will carry forward to our broader discussions, so let's summarise them.  Some key concepts:
\begin{itemize}
\item  \textbf{Scale separation.}  When you are thinking about the physics of some object, in this example a molecule, on length scales much greater that its size one can view the physics dealing in terms of its various constituents and their arrangement, however that approach is cumbersome and unnecessary.  Alternatively, one can build an effective description which treats the object as a point particle with non-vanishing `moments'.  An effective description should work with the objects required for that description and no more.  It would be madness to calculate the Hydrogen electron orbitals by solving for the nuclear part in terms of quarks and gluons; it is much better to approximate the proton as a point particle with spin and charge, since the length scales concerning Hydrogen orbitals are far far greater than those associated with QCD dynamics.  This is efficient and also powerful, since the effective approach will accurately describe the physics of vast varieties of objects which may have very different microscopic configurations yet give rise to the same family of multipoles.  This latter aspect is known as universality and the classes of objects with the same effective description are known as `universality classes'.
\item  \textbf{Symmetry.}  This was a key tool:  The different universality classes may be organised by their symmetry properties.  For Buckyballs the first non-vanishing multipole is the 64-pole, but for Methane it is the Octopole ($l=3$).  These two facts are immediately understood through symmetry, since Methane only respects the less-constraining Tetrahedral point group, not the large Icosahedral group.
\item  \textbf{Dimensional Analysis.}  This is a tremendously powerful tool in the presence of a scale separation.  Dimensions of length and other quantities must always be commensurate, hence if one understands the dimensionality of parameters then the overall scaling of physical observables can be determined.
\item  \textbf{The Cutoff.}  An EFT, by definition, has a physical cutoff beyond which the EFT loses validity.   We usually refer to this scale as $\Lambda$, meaning the mass scale of new states beyond the EFT.  We may or may not know specifically what exists at that scale, but we can capture its role in the EFT via dimensional analysis, up to $\mathcal{O}(1)$ coefficients.  Note that sometimes we (more precisely; not me) may \emph{estimate} the magnitude of effects from states at the cutoff by performing loop integrals and seeing how the integral cutoff enters.  This should not be taken overly seriously.  It is a sloppy means of estimating effects beyond the regime of validity of the EFT but is not physically quantitative in and of itself.  As such, sensitivity of some IR quantity to the cutoff in some scheme in a loop integral is not a problem \emph{per se}, but just a sloppy means for estimating effects which ought otherwise be better estimated through dimensional analysis and symmetries.
\item  \textbf{Totalitarian Principle.}  The phrase `Everything not forbidden is compulsory' is accredited to Murray Gell-Mann.  It takes an important place in the conceptual structure of effective theories, in the sense that typically everything not forbidden by symmetry is generated, in our case this would be the allowed terms in the electrostatic potential.  If this is not the case then either a) there is a symmetry at play that hasn't yet been identified or b) there is some fine-tuning of the microscopic constituents which forces some term in the potential to zero, or small values, even if symmetry allows it.
\item  \textbf{Accidental Symmetry.}  As one goes to longer and longer distances the electrostatic potential becomes more and more $\SO(3)$-symmetric, such that in the infinitely long distance scale it becomes exactly $\SO(3)$-symmetric.  It also vanishes, but the relevant question is how quickly it vanishes to become $\SO(3)$-symmetric.  Note that for methane, the approach to $\SO(3)$-symmetry is much slower, scaling as $r^{-4}$ rather than $r^{-7}$.  There are two related lessons here.  a)  As one goes to longer and longer distance scales symmetries often become enhanced, with more symmetry emerging as symmetry-breaking terms become irrelevant.  We often call such an emergent symmetry `accidental' since the microscopic physics doesn't reflect that full symmetry, yet the long-distance physics does.  b)  A football is closer to a sphere than a tetrahedron, thus at long distances a Buckyball looks more $\SO(3)$-symmetric than a methane molecule does.
\item  \textbf{Clues.}  Everything discussed here has been phrased in terms of the `matching' from the known short-distance physics to the appropriate effective long-distance description.  However, what if you are performing real experiments that only work at long distances and cannot access the short distance physics?  Then you have to try and guess at the short distance physics by studying the structure of the long-distance effective description.  By measuring the various terms in the electrostatic potential one can hope to gather more clues.  However, this is easier said than done.  Often there are only one or two leading terms, but the other terms are suppressed by increasingly larger powers of the small ratio $r_0^\mathcal{N}/r^\mathcal{N}$.  There are then two options: a) Perform increasingly higher precision measurements, in order to be able to measure increasingly tiny effects, or b)  Perform measurements at smaller distance scales to amplify this ratio and the size of the effects making them easier to observe.
\end{itemize}
To understand what we're really up to in the grand effort of particle physics, and reductionism, both experimentally and theoretically, it's very important that all of the above concepts are clear to you.  Now to an example phrased in terms much more akin to particle physics.

\subsubsection*{Towards EFT in QFT.}
\label{sec:EFTinQFT}
Let us now construct a particle physics model by leaning on the analogy we just discussed.  Suppose in nature, in an abstract sense, there is a triplet of real scalar fields $\bold{\phi} = (\phi_1,\phi_2,\phi_3)$ whose interactions respect the Icosahedral group which acts not as a spatial symmetry as in the previous example, but instead acts on this triplet of scalars.  We typically refer to a symmetry like this as an `internal' symmetry, since it does not involve spacetime but instead involves an action on the fields themselves.   In fact, we can translate the symmetry of the Icosahedral group into particle physicist using a table connecting `Symmetry Groups' to `Abstract Groups' found, for instance, on \href{https://en.wikipedia.org/wiki/Point_groups_in_three_dimensions#The_groups_arranged_by_abstract_group_type}{more Wikipedia}.  We see that $I_h$ corresponds to the discrete $A_5$ subgroup of $\SO(3)$.  You're welcome to study this symmetry more closely, however it's probably easier to think of it acting as a subset of the transformations $\bold{\phi} \to O \bold{\phi}$, where $O$ is an orthogonal matrix.

Since, if an interaction respects $\SO(3)$ it automatically respects $A_5$, we may write the Lagrangian for this scalar in terms of two pieces $\mathcal{L}_{\SO(3)}$ and $\mathcal{L}_{\cancel{\SO(3)}}$, where the latter respects $A_5$.  In this EFT we thus have
\be
\mathcal{L}_{\SO(3)} =- \frac{1}{2} \phi_i  \Box \phi_i - \frac{1}{2} m^2 \phi^2 - \frac{\lambda}{12} \left(\phi^2 \right)^2 + \frac{c_6}{M^2} \left(\phi_i \Box^2 \phi_i + \lambda_6 \phi^2 \Box \phi^2+ \kappa_6 \left(\phi^2 \right)^3 \right) +...
\ee
where $\Box = \partial_\mu \partial^\mu$, $\phi^2 = \phi_i \phi_i$, $c_6$ is a number, $\lambda_6$ and $\kappa_6$ are couplings and $M$ is the mass scale at which those heavier, microscopic states can no longer be ignored and this description of the physics breaks down.  This can be thought of very analogously with the Buckyball, wherein we would identify $M \sim 1/r_0$.  The breakdown of the EFT does not mean anything radical, it just means that the effective description is no longer appropriate.  Imagine what a mess it would be to try and describe a Buckyball as a perfect sphere with multipole moments when working at length scales much smaller than its physical size.

If we don't know $M$ then we don't know precisely when the EFT will break down from within the EFT itself.  However, we can establish at which scale the new microscopic must have shown up by.  We can see this, for instance, because scattering amplitudes with an internal scalar propagator will have contributions scaling as $ \propto \frac{1}{p^2-m^2} + \frac{c_6}{M^2}$ and for $|p^2| \gg M^2$ the dimension-6 contribution has become greater in magnitude than the leading term, meaning that the microscopic `UV' physics which generated the higher dimension operators can no longer be considered a small perturbation of the low energy `IR' description at this energy.  Similar arguments can be made for the other two terms in $2\to2$ and $2\to4$ scattering. The omitted terms arise at dimension-8 and beyond.

Now consider that, fundamentally, the full $\SO(3)$ isn't a symmetry of the theory, only $A_5$.  This means that a term such as
\bea
\mathcal{L}_{\cancel{\SO(3)}} & = & \frac{\kappa_\mathcal{R}}{M^2} \mathcal{R}_6^{a1,...,a_6} \phi^{a_1} ... \phi^{a_6}+... \\
& = & \frac{\kappa_\mathcal{R}}{M^2} (\phi_1^2-\phi_2^2) (\phi_2^2-\phi_3^2) (\phi_3^2-\phi_1^2)
\eea
is allowed in the low-energy EFT, where $\mathcal{R}$ is the same tensor as we considered for the Buckyball and $\kappa_\mathcal{R}$ is a coupling.  We thus see that the first order at which the true symmetry is revealed is dimension-6.  At lower dimensions, hence in $2\to2$ scattering, for instance, at tree-level one will have scattering amplitudes which respect the larger $\SO(3)$ symmetry.  Hence this $\SO(3)$ symmetry is `accidental', or `emergent', in some sense.  Only at higher orders in perturbation theory or with higher multiplicity scattering will we learn about the full underlying symmetry.

There is one last aspect I would like to introduce which can be tremendously useful, but is often overlooked.  This concerns $\hbar$ counting.  If one considers the path integral and treats $\hbar$ as though it is dimensionful, we see that due to the kinetic terms all fields have dimension of $[ \phi, \psi, A_\mu, h_{\mu\nu} ] =[ \hbar^{1/2}]$.  This is a useful tool for keeping track of loop factors, with each one coming accompanied by a factor of $\hbar$.  We also see that a Yukawa coupling and gauge coupling must have dimensions of $[ y,g] = [ \hbar^{-1/2}]$, whereas a quartic scalar interaction has units of $[ \lambda] = [ \hbar^{-1}]$.   Furthermore, we see that $[ \kappa_\mathcal{R} ] = [ \hbar^{-2}]$ and so on.

Lets see how this perspective can be put to work.  Consider the kinetic term for $\phi$.  Its coefficient is dimensionless, thus any wavefunction renormalisation must also be dimensionless.      In principle, on dimensional grounds one could have two-loop wavefunction renormalisation proportional to $\kappa_\mathcal{R}$ and one insertion of $\mathcal{R}_6$.  However, $\mathcal{R}_6$ is traceless, and the only way to reduce the number of indices would be to take a trace.  So we would need at least two insertions of $\mathcal{R}_6$ to get something non-zero, where indices from each tensor insertion are summed over.  But this means that the $\phi$ wavefunction renormalisation can only be affected by $\mathcal{R}_6$ at the four-loop level, seen simply by inserting the number of $\hbar$'s required for a dimensionless coefficient!  One can confirm this by drawing diagrams, but in more general settings $\hbar$ counting can be much quicker and more failsafe as a method to ascertain how different interactions enter quantum-mechanically into different observables.

\section{Why Pursue New Paradigms?} 
Why would anyone look for new paradigms, new EFTs?  A dreamer would do this just to know what is possible.  There is great value in expanding our knowledge and database of interesting EFTs.  A pragmatist would want to discover a paradigm that could shed light on our understanding of nature, of the world we inhabit, ideally answering presently unanswered conceptual puzzles and also making falsifiable predictions, in the tradition of the scientific method.  These lectures focus on the latter, albeit with multiple nods to the dreamers.

Arguably the most powerful EFT we have ever known is the Standard Model (SM) of particle physics, making it a natural target for investigation.  Now, many of you will be aware already that no evidence for those higher-dimension operators encoding the microscopic physics has shown up yet.  In rather old-fashioned lingo we would say that no evidence for non-renormalisable interactions has shown up.  Maybe, then, the SM is not even an EFT and there is no further microscopic physics?  This would be a welcome simplification of our enterprise, but unfortunately it cannot be the case.  There are two reasons for this.  The first is that at extremely high energies the hypercharge gauge interactions become strongly coupled, known as a Landau Pole, and the SM can no longer be used to make physical predictions.  In other words, another effective description of nature is required beyond those energies.   The second reason concerns gravity, wherein above the Planck scale gravitons also become strongly interacting, again meaning that it can no longer be used to describe nature.  So the SM is an EFT and that is not a matter of opinion.

So what replaces the SM at microscopic distance scales?  To figure that out we need to get into the habit of speculating as to what nature looks like at microscopic distances and, consequently, how it may be described.  Such speculation is not new.  In fact, that's how the SM was constructed in the first instance.  This speculation should not be completely wild though; it should be guided by the clues hidden in the structure of the SM itself.  I am aware of a few such clues.  One concerns neutrino masses, another concerns the flavour structure of the SM, yet another the nature of CP-violation in the QCD.  Sadly we don't have time to discuss all of these interesting things.  Instead I will focus on one particular set of clues, which are the mass of the Higgs Boson and the EW scale.

Before, in the context of the Buckyball and the scalar model, I noted that the dimensionful scales found in the EFT were characteristic of the microscopic substructure they are associated with.  It turns out that this holds true in a great many known scenarios.  You can see from the Buckyball example why it would be.  The only ways to generate multipole moments much smaller than the length scale of a molecule would be a) if some multipole is forbidden by a symmetry, as in the case of the vanishing dipole moment for a Buckyball, or b) if somehow the microscopic configuration is fine-tuned.

\begin{figure}[t]
\centering
\includegraphics[height=2.0in]{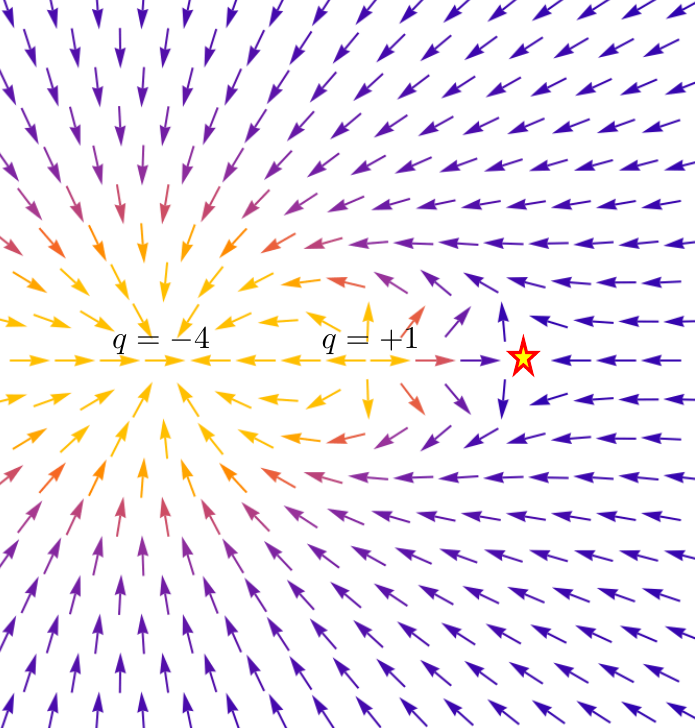}  
\caption{Field around two charges, special point indicated with a star.}
\label{fig:tuning}
\end{figure}

What do I mean by a `fine-tuned' microscopic configuration?  This is relatively straightforward to describe, but the main element to bear in mind is that it ought not to be secretly a configuration of enhanced symmetry.  Let us go to the simplest multipole; the monopole.  Suppose there are two ions, one of charge $q_1=+1$ and the other of $q_2=-4$, as depicted in \Fig{fig:tuning}.  Some external physicist places the former at a distance $1$m from you and the latter a distance $2$m, both along the line of sight.  You measure the nature of the electrostatic field with your own probe charge and find that the force on it vanishes at your location and is very weak in the nearby vicinity.  You might first think that there is only either a very small charge nearby or a large one very far away.  However you then move further away and see that a monopole-like field appears and, once you get very far away, much further than the separation of the charges, it looks very much like a charge $-3$ point object.  The explanation for the vanishing monopole at the beginning is fine-tuning: You were simply at a special location where an accidental cancellation occurs, thus the measurements performed there did not truly reflect the underlying microscopic nature of the setup.  Furthermore, if you displace your probe charge even by a small amount it will quickly move away into a region of much stronger field, as depicted by the arrows in \Fig{fig:tuning}.  In other words, the region of fine-tuning is unstable against small changes in the fundamental parameters; in this instance the parameter is the location of your probe charge.

Something very similar could, and occasionally does, occur in a quantum field theory.   Suppose there is some observable quantity, such as a mass or coupling, that a) one can measure experimentally and b) that is predicted within the more fundamental microscopic theory.  You measure this parameter at some fixed energy scale and find it is small.  However, the value of that parameter is renormalisation-group dependent, in that it receives contributions from physics over a range of scales.  You go ahead and measure this parameter at very short distance scales and find it is much much larger.  In fact, you see that the original parameters were finely balanced such that, RG-evolving over scales, various contributions cancelled out to leave a result that is much smaller than `typical'.  This is very analogous to the probe charge in the background field of \Fig{fig:tuning}, where the RG evolution is like moving this probe along contours in this plane.

Let's bring all of this back to the Higgs mass.  The SM is an EFT with some as-yet-unknown UV-completion within which the SM parameters are calculable in terms of more, or less, other fundamental parameters, just like the pion mass is calculable in terms of the quark masses and QCD gauge coupling at some scale.  What is now puzzling, after LEP and in the face of present day measurements at the LHC, is that it seems the scale of the `UV-completion', let's call it $\Lambda$, is well above the Higgs mass.  The reason being that nothing else has shown up yet at accessible energies.  So we appear to have $m_h^2 \ll \Lambda^2$.

We know that one can have great separation in dimensionful scales between IR and UV parameters in two instances.  The first requires a symmetry.  This doesn't seem to be the case though as there is no symmetry restored in the EFT when we take $m_h^2 \to 0$ (the cutoff is a part of the EFT remember).  The only other option left is fine-tuning.  The question then is why nature would have fine-tuned the Higgs mass?  For sure fine-tuning occurs all over the place in the natural world, however it usually comes at the price of some sort of plenitude.  For instance there are many elements, and so statistically one would expect some of them to have unexpectedly low excited energy levels.  One example of this is the Thorium-229 nucleus, with an excited \emph{nuclear} state which, due to accidental cancellations, only has an excitation energy of $8.3$ eV!  Viewed in the context of a single nucleus this fine-tuning at the $\mathcal{O}(10^{-3})$-level seems very unlikely to have happened.  Viewed in the context of the $\mathcal{O}(10^{2})$ or so elements of the periodic table it doesn't seem all that outrageous given that with so many elements one would have expected occurrences at the $\mathcal{O}(10^{-2})$-level in any case.  Better known, on a much larger scale, there are many planets and moons in the solar system/galaxy/Universe, so it wouldn't be surprising if some of them, like ours, exhibited a lunar angular diameter matching the host star, such that perfect solar eclipses are possible.  In both contexts the fine-tuning is explained in terms of a landscape of examples.  However, the SM is but one...

So, why is the Higgs boson apparently significantly lighter than the states present at the UV-completion of the SM?  This is variously known as the `hierarchy problem', `Higgs Naturalness', `Higgs fine-tuning', and so on.  All refer essentially to the same puzzle.  So, is it due to a symmetry?  Then which symmetry and what are its predictions?  Is it due to fine-tuning of parameters?  Then how did that fine-tuning come about?  What is its rationale?  Or, maybe, our picture of the SM as an EFT is misguided, or maybe EFT itself has broken down?  If so, then what \emph{is} happening?

I wish I could answer these questions.  Sadly, all I can do is sketch some \emph{possible} answers...

\section{A Pion-like Higgs?}
Since discovering the Higgs boson we have cornered the `How?' of EW symmetry breaking.  However, we don't yet have the `Why?'  There is a strong analogy between the Higgs sector of the SM and the pions of QCD which I think will be useful to add context to this question.

As we know, in the `UV' (i.e.\ short distance physics) the quarks and gluons of QCD have perturbative interactions amongst themselves.  However, in the `IR' (i.e.\ long distance physics) this coupling becomes strong and the quarks and gluons cannot be considered asymptotic states on long distance scales.  Instead, they are bound into hadrons which become the true asymptotic states on distance scales greater than the size of the proton, for example.  In this sense, the hadrons (especially the pions) and their associated `IR' theory description known as the `Chiral Lagrangian' are the `How' of chiral symmetry breaking and confinement, and QCD is the `Why'.

To study this more closely let's go down to life below $1$ GeV.  Working below this energy scale we observe that there are three pion degrees of freedom, packaged into a neutral pseudoscalar field $\pi^0$ and a charged field $\pi^\pm = \pi_1 \pm i \pi_2$, with masses $m_0 = 135$ MeV and $m_\pm = 140$ MeV.  Clearly they are very close in mass, so one might assume there is some symmetry that almost enforces their mass to be equal.  In fact, it is a good idea to think of these pions to be packaged into the adjoint representation of $\SU(2)$, as $\Pi = e^{\sum_i \pi_i \sigma_i/f_\pi}$, where the latter are simply the Pauli matrices, and an $\SU(2)$ transformation takes $\Pi \to U \Pi U$, where $U$ is a unitary $2\times2$ matrix.

Now we may trivially write their mass in an explicitly symmetry-invariant manner
\be
\mathcal{L}_{Mass} = \frac{1}{2} m_\pi^2 f_\pi^2 \text{Tr} \Pi \to \frac{1}{2} m_\pi^2 (\pi_0^2 + \pi_1^2 + \pi_2^2) +... = \frac{1}{2} m_\pi^2 \pi_0^2 + m_\pi^2 \pi^+ \pi^- +... ~~.
\ee
Lets do a spurion analysis.  The parameter $m_\pi$ is the only spurion that breaks a shift symmetry acting on the pions (which to lowest order in some constant $\epsilon$, acts like $\pi \to \pi +\epsilon$), thus if we think of this as an EFT then UV effects will not generate large corrections to their mass.  Furthermore, this parameter respects the $\SU(2)$ symmetry, thus all UV corrections will respect the symmetry and the pions will continue to have the same mass.

So far so good.  We have a pretty decent theory for the pions.  However, there is an elephant in the room.  The charged pions interact with the photon through the kinetic terms
\be
\mathcal{L}_{Kin} =  \frac{1}{2} (\partial_\mu \pi_0)^2 +| (\partial_\mu + i e A_\mu) \pi^+ |^2 ~~.
\ee
This interaction not only breaks the $\SU(2)$ symmetry, since it only affects the charged pions, but it also breaks their shift symmetry!  Although it may look innocuous, this is not some minor modification of the theory.  In fact, it \emph{completely} destablises the entire setup.  Even without performing any calculations we know we now have a spurion parameter $e$ that breaks these symmetries, thus if we consider this as an effective field theory, which we should, then there is absolutely nothing to forbid corrections arising at the quantum level which scale as
\be
\delta \mathcal{L}_{Mass} \sim \frac{e^2}{(4 \pi)^2} \Lambda^2 \pi^+ \pi^- ~~,
\ee
where the $4 \pi$ factor is typical for a quantum correction.  Now we have a hierarchy problem, since if $\Lambda \gtrsim 750$ MeV then we would have a big puzzle, as these corrections would be greater than the observed mass splitting.  How can we address this puzzle?  The most obvious answer is that it must be the case that $\Lambda \lesssim 750$ MeV, thus the `UV' begins to show up at that scale.  In other words there \emph{must} be new fields and interactions that become relevant at a scale of $E\sim750$ MeV that will somehow tame these corrections.  It turns out that nature did indeed choose this route, and in fact the $\rho$-meson shows up, alongside all the rest of the fields associated with QCD, and then eventually at higher energies the quarks and gluons themselves.  All of this physics at the cutoff and above then explains why the pion mass splitting is what it is (see \cite{Das:1967it}).  The actual correction is
\be
m_{\pi^\pm}^2-m_{\pi^0}^2 \approx \frac{3 e^2}{(4 \pi)^2} \frac{m_\rho^2 m_{a_1}^2}{m_\rho^2 + m_{a_1}^2} \log\left(\frac{m_{a_1}^2}{m_\rho^2 }  \right)
\ee
where $\rho$ and $a_1$ are the lightest vector and axial vector resonances.  So this hierarchy problem is resolved very clearly in QCD.  The quadratic correction from electromagnetism very much exists and is calculable.  New composite resonances kick in to tame these quadratic corrections, and soon after that, above the QCD scale, the pion itself is no longer a physical state as it is a composite made up of fermions.  Fermions do not receive quadratic corrections to their mass, so we can understand why the pion mass splitting is not sensitive to physics at, for example, the Planck scale!

Imagine, however, that the expected new physics had not shown up at $E\sim750$ MeV.  We would have a really big puzzle and we would have to try and understand what is going on.   We could simply add an additional parameter to our action 
\be
\delta \mathcal{L}_{Tune} \sim \delta_m^2 \pi^+ \pi^- ~~,
\ee
and then fine-tune this against the other corrections to keep the sum small, however this would seem very ad.\ hoc.  Nature did not choose this route.  Instead, nature chose for the mass splitting to be natural ($=$ not fine-tuned).  In essence, the requirement of naturalness is satisfied precisely as we would expect from taking the measured mass splitting and turning it around to predict new fields at some energy scale!

Nowadays with the Higgs boson we are in a similar situation.  We have a scalar field, the Higgs.  If the Lagrangian were simply the kinetic terms and its mass, then we would have no problem at all, because the mass would be the only parameter that breaks a shift symmetry for the Higgs, hence it would be plausibly stable against UV quantum corrections.  However, we also have the gauge interactions that break any shift symmetry, just like the pions, but also more importantly the Yukawa interactions
\be
\mathcal{L}_{Yukawa} =  \lambda H Q U^c + h.c. ... ~~,
\ee
which also also break the shift symmetry, where the top Yukawa is the most significant breaking term.   We may thus pursue exactly the same reasoning as for the pions.  Whatever the UV-completion of the Higgs sector, at the quantum level there should arise corrections to the Higgs mass which scale as
\be
\delta \mathcal{L}_{Mass} \sim \frac{6 \lambda_t^2}{(4 \pi)^2} \Lambda^2 |H|^2 ~~.
\ee
In natural units $\lambda_t \approx 1$, thus for these mass-squared corrections to remain below the EW scale we require $\Lambda \lesssim 500$ GeV.  Just as for the pions, unless some new physics kicks in around this scale we have an issue, which is that if the cutoff of the SM exceeds $500$ GeV, then there must be some sort of fine-tuning taking place.

So, we see that the hierarchy problem is not some wishy-washy notion, but is in fact very crisp and familiar.  Furthermore, it points directly to the $\sim$ TeV scale as somewhere where \emph{something} ought to be going on.  For the pions the reasoning of EFT worked beautifully, so what is going on with the Higgs?  Let's try and figure it out.

\subsection{pNGB Higgs}
The first and perhaps most obvious possibility to consider is whether the Higgs boson is really just like the pions of QCD.  This is something that has been considered for some time because it is a very interesting and well-motivated possibility.

When discussing how a scalar mass might be kept small I have frequently referred to the scalar enjoying a shift symmetry.  There is in fact a natural setting in which such symmetries arise and may even be generalised beyond to non-Abelian shift symmetries that include non-linear interactions.  It is a deep and very beautiful theorem, proven by Jeffrey Goldstone and others \cite{Nambu:1960tm,Goldstone:1961eq}, that when an exact global symmetry is spontaneously broken this gives rise to massless scalar bosons.  More specifically, if a global symmetry $\mathcal{G}$ is spontaneously broken to a smaller symmetry $\mathcal{H}$ then the theory will contain massless Nambu-Goldstone bosons living in the coset space, $\mathcal{G}/\mathcal{H}$.  In fact, the theorem really goes beyond masslessness.  The Nambu-Goldstone bosons will have no scalar potential at all and you can picture their vacuum manifold as being a space along which the vacuum energy is always the same.

Now, calling it a broken symmetry is actually a bit of a misnomer, because the entire symmetry $\mathcal{G}$, is actually always there, however in the Lagrangian we will see the remaining symmetry $\mathcal{H}$ very explicitly as a linearly realised symmetry, with fields transforming in the usual way, whereas the symmetry for the other generators of $\mathcal{G}$, described by generators living in $\mathcal{G}/\mathcal{H}$, will actually be less apparent, and we only see it by its `non-linear' action on the Goldstone bosons, which to linear order will correspond to the shift symmetry we desire.

If the global symmetry is also explicitly broken then the fields are not true Goldstone bosons, but if this explicit breaking is small then they may still be much lighter than the other scales in the theory.  Now, since the symmetry is explicitly broken, we call them `pseudo-Nambu-Goldstone bosons' (pNGBs).

With this in mind, we can see why the pions were light.  In the UV theory, which is QCD, the up and down quark masses are much smaller than the strong coupling scale, and these are the only parameters that explicitly break an $\SU(2)_A$ chiral symmetry acting on the up and down quarks.  If we assume these mass parameters are the same, then the approximate action is
\bea
\mathcal{S} = -\int d^4 x\,   \left[  \mathcal{L}_{Kin} +  M (\bold{q}_L^\dagger \cdot \bold{q}_R + h.c)  \right]  ~~.
\eea
where $M \ll \Lambda$ and the quarks are in doublets.  This action respects an $\SU(2)_V$ vector flavour symmetry $\bold{q}_L \to U \bold{q}_L$,  $\bold{q}_R \to U \bold{q}_R$, however in the limit $M\to 0$ the symmetry is doubled to two independent symmetries $\bold{q}_L \to U_L \bold{q}_L$,  $\bold{q}_R \to U_R \bold{q}_R$.  This means that the mass parameter explicitly breaks an $\SU(2)_A$ axial global symmetry.  When the quarks condense due to QCD $\langle \bold{q}_L^\dagger \cdot \bold{q}_R \rangle \propto \Lambda^3_{QCD}$ the axial symmetry is spontaneously broken $\SU(2)_A\to 0$, hence we get $(2^2-1)$ Goldstone bosons.  These Goldstone bosons are the three pion degrees of freedom we have been discussing all along.  When the quark masses are turned on the symmetry is explicitly broken, thus the pions become massive, however they can be \emph{naturally} lighter than the cutoff as the quark mass is the only parameter that breaks the shift symmetry, which is the IR manifestation of the UV chiral symmetry!

So the obvious question is:  Could the Higgs be a pNGB and even perhaps composite, emerging from some strongly coupled gauge theory in the UV, just like the pions?  This has been an extremely active area of investigation and the answer is yes, the Higgs could be just like the pion and, just as the charged pions have gauge interactions that break their shift symmetry, so too can a pseudo-Goldstone Higgs boson.  The top quark interactions which explicitly break the global symmetry, and hence the shift symmetry, are very large however, so some work is required to have them not lead to very large Higgs mass corrections.

Let's see how this goes.  Since the Higgs is not massless it is not really a Goldstone, but could be a pseudo-Goldstone boson, just like the pions, thus I will refer to models of this class as pNGB Higgs models, (pseudo-Nambu-Goldstone boson Higgs).  The main classes of models of this class are composite Higgs models (just like pions), Little Higgs models (similar technology, but with machinery that can protect the Higgs mass to higher loop orders), and the more recently popular Twin Higgs models.  In the next section I will sketch the main ideas common to composite and Little Higgs models.  However, if you wish to know more the lectures by Contino \cite{Contino:2010rs} not only beautifully explain the field theory behind composite Higgs models, but also present the types of models more commonly considered for vanilla composite Higgs scenarios.  The review by Schmaltz and Tucker-Smith on the Little Higgs models is a superb starting point for these models \cite{Schmaltz:2005ky}.

\subsubsection{Towards the Higgs}
The basic recipe is the following.  Let us take some symmetry $\mathcal{G}$ with a gauged subgroup $\tilde{\mathcal{G}}$, spontaneously broken to $\mathcal{H}$ with a gauged subgroup $\tilde{\mathcal{H}}$.  Now, we have have $N_G=\text{dim}(\mathcal{G})-\text{dim}(\mathcal{H})$ Goldstone bosons, of which $N_g = \text{dim}(\tilde{\mathcal{G}})-\text{dim}(\tilde{\mathcal{H}})$ are eaten by the gauge bosons, leaving $N=N_G-N_g$ massless scalars at tree-level.

Thus we see that in order to fit a Higgs doublet into this concoction we must have at least $N\geq 4$ and $\tilde{\mathcal{H}} \supset \SU(2)\times\U(1)$.  A great deal of effort has gone into enumerating the possibilities, but let us study the absolute simplest one.  This model has $\mathcal{G} = \SU(3)$, $\mathcal{H}=\tilde{\mathcal{H}}=\SU(2)$, thus the number of Goldstone bosons is $N=8-3=5$.  This model in fact does not respect custodial symmetry, which means that the dangerous operator
\be
\mathcal{O}_T = \frac{1}{\Lambda^2} (H^\dagger D_\mu H)^2
\label{eq:Tparameter}
\ee
that modifies the SM prediction for the W to Z-boson mass ratio can be generated by the physics at the UV scale, so this model is actually very much disfavoured by the precision LEP measurements.  Nonetheless, this model is so simple that it serves well as a straw man for pNGB scenarios, so we will study it here.

The low energy dynamics of the pNGBs are described in full generality by the CCWZ construction \cite{Coleman:1969sm,Callan:1969sn}, that I encourage you to study, however for these lectures it suffices that we may capture the relevant operators by considering what is generally known as a non-linear sigma model, with the field parameterisation
\be
\Sigma = e^{i \Pi/f} \Sigma_0, ~~~~~~ \Pi = \pi^a T^a,
\label{eq:sigma}
\ee
where $\pi^a$ are the pNGBs, $T^a$ are the broken generators of $G$, and $\langle \Sigma \rangle = |\Sigma_0| = f$. The global symmetry breaking is induced by a scalar field, $\Sigma$, transforming as a $\mathbf{3}$ under $\SU(3)$, which acquires a vacuum expectation value 
$\Sigma_0 = (0,0,f)$. The pNGBs can thus be parameterized by the non-linear sigma field as in Eq.~(\ref{eq:sigma}),
with
\begin{equation}
\Pi = \pi^a T^a = 
\left( 
\begin{array}{ccc}
0 & 0 & h_1\\
0 & 0 & h_2\\
h_1^\dag & h_2^\dag & 0
\end{array}
\right) + \dots,
\label{eq:pion}
\end{equation}
with $T^a$ the broken generators of $\SU(3)_W$ and we have not included the additional singlet pNGB corresponding to the diagonal generator.  We may write the sigma field explicitly in terms of the Higgs doublet as
\begin{equation}
\Sigma  =  
\left( 
\begin{array}{c}
i h_1 \, \displaystyle{ \frac{\sin( |h|/f)}{ |h|/f}}\\
i h_2 \, \displaystyle{ \frac{\sin( |h|/f)}{|h|/f}} \\
f\, \displaystyle{ \cos( |h|/f)}
\end{array}
\right),
\end{equation}
where $|h|\equiv \sqrt{h^\dag h}$. 

\subsubsection{Gauge Interactions}
The gauge interactions can be added in the usual way.  If we wish, we can add them in an $\SU(3)$-invariant manner, with the covariant derivative
\be
D_\mu \Sigma ~~,~~ D_\mu = \partial_\mu + i g \sum_a W^a_{\mu} \lambda^a ~~,
\ee
where $\lambda^a$ are the $\SU(3)$ generators.  Then we simply set all but the $\SU(2)$ gauge fields to zero.  Note that after electroweak symmetry breaking the interaction strength of the physical Higgs boson with the SM gauge fields is suppressed by a factor\footnote{This may be found from the usual relation $c_{hVV}=\frac{1}{g m_V} \frac{\partial m_V^2}{\partial h} $.}
\be
\cos (v/f) \approx 1-\frac{1}{2} \frac{v^2}{f^2}  ~~,
\ee
thus we can test a pNGB scenario like this by looking for modified Higgs interactions!

Gauging a subgroup of the full global symmetry is an explicit breaking of the global symmetry, thus the pNGB Higgs mass is not protected against quadratic corrections in the gauge sector.   Indeed, in analogy with the pion mass corrections from before, at one loop gauge interactions will generate a Higgs mass-squared proportional to
\be
\delta m_h^2 \sim \frac{g^2}{16 \pi^2} \Lambda^2
\ee
where $\Lambda$ is the UV cutoff.  In a pNGB model where this is the full story then one must follow calculations such as for the pion mass splitting, which include a priori unknown form factors, in order to estimate the correct magnitude of the correction.  Note that in order for these corrections not to be too large one requires that the cutoff is not too far away, and thus the full-blown dynamics of the composite sector, including heavy vector mesons, should/could be accessible at the HL-LHC.

We may also employ additional tricks to suppress these corrections.  Imagine we didn't switch off the additional gauge bosons.  Then we would have the full $\SU(3)$ symmetry, however we wouldn't have any leftover Goldstone bosons to play the role of the Higgs doublet.  Then let us instead take two separate $\Sigma$ fields, each with their own $\SU(3)$ global symmetry, but we gauge the diagonal combination of these symmetries, such that both fields are charged under the $\SU(3)$ gauge symmetry.   When both fields get a vev we get two sets of $\SU(3)/\SU(2)$ Goldstone bosons.  One set is eaten, but the other set remains light.  Since the original theory was written in a fully $\SU(3)$-invariant manner, no quadratic divergences arise.  At worst, at one loop the gauge interactions will induce dangerous interactions such as $(\Sigma_1 \cdot \Sigma_2)^2$, but this is suppressed by a loop factor, such that the resulting correction to the Higgs mass is
\be
\delta m^2 \sim \frac{g^4}{16 \pi^2} \log\left( \frac{\Lambda^2}{\mu^2} \right) f^2
\ee
which is significantly smaller than the correction in the simplest model!  This is the essence of the Little-Higgs trick for the gauge sector, and it can be extended to a greater number of symmetries to further suppress these corrections.

\subsubsection{Top Quark Interactions}\label{sec:toppart}
We must also accommodate the top quark Yukawa.  The simplest way to do this is to work in analogy with the gauge sector.  With the gauge sector we start with a full $\SU(3)$ gauge multiplet, and set some fields to zero, which explicitly breaks the symmetry.  Here we may do the same, by introducing the incomplete $\SU(3)$ quark multiplet $Q = (t,b,0)$, alongside the usual right-handed top $t_R$.  Then the Yukawa interaction can be written in an $\SU(3)$-invariant manner
\be
\mathcal{L}_\lambda = \lambda_t Q \cdot \Sigma t_R
\ee
Of course, just as with the gauge sector, at one loop a quadratically divergent Higgs mass correction will be generated
\be
\delta m^2 \sim \frac{3 \lambda_t^2}{16 \pi^2} \Lambda^2 ~~.
\ee
One may wish to simply tolerate this, and thus place strong limits on how large $\Lambda$ can be for the solution to the hierarchy problem to really hold.  Other options include adding `top partner fields'.  For composite Higgs scenarios there are numerous possibilities, thus I refer the interested reader to \cite{DeSimone:2012fs} for an overview.  I'll just sketch a basic example showing how these extra fields may cancel quadratic divergences.

\begin{figure}[t]
\centering
\includegraphics[height=1.3in]{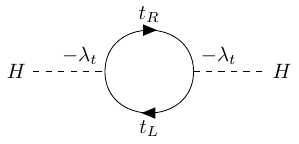} ~~\includegraphics[height=1.3in]{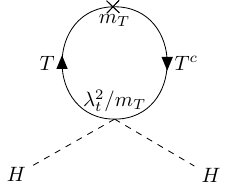} 
\caption{The cancellation between quadratic divergences from the top quark loop and fermionic top-partner loops.}
\label{fig:toppart}
\end{figure}

Let us really make the interaction $\SU(3)$-invariant by putting the missing field back in $Q = (t,b,T)$, and also add a little bit of explicit breaking of $\SU(3)$ by coupling $T$ to a conjugate fermion to give it a Dirac mass $M_T \ll \Lambda$.  Thus we have
\be
\mathcal{L}_\lambda = \lambda_t Q \cdot \Sigma t_R + M_T T^c T ~~,
\label{eq:toppart}
\ee
where now the SM right-handed top quark will in general be a linear combination of $t_R$ and $T_c$.  At high energies $M_T$ is just a small perturbation, and the Yukawa is fully $\SU(3)$ symmetric, thus based on symmetry reasons alone the largest quadratic correction we can generate for the Higgs mass can at most be of the order
\be
\delta m^2 \sim \frac{3 \lambda_t^2}{16 \pi^2} M_T^2 ~~.
\ee
If $M_T \ll \Lambda$ then we have tamed, to some degree, the quadratic corrections to the Higgs from the top sector.  One can show that this setup leads to a diagrammatic cancellation of the form shown in \Fig{fig:toppart}.  However, we now have an additional coloured fermion that we can search for at the HL-LHC.

%Alternatively, the little Higgs recipe is very simple.  One again writes the theory in a manifestly $\SU(3)$ manner, as with the gauge interactions, by completing the quark multiplet $Q = (t,b,T)$ and adding an extra right-handed top quark, such that we now have two $t_{R_1},t_{R_2}$.  The Yukawa interaction is now
%\be
%\mathcal{L}_\lambda = \lambda_1 Q \cdot \Sigma_1 t_{R_1} + \lambda_2 Q \cdot \Sigma_2 t_{R_2} ~~.
%\ee
%In the mass-eigenstate basis we see we have an action that is actually free of quadratic divergences at one-loop, with the relevant diagrams shown in \Fig{fig:toppart}!  \MCcomment{This is set as homework.}  Of course, as before, the dangerous log-divergent quartic shows up, but this is tolerable.

\subsubsection{Fine-Tuning}
\label{sec:FT}
This concludes a summary of the basic features of pNGB Higgs models.  The details are much more involved than I have sketched here, however these basic building blocks should provide enough coverage to delve into the literature!  The last thing to consider is the fine-tuning in these theories.  This stems from two angles.  In any microscopic theory explaining the `Why?' of EW symmetry breaking there are two observable parameters that must be \emph{predicted} from within the UV theory.  These are the scale of EW symmetry breaking (Higgs vev) and the Higgs mass.  Let us focus on the former.  An excellent reference for this discussion is \cite{Panico:2015jxa}.

Due to the nature of the global symmetry if one assumes all of the explicit symmetry breaking arises due to the gauge and Yukawa couplings then the leading contributions to the Higgs potential are of the form
\be
V(h) \approx \beta f^2 \Lambda^2 \left(-r \sin^2 \frac{h}{f}+\frac{1}{2}\sin^4 \frac{h}{f} \right) ~~,
\label{eq:pngbtune}
\ee
where $\beta$ and $r$ are expected to be $\mathcal{O}(1)$ coefficients.  The minimisation condition for this scalar potential doesn't care about the overall prefactor, thus we find that we require
\be
r = \sin^2 \frac{v}{f} ~~.
\ee
There is no symmetry in this theory that can suppress the coefficient of one term over the other, thus to have small $v/f$ requires that parameters in the UV theory are fine-tuned so as to largely cancel and give a small value for $r$.  But, we need $v/f \ll 1$ in order to accommodate the present constraints on Higgs coupling modifications.

This fine-tuning issue is a generic problem for pNGB Higgs models and is known as `$v/f$-tuning'.  It exists independently of the tuning require to obtain a small Higgs mass, i.e.\ small $\beta$ and so is, in some sense, a lower limit on the amount of fine-tuning that exists in this class of theories.\footnote{There are exceptions if one relaxes the assumption that all explicit symmetry breaking comes from the gauge and Yukawa interactions \cite{Durieux:2021riy,Durieux:2022sgm}.}

Often we attempt to quantify the fine-tuning, not as an exact science, but as a means to understand how plausible a physical theory is, assuming nature doesn't arbitrarily fine-tune parameters just for our amusement.  We typically call this parameter $\Delta$ and, if large, the theory is fine-tuned.  So here we see that
\be
\Delta \gtrsim \frac{f^2}{v^2} ~~.
\ee
Note, however, that the corrections to the Higgs couplings scaled similarly, thus we have that
\be
|\Delta| \gtrsim \frac{1}{2 \delta_{hVV}} ~~.
\ee
Hence in minimal incarnations of pNGB (or pion-like) Higgs models one find a direct connection between the magnitude of modifications to Higgs couplings and the amount of fine-tuning present in the theory.  The more SM-like, the more fine-tuned!   This will inform our perspective on these models at the HL-LHC, as we now discuss.

\subsubsection{HL-LHC Prospects}
Near-future experimental prospects for a pNGB Higgs are interesting and come from three angles.

\begin{center}
\textbf{Higgs Couplings}
\end{center}
As we saw, one expects modifications to Higgs couplings in pNGB scenarios and these modification are directly tied to the amount of fine-tuning, hence they are indicative of the `plausibility', by some metric, of the theory.  Presently we have constraints on Higgs couplings with a precision of around $7-8\%$, or globally around the $6\%$ level \cite{ATLAS:2022vkf,CMS:2022dwd}.  In fact, in the vector couplings ATLAS and CMS both have a slight preference for an enhanced coupling, although this is not significant, placing even more stringent constraints on pNGB models at present, since coupling reductions are more difficult to accommodate with the measured data.

Either way, for sake of comparison with HL-LHC expectations I will assume a SM-like central value, and present constraints at around the $7\%$ level.  At HL-LHC we expect the coupling precision to increase significantly, down to the $1.5\%$ level \cite{deBlas:2019rxi}.  Note that this is consistent with a $\sqrt{\mathcal{L}}$ scaling.  There are two ways to look at this.  What is presently a $\sim 1.5 \sigma $ fluctuation could grow to a $5\sigma$ discrepancy in Higgs couplings at the HL-LHC!  There really is plenty of room left for surprises to show up in Higgs couplings.

For another perspective, a SM-like Higgs with present coupling sensitivity corresponds to fine-tuning around the $14\%$ level at the least.  On the other hand, if the Higgs coupling measurements persistently remain SM-like, to within precision, at the HL-LHC then in minimal pNGB-like Higgs scenarios the fine-tuning would grow considerably, to around the $3\%$ level at a minimum.

\begin{center}
\textbf{Vector Resonances}
\end{center}
You will recall that in QCD a great number of resonances, beyond the lightest pions, arise as a consequence of confinement in QCD.  Thus, similarly, were the Higgs to be pion-like, arising as a composite from some strongly-coupled high energy sector, we should also expect a slew of heavy resonances to come along for the ride.  In particular, vector counterparts of the pions, such as the $\rho$-meson in QCD, ought to arise.  Such resonances are thus characteristic, and largely unavoidable, expectations for a pion-like Higgs.

\begin{figure}[t]
\centering
\includegraphics[height=3in]{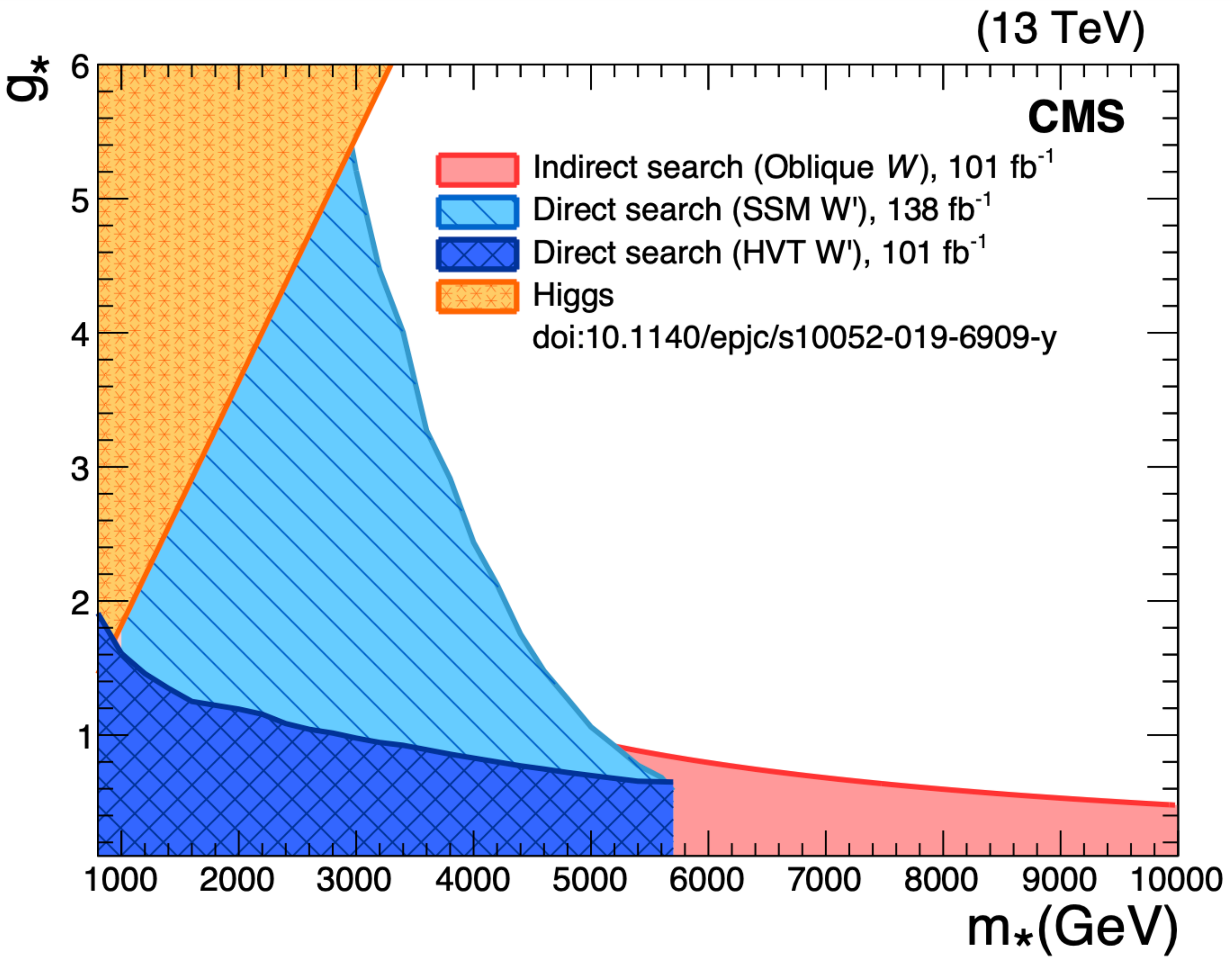} 
\caption{CMS limits, taken from \cite{CMS:2022krd}, on the heavy vector resonances expected in pNGB-like Higgs scenarios.}
\label{fig:cmscomp}
\end{figure}

These resonances would couple to the EW gauge sector and essentially mix with the EW gauge bosons in the UV.  As a result they would inherit couplings to fermions and could thus arise with very clear signatures such as dilepton or diboson resonances.  Present limits on dilepton resonances are coupling-dependent, however, under certain assumptions they can already exceed $5$ TeV (see \Fig{fig:cmscomp}).

What are the prospects for the HL-LHC?  Since, as a function of mass, the production cross section for a resonance depends on the parton PDFs one can not perform an easy rescaling to determine the reach.  However, at this point I introduce you to the \href{http://cern.ch/collider-reach}{Collider Reach} tool.\footnote{Be careful to use only http, not https, which won't open.} This tool makes simple assumptions and rescales according to pdfs.  Using it one finds that the mass reach should extend from $5$ TeV to above $6.5$ TeV at HL-LHC, which is indeed consistent with more dedicated studies \cite{Thamm:2015zwa}.
%XXX Note that \cite{Panico:2021vav} gives more than a factor 2 improvement between 300 and 3000, meaning 2.2 sigma to more than 5 sigma.  It is essentially $7 \times 10^{-5}$ to $3 \times 10^{-5}$.

\begin{center}
\textbf{Coloured Resonances}
\end{center}
We saw in \Sec{sec:toppart} that we required coloured (QCD-charged) fermionic states to complete the multiplets in our example pNGB Higgs scenario.  It turns out that this is generically the case and so a typical prediction for a pNGB-like Higgs scenario is the existence of such coloured fermions around the TeV scale, with a rich and interesting phenomenology \cite{DeSimone:2012fs}.

At present, we have not observed such states at the LHC, up to around $1.5$ TeV \cite{CMS:2022fck}.  This is already starting to put fine-tuning pressure on vanilla pNGB models since $\frac{3}{16 \pi^2} (1.5\text{ TeV})^2 \sim (200\text{ GeV})^2$, typically leading to fine-tuning contributions at the $\mathcal{O}(10\text{'s})~ \%$ level.  As we will soon see, this leads model builders to consider how `generic' the requirement for such top partners is.  How will things change at the HL-LHC?

\begin{figure}[t]
\centering
\includegraphics[height=3in]{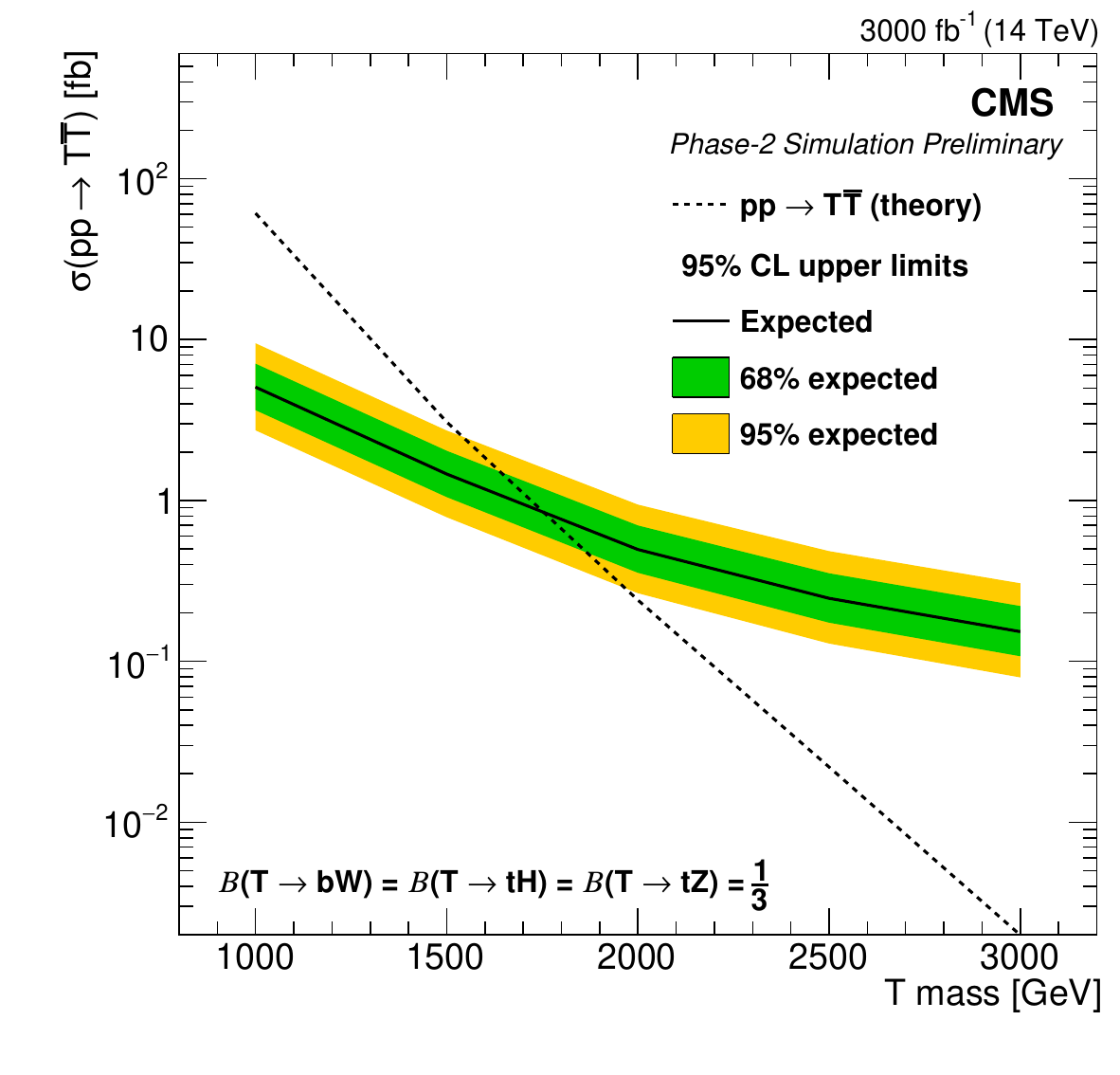} 
\caption{CMS projection, taken from \cite{CMS:2022jeb}, of the HL-LHC reach for vector-like T quarks.  Note the slope of the cross section with mass.}
\label{fig:cmsestimate}
\end{figure}

For broad consideration of BSM resonances at HL-LHC I draw your attention to \cite{CidVidal:2018eel}.  For the case at hand, one does not expect the reach to increase significantly as compared to the present day.  The reason is made clear in \Fig{fig:cmsestimate}.  The dashed line shows the pair production cross section for a particular type of heavy coloured fermionic resonance, alongside the projected reach.  With present limits around $1.5$ TeV, based on $\sqrt{\mathcal{L}}$ scaling we should expect to be able to access a cross section which is a factor $\sim4.6$ or so smaller.  From the dashed line of \Fig{fig:cmsestimate} we see this corresponds to masses around $\lesssim 1.8 $ TeV, which is indeed where the expected limit line for HL-LHC crosses the predicted cross section.

As a result, we see that, due to a production cross section which falls steeply with mass, largely because of the decrease of PDFs at larger $x$, for fixed $\sqrt{S}$ we ought not to expect the limit on the coloured fermionic resonances typical of pNGB models to change significantly between now and the end of the HL-LHC.

\subsection{Twin Higgs}
We have seen that the non-observation of coloured resonances at the LHC pushes their mass upwards to a point at which a vanilla pNGB-like Higgs scenario is becoming fine-tuned.  Qualitatively this essentially implies that a pNGB-like Higgs scenario is less likely to be truly realised in nature.  After all, why would nature have arbitrarily fine-tuned microscopic parameters such that macroscopic observables such as the Higgs boson mass are especially small?

There are three plausible logical conclusions:
\begin{itemize}
\item  The Higgs is pNGB-like, in a vanilla scenario, but the theory is a bit fine-tuned.
\item  The Higgs is not pNGB-like.  Look elsewhere.
\item  The Higgs is pNGB-like, but we may have made a critical assumption that leads to apparent fine-tuning.
\end{itemize}
In any healthy scientific process the latter would deserve further scrutiny.  The theory community has undertaken this exercise and essentially `rediscovered' a class of models which indeed expose a key assumption that is not necessary in all cases for a pNGB-like Higgs.  These models are known as `Twin Higgs' models and, importantly, they reveal a possibility in which the connection between fine-tuning and the mass of new coloured resonances is broken or, to some extent, delayed.  In this section I will try to describe, in the simplest terms I can, the theoretical structure underpinning these models and their phenomenology at the HL-LHC.

\subsubsection*{Theoretical Structure}
At its heart the Twin Higgs \cite{Chacko:2005pe} is a pNGB Higgs model.  One begins with a scalar multiplet $H$ transforming as a fundamental under a global $\SO(8)$ symmetry.  The renormalizable potential for this theory is
\be
V = - m^2 |H|^2 + \frac{\lambda}{2} |H|^4
\label{eq:th}
\ee
where we have intentionally written a negative mass-squared.  In the vacuum the global symmetry breaking pattern is $\SO(8) \to \SO(7)$, thus irrespective of the magnitude of $m$ there will exist $7$ massless Goldstone bosons.  It is important to keep in mind that $m$ could be very large and in a theory with new physics scales, $m^2$ will contain all of the UV contributions.  For example, if there are new states of mass $\Lambda$ we expect contributions $m^2 \sim \text{loop} \times \Lambda^2$.  This does not introduce additional quadratic divergences to the mass of the Goldstone bosons since these contributions are $\SO(8)$ symmetric and thus the Goldstone boson masses are still protected by Goldstone's theorem.

Let us break up $H$ into a representation of $\SU(2)_A \times \SU(2)_B \subset \SO(8)$ as
\be
H = \left( \begin{array}{c} H_A \\  H_B  \end{array} \right)   ~~.
\ee
We may rewrite \Eq{eq:th} as
\be
V = - m^2 \left( |H_A|^2 + |H_B|^2 \right) + \frac{\lambda}{2} \left( |H_A|^2 + |H_B|^2 \right)^2 ~~,
\label{eq:symm0}
\ee
which is precisely the same as \Eq{eq:th}, but written in a different manner.   We may also write this as 
\be
V =  \frac{\lambda}{2} \left( |H_A|^2 + |H_B|^2 - \frac{f^2}{2} \right)^2 ~~,
\label{eq:symm}
\ee
where, ultimately, $f^2 = v_A^2 + v_B^2$.

\subsubsection{Gauge Interactions}
We now augment the theory by gauging the two $\SU(2)_A \times \SU(2)_B$ subgroups.  If the vacuum expectation value for $H$ lies completely in the $H_B$ field then the three Goldstone bosons from $H_B$ will be eaten by the $\SU(2)_B$ gauge bosons, to become their massive longitudinal components.  The four Goldstone bosons from $H_A$ will remain uneaten because the off-diagonal gauge bosons of $\SO(8)$ which would have eaten these degrees of freedom were explicitly removed from the theory when we chose not to gauge the full $\SO(8)$ symmetry.  Thus we have four light scalars charged under the unbroken $\SU(2)_A$ gauge symmetry.  It is apparent that if $\SU(2)_A$ could be identified with the SM weak gauge group, and if $H_A$ could be identified with the SM Higgs doublet, then we have a candidate solution of the hierarchy problem!  However there are some further complications which must first be overcome.

The first point to note is that by coupling the scalars to gauge bosons we have introduced a new source of quadratic divergences.  Regularising the theory with a cutoff $\Lambda$ we generate terms such as
\be
V \sim \frac{g_A^2}{16 \pi^2} \Lambda_a^2 |H_A|^2 + \frac{g_B^2}{16 \pi^2} \Lambda_a^2 |H_B|^2  ~~,
\label{eq:gaugeloop}
\ee
where as yet there is no reason to believe the effective cutoff is the same for each field.  However, if we impose an exchange symmetry on the entire theory $A \leftrightarrow B$ then $g_A = g_B$, and assume the UV physics respects this exchange symmetry, such that $\Lambda_a=\Lambda_b$, then the contributions in \Eq{eq:gaugeloop} are equal.  Furthermore, because they are equal they respect the $\SO(8)$ symmetry, thus they do not actually introduce any new quadratically divergent contributions to the Goldstone boson masses.  Hence these dangerous contributions have been ameliorated by a combination of Goldstone's theorem and the fact that an exchange symmetry accidentally enforces an $\SO(8)$-invariant structure on the quadratic part of the action.  This is worth reemphasising: quadratic divergences have not been removed from the theory, but the sensitivity of the Goldstone boson masses to those divergences has been removed by Goldstone's theorem.

Unfortunately at the level of the quartic couplings the picture is not as clean.  The scalar quartic couplings will run logarithmically due to the gauge interactions.  This running must only respect the exchange symmetry and the $\SU(2)_A \times \SU(2)_B$ symmetry, but not necessarily  the full $\SO(8)$ symmetry.  In practice, even if we enforce an $\SO(8)$ symmetric scalar potential in \Eq{eq:symm} at a scale $\Lambda$, at the lower scale of symmetry breaking $m$ we expect additional contributions to the effective potential
\be
V_{BR} \sim \frac{g_A^4}{16 \pi^2} \log \left( \frac{m}{\Lambda_a} \right)  |H_A|^4 + \frac{g_B^4}{16 \pi^2}  \log \left( \frac{m}{\Lambda_b} \right) |H_B|^4 ~~.
\label{eq:TH}
\ee
Even when we impose the exchange symmetry, $g_A = g_B$, $\Lambda_a=\Lambda_b$, these terms explicitly break the $\SO(8)$ symmetry, thus they will in general lead to a non-zero mass-squared for the now pseudo-Goldstone bosons
\be
m_{pNGB}^2 \propto  \frac{g_A^4}{16 \pi^2} m^2 \log \left( \frac{m}{\Lambda} \right) ~~.
\label{eq:BR} 
\ee
This tells us that in this theory we may only hope to have a loop factor in the hierarchy $m_h^2 \sim  \frac{g_A^4}{16 \pi^2} m^2$, and, as $m$ is quadratically sensitive to the cutoff, a loop factor in the hierarchy $m^2 \sim  \frac{g_A^2}{16 \pi^2} \Lambda^2$.  In the end of the day with this mechanism we expect the cutoff scale of the full theory to be an electroweak loop factor above the weak scale, demonstrating that the Twin Higgs can only be a solution to a little hierarchy problem and to solve the full hierarchy problem this theory must be UV-completed.

A final issue is that the theory presented above respects the exchange symmetry $A \leftrightarrow B$.  This implies that the vacuum will also respect this symmetry, with $v_A = v_B$.  Amongst other things, this predicts that the SM Higgs boson would be a perfect admixture of $H_A$ and $H_B$ and would couple to the SM gauge bosons with a suppression factor $\cos \theta_{AB} = 1/\sqrt{2}$.  Clearly this is at odds with observations.  To resolve this issue we can, for example, introduce a small soft symmetry breaking term
\be
V_{SB} = -m_B^2 |H_B|^2 ~~.
\label{eq:break}
\ee
This term explicitly breaks the global symmetry and even the exchange symmetry.  It is important to note that since $m_B$ breaks the exchange symmetry it may be small in a technically natural manner.  Even though the Goldstone bosons have obtained mass from this operator, this mass is insensitive to the cutoff and can be naturally small: $m_{GB} \ll \Lambda$.  Importantly, this exchange symmetry breaking can align most of the vacuum expectation value into the $B$ sector, realising $v_A \ll v_B$.  This will suppress the Higgs mixing with the other neutral scalars and will also allow a hierarchical structure $v_A \ll v_B \ll \Lambda$, at the cost of a tuning comparable to $v_B^2/v_A^2$, which is the Twin Higgs analogue of the $v^2/f^2$ tuning we encountered before.

\subsubsection{Yukawa Interactions}
As far as the scalar fields and the gauge interactions are concerned, this is essentially all that is required of the Twin Higgs model.  Hypercharge may be trivially included in this picture.  The last step is to couple the SM Higgs to fermions.  If we add Yukawa couplings of $H_A$ to fermions, for example the up quarks, as
\be
\mathcal{L} \supset \lambda_A H_A Q_A U^c_A ~~,
\ee
then we see an immediate problem.  The top quark loops lead to $\SO(8)$-violating quadratic divergences
\be
m_A^2 \propto \frac{\lambda_t^2}{16 \pi^2} \Lambda^2
\ee
and the solution has been destroyed.  However, the resolution is immediately apparent.  We enforce the exchange symmetry $A \leftrightarrow B$ by introducing Twin quarks with identical couplings, such that the Yukawa couplings are now
\be
\mathcal{L} \supset \lambda_A H_A Q_A U^c_A +\lambda_B H_B Q_B U^c_B ~~,
\ee
and the quadratic divergences are once again $\SO(8)$-symmetric
\be
V \sim \frac{\lambda_A^2}{16 \pi^2} \Lambda_a^2 |H_A|^2 + \frac{\lambda_B^2}{16 \pi^2} \Lambda_b^2 |H_B|^2  ~~,
\label{eq:toploop}
\ee
since $\lambda_A = \lambda_B$ and $\Lambda_A = \Lambda_B$.  Thus, once again, now in the matter sector the theory at the scale $\Lambda$ is approximately $\SO(8)$ symmetric and the SM Higgs boson is realised as a pseudo-Goldstone boson of spontaneous global symmetry breaking.  At one-loop, as for the gauge sector, a potential is generated for the pNGB from the term
\be
V_{BR,t} \sim \frac{3 y_t^4}{32 \pi^2} \left( \log \left( \frac{m}{\Lambda_a} \right)  |H_A|^4 + \log \left( \frac{m}{\Lambda_b} \right) |H_B|^4 \right) ~~.
\label{eq:THt}
\ee
This gives the dominant contribution to the Higgs potential.

\begin{figure}[t]
\centering
\includegraphics[height=1.0in]{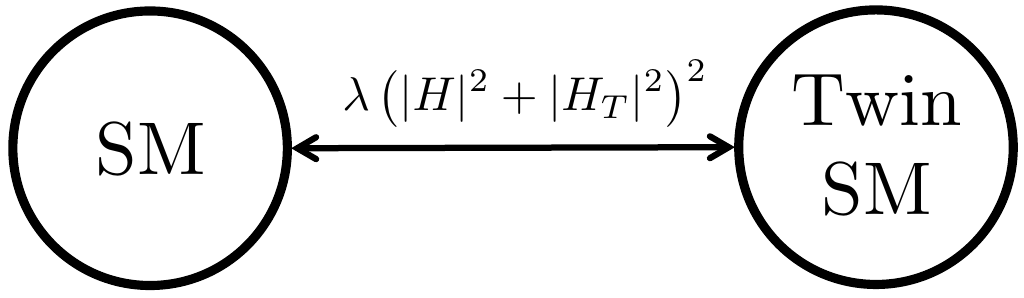} 
\caption{The structure of the Twin Higgs model.  The SM and an entire copy are symmetric under a complete exchange of all fields.  This ensures that the quadratic scalar action respects an accidental $\SO(8)$ symmetry, of which the SM Higgs is a pNGB.  All interactions between the SM and Twin SM are through this single Higgs potential term.}
\label{fig:twin}
\end{figure}

We can also see that if the Twin symmetry is imposed for all degrees of freedom, including gluons and leptons, then at any loop order the Higgs mass will still be free of quadratic sensitivity to the cutoff.  This is the essence of the Twin Higgs mechanism which, in the simplest incarnation, requires an \emph{entire copy of the SM} which is completely neutral under the SM gauge group, but with its own identical gauge groups.  The only communication between the SM and the Twin Sector is through the Higgs boson.  This is depicted in \Fig{fig:twin}.

The presentation of the Twin Higgs mechanism may appear somewhat backwards and a little laborious in comparison to other possible presentations.  This has been intentional, in the hope that it may anticipate a potential misconception for those not familiar with Twin Higgs model.  It is sometimes considered that it is seemingly ad hoc or arbitrary to add an entire copy of the SM for the Twin Higgs mechanism to work.  Hopefully this section has made it clear that there is nothing arbitrary about the introduction of the new fields.  The mechanism is not justified by adding an entire copy of the SM and then proving a diagram-by-diagram, and loop-by-loop, cancellation of quadratic divergences.  Rather, the new fields are introduced in order to realise an exchange symmetry $A \leftrightarrow B$.  The exchange symmetry ensures that at the quantum level the quadratic part of the scalar potential respects an accidental $\SO(8)$ symmetry, even with quadratic divergences included.  The observed Higgs boson mass is insensitive to this $\SO(8)$-symmetric quadratic divergence because it is a pseudo-Goldstone boson of spontaneous $\SO(8)$ breaking.\footnote{It is also possible to see a diagram-by-diagram cancellation of quadratic divergences rather than relying on the symmetry-based argument here, however this is less illuminating.}

Another way to see this insensitivity of the Higgs mass to the UV cutoff is to work in Unitary gauge in what as known as the `nonlinear' representation.  This way we see that all quadratic terms scale as $\sin^2 (h/f)+\cos^2(h/f) = 1$, thus they don't influence the Higgs mass.

\subsubsection{Fine-Tuning}
Thus far the scalar potential is exchange symmetric, meaning that one obtains a pattern of symmetry breaking with $v_A = v_B$ or $v_A=0$, $v_B=f$.  Both options are phenomenologically unviable, the former because the Higgs would not be SM-like, and the latter as it has no EW symmetry breaking.  Thus an additional source of exchange symmetry breaking must exist in order to realise $v_A=v \ll v_B$.

In the end one obtains the same form of scalar potential as in \Eq{eq:pngbtune}, with the exception that the Twin structure leads to the replacement $h^2/f^2\to2 h^2/f^2$, thus the degree of fine-tuning is ameliorated, by a factor 2.  The reason for this is that one always has a `Twin' contribution to the potential, thus the quartic term takes the form
\be
\sin^2 (h/f) \to \sin^2 (h/f)-\cos^2(h/f) ~~,~~ \sin^4 (h/f) \to \sin^4 (h/f)+\cos^4(h/f)
\ee
thus a similar analysis reveals that
\be
\Delta \gtrsim \frac{1}{2} \frac{f^2}{v^2} ~~,
\ee
a factor 2 less than a standard pNGB Higgs model.  This is an improvement, however it is clear that obtaining a SM-like Higgs ($v\ll f$) will still come at the cost of some fine-tuning.

\subsubsection{Phenomenology}
Unlike in standard composite Higgs models, where the copious production of new coloured particles at the LHC is a generic prediction, the collider signatures of the Twin Higgs are thin on the ground.  In both theories a key prediction is the existence of so-called `top partners' which regulate the quadratically divergent top quark loops contributing to the Higgs mass-squared.  In the Twin Higgs these are fermions charged under Twin QCD$_T$ but not under SM QCD.  They are in fact the first known example of a theory with the moniker ``Neutral Naturalness'', used to describe theories in which the top-partners are not charged under QCD.  This drastically suppresses top-partner production at the LHC since the only coupling to the SM is through the Higgs and any top-partner production must go through an off-shell Higgs boson.  The most promising approaches to test the Twin Higgs lie elsewhere.

One very robust prediction of the Twin Higgs scenario is a universal suppression of Higgs couplings to SM states.  The reason for this is that the Higgs bosons from both sectors, $h_A$ and $h_B$, have a mass mixing controlled by the hierarchy of vevs $v_A^2/v_B^2$.  As $h_B$ is a SM singlet this is equivalent to the previous pNGB-like Higgs scenario where all Higgs couplings are diluted by a factor $\cos v/f$.  This mixing may be constrained by searching for an overall reduction in Higgs signal strengths at the LHC and, since the ratio $v_A^2/v_B^2$ is a driving indicator of the tuning in the theory, Higgs measurements directly probe the tuning of the Twin Higgs scenario.  In fact, as we already saw, present constraints on modified Higgs couplings already push this tuning to the $\sim 10 \%$ level.

Another possibility is that due to the Higgs Portal mixing the heavy Higgs boson may be singly produced at the LHC and it could decay to SM states with signatures, but not signal strengths, identical to a heavy SM Higgs boson.  It may also decay to pairs of Higgs bosons, leading to resonant di-Higgs production.

\begin{figure}[t]
\centering
\includegraphics[height=1.3in]{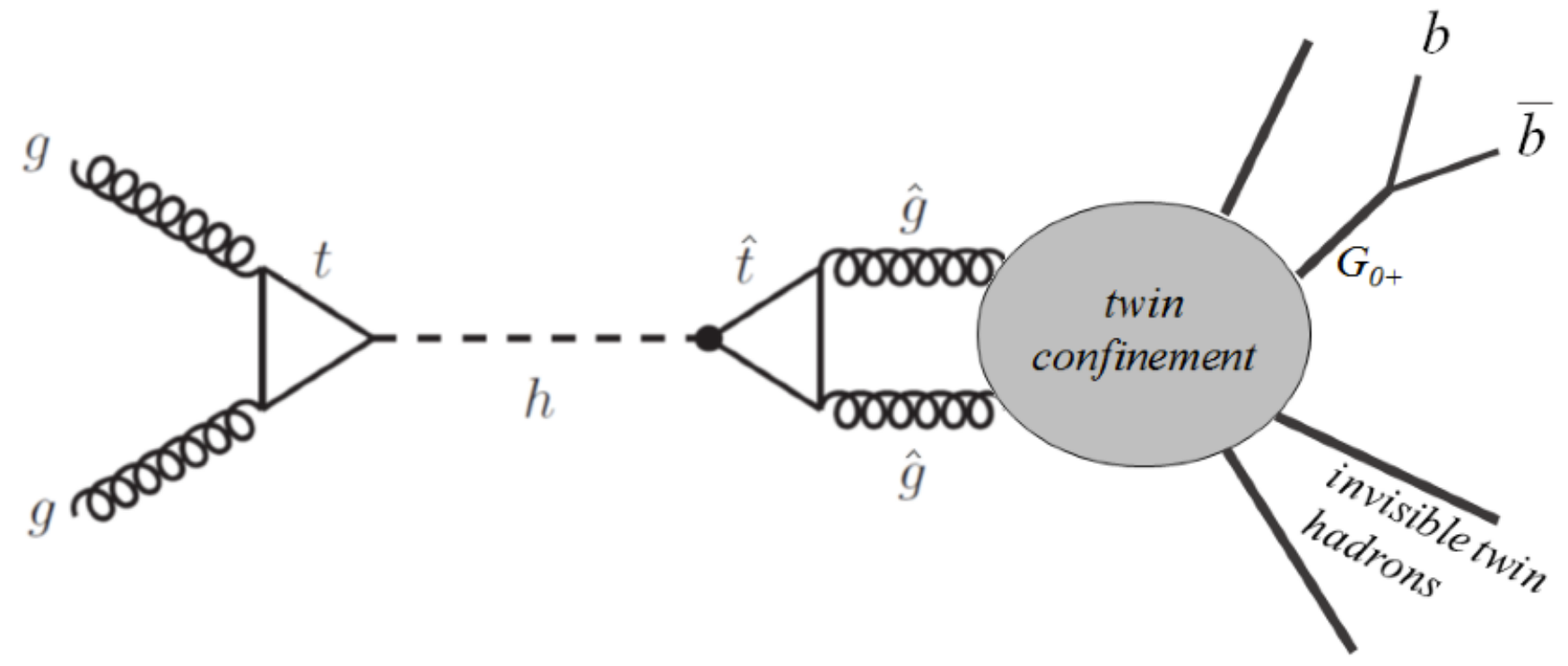} 
\caption{Twin glueball production and decay through the Higgs portal.  Figure taken from \cite{Craig:2015pha}.}
\label{fig:glueball}
\end{figure}

The most exciting aspect for HL-LHC is that more exotic signatures arise once the Twin sector is considered in full.  If Twin sector states are produced through the Higgs Portal they may decay into lighter Twin sector states, eventually cascading down to the lightest states within the Twin sector.  These lightest states may then decay back into SM states, leading to a huge variety of exotic signatures.  In essence, the Twin Higgs scenario provides a framework in which many so-called `hidden valley' signatures \cite{Strassler:2006im,Strassler:2006ri,Han:2007ae} may be realised.  As the motivation comes from the hierarchy problem, it is necessary that the new states must lie within some proximity to the weak scale.  Taking naturalness as a guide there are many possibilities for the spectrum in the Twin sector since it is possible that the lighter states which are less relevant for Higgs naturalness may have modified couplings to the Twin Higgs or may even not exist.

\begin{figure}[t]
\centering
\includegraphics[height=2.5in]{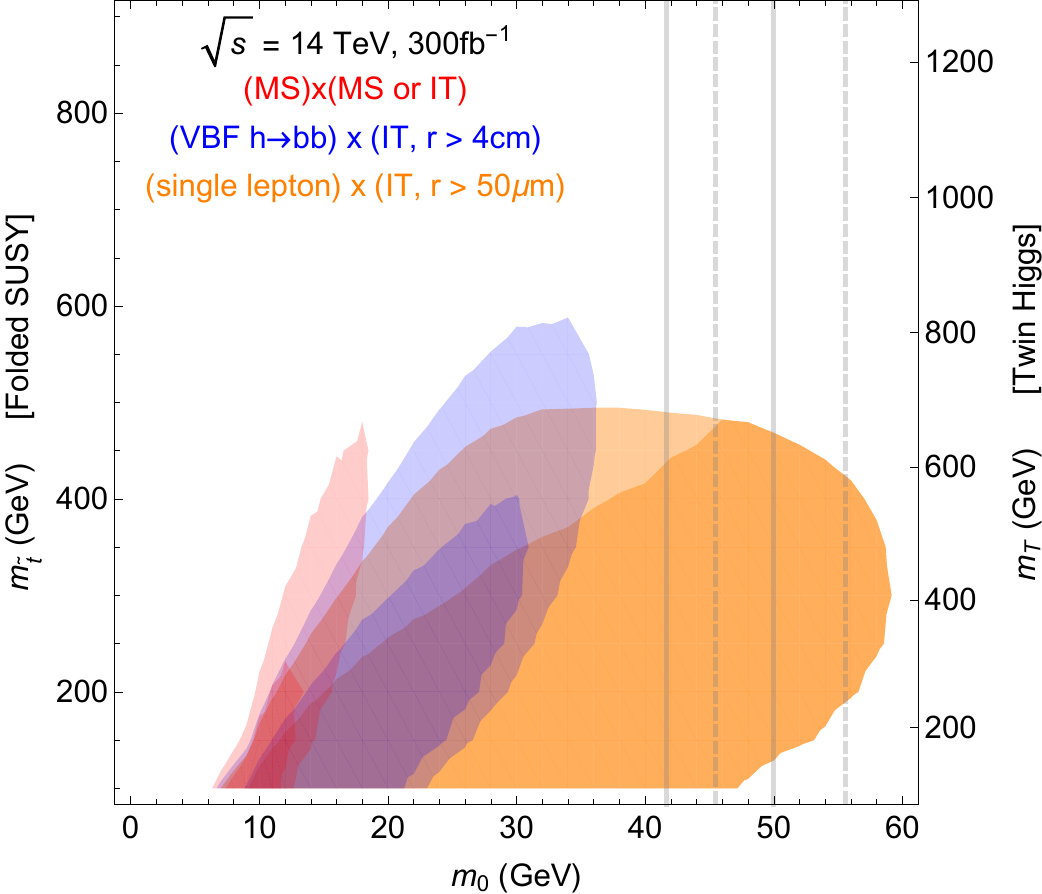} ~~ \includegraphics[height=2.5in]{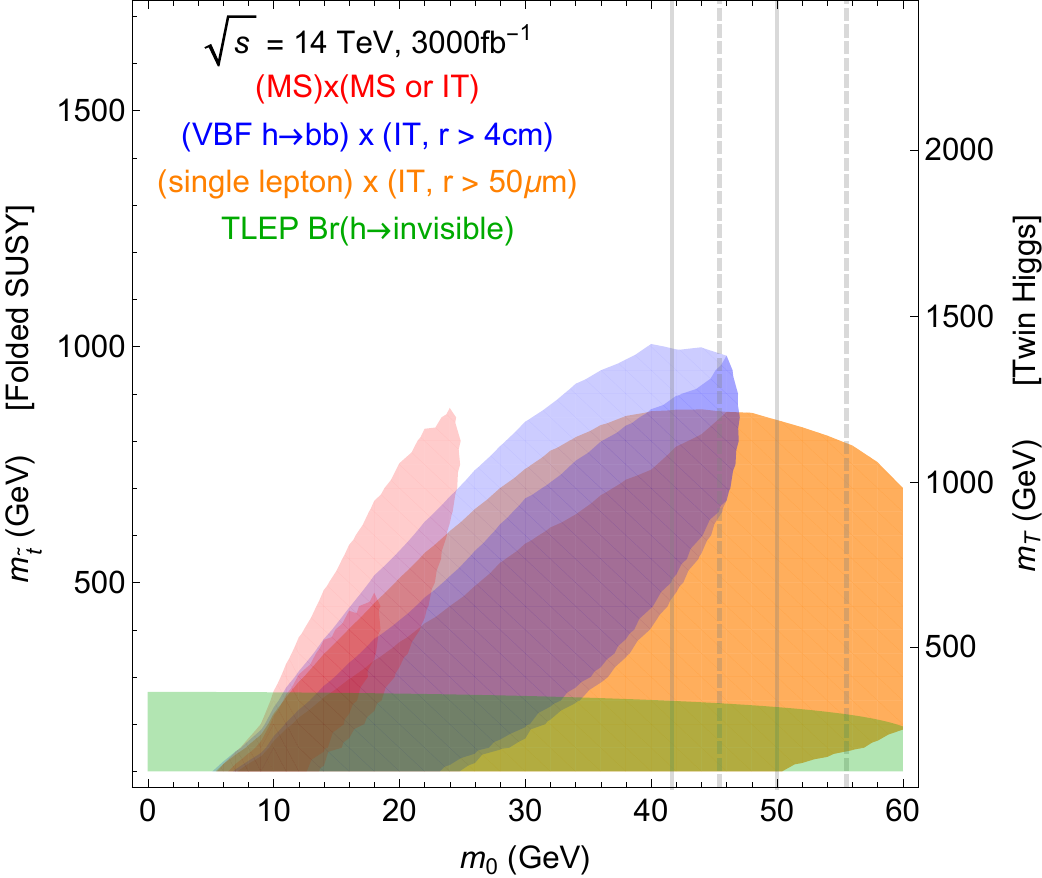}
\caption{Expected collider limits on the parameter space of the Twin Higgs model (far right axes) from constraints on exotic Higgs decays, as a function of the Twin Glueball mass.  Figure taken from \cite{Curtin:2015fna}.  A significant inprovement in reach is observed as one goes from the LHC to the HL-LHC.}
\label{fig:limits}
\end{figure}

A particularly interesting example is for exotic Higgs decays into Twin glueballs, as depicted in \Fig{fig:glueball}.  This is possible because the Higgs couples to the Twin Top quarks, leading (at one loop) to a coupling to Twin gluons.  The Higgs may thus decay to the Twin glueballs, which then decay, after some displacement, back through an off-shell Higgs to SM states, including bottom quarks.  Such an exotic Higgs decay signature can be used to search for the Twin sector states.  The expected reach for scenarios like this, taken from \cite{Curtin:2015fna}, is shown in \Fig{fig:limits}.

Due to the displaced vertices and high multiplicity final states we would expect a very low background to these searches, and hence na\"ively a significant leap in sensitivity as one goes to the HL-LHC.  Indeed, in \Fig{fig:limits} this is precisely what we see.  In fact, the branching ratio for the Higgs into the Twin Glueballs scales proportional to $\propto 1/m_T^4$, thus an improvement in reach of say $\sim600$ GeV to $\sim 1$ TeV corresponds to an improvement in reach for rate of a factor $\sim 8$.  Thus we see we are doing much better than $\sqrt{\mathcal{L}}$ scaling here, which would give a factor $\sqrt{3000/300} \sim 3$.  In fact, the improvement scales much more like $\mathcal{L}$, which would give a factor $\sim 10$.  Hence this is an phenomenological scenario, motivated by trying to understanding the microscopic origins of the EW scale, in which the HL-LHC offers a significant leap due to an exotic low-background signature.  Interestingly, exotic Higgs decays play a crucial role.

\subsection{Gegenbauer Higgs}\label{chap:Geg}
The Twin Higgs models arose by reconsidering the assumption that the top partners were coloured, or, more accurately, that the global symmetry in a pNGB Higgs model commutes with the SM gauge symmetries.  Are there any other assumptions the warrant reconsideration?  In \cite{Durieux:2021riy,Durieux:2022sgm} it was argued that yes, there is one assumption in pNGB models that has essentially shaped our perspective on that broad possibility.\footnote{I should declare a conflict of interest here, since myself and my collaborators proposed this class of models and I may be biased, one way or the other.}  This assumption is that all of the explicit symmetry breaking originates with the Yukawa and gauge interactions of the Higgs.  Recall that any source of explicit symmetry breaking generates a potential for the pNGB, thus it is essentially a minimality argument that the potential arises only due to these sources.

Is this a good assumption?   I would argue that it may not be.  After all the charged pion experiences explicit symmetry breaking originating from both the QED gauge interaction \emph{and} the explicit quark masses.  These two contributions are entirely different in nature and origin, the latter being essentially a `UV' source and the latter existing also in the `IR', in the sense that its contribution to the potential receives contributions over a range of scales from the UV cutoff down to the IR, revealed by the logarithm.

So, is the `minimality' assumption typically imposed on pNGB Higgs models an important one?  Essentially, this question is academic unless it strongly impacts the qualitative picture for pNGB Higgs models.  In \cite{Durieux:2021riy,Durieux:2022sgm} it was argued that this is indeed the case.  Recall in \Sec{sec:FT} that our estimate of how fine-tuned the theory needed to be was rooted in the explicit form of the pNGB potential which was, itself, rooted in the nature of the explicit symmetry breaking.  To understand the impact of the minimality assumption we thus need to generalise the general form the scalar potential could take if we allow for additional sources of explicit symmetry breaking.

Reconsider the structure of the Twin Higgs model.  I will employ it because this model addresses, to a large extent, the Higgs mass tuning arising due to the apparent absence of coloured top-partner fields, whereas it does not ameliorate the `$v/f$' tuning, so perhaps relaxing the minimality assumption might help in that latter respect.

Let us retain all the features of the original Twin Higgs model, but add to the Top-generated Higgs potential of \Eq{eq:THt} a new potential which arises from a new source of explicit symmetry breaking.  It will be useful to work with all the Higgs fields packaged as $8$ real scalar degrees of freedom in an $\mathbf{8}$ of $\SO(8)$ denoted $\boldsymbol{\omega} = (f+\rho) \boldsymbol{\phi}\,$. Here $\rho$ is the radial mode of the spontaneous symmetry breaking and $\boldsymbol{\phi}$ parameterises the vacuum manifold $\boldsymbol{\phi}\cdot \boldsymbol{\phi} =1$, 
\beq
\boldsymbol{\phi} =\frac{1}{\Pi} \sin \frac{\Pi}{f} \begin{pmatrix}
           \Pi_{1} \\
           \vdots \\
           \Pi_{7} \\
          \Pi \cot \frac{\Pi}{f}
         \end{pmatrix}\;,
         \qquad\text{with } \Pi = \sqrt{\boldsymbol{\Pi} \cdot \boldsymbol{\Pi}}~~~.
\eeq
The first $4$ components of $\boldsymbol{\omega}$ comprise the SM Higgs multiplet and the latter $4$ the Twin Higgs multiplet.  The gauging of the SM and Twin electroweak (EW) groups leads to $6$ pNGBs being eaten by massive gauge bosons, leaving only the Higgs field $h$ as physical scalar degree of freedom, as manifest in the unitary gauge where $\Pi_{i} = \delta_{i4} h$.

\emph{Any} source of explicit symmetry breaking can be written as the sum of spurions irreps of $\SO(8)$.  Enforcing that the two $\SU(2)\times\SU(2)$ subgroups remain unbroken restricts the values these irreps can take.  Each irrep spurion would contribute to the scalar potential as \cite{Durieux:2022sgm}
\bea
K_{2n}^{\,i_1 \ldots i_{2n}}  \phi_{i_1} \ldots \phi_{i_{2n}} =&\; G_n^{3/2} \left( \cos 2 h/ f \right)~~,
\eea
where $G_n^{(N-1)/2}$ is a `Gegenbauer Polynomial'.  This may sound exotic, but these are simply the spherical harmonics in $N$ dimensions.  For instance, you will be familiar with them as the Legendre polynomials you came across when you solved for the Hydrogen orbital wavefunctions.  Furthermore, the UV-dominated part of \Eq{eq:TH} is simply the case for $n=2$ , up to an overall unphysical additive constant, as it ought to be, since any potential generated by a source of explicit symmetry breaking can be captured in this way.

What impact does this additional contribution to the pNGB potential have?  Well, interestingly these Gegenbauer polynomials naturally have minima at small field values, as shown in \Fig{fig:geg}.  Thus a significant benefit that arises when one considers additional sources of explicit symmetry breaking is that the na\"ive `$v/f$'-tuning relationship, following from \Eq{eq:pngbtune} for a standard scenario which scaled as $v^2/f^2$, is broken!  Significantly less fine-tuned models are possible, since a contribution to the potential of this form realises a global vacuum with $v\ll f$ without introducing fine-tuning.  See \cite{Durieux:2022sgm} for more on this.

\begin{figure}[t]
\centering
\includegraphics[height=2.2in]{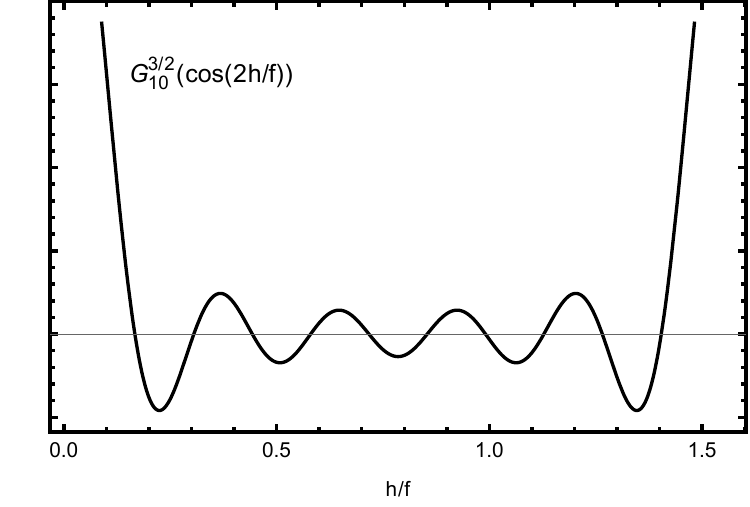} 
\caption{An example Gegenbauer potential arising from explicit symmetry breaking by a $20$-index irrep.}
\label{fig:geg}
\end{figure}

However, there is no free lunch.  Such a scenario is natural, since the form of the scalar potential is stable under radiative corrections from the UV and IR.  Still, an important question arises which is why such a non-minimal source of explicit symmetry breaking would arise in the fundamental theory, without being accompanied by additional, more minimal, sources.  Recall the Buckyball?  There a symmetry enforced that the lowest multipole possible was the 64-pole, corresponding to an $n=6$ irrep.  Na\"ively this lends support to the possibility of non-minimal irreps explicitly breaking a global symmetry associated with the SM in the UV.  However, what is different is that there was a symmetry, the Icosahedral group, which forbade the lower irreps.  In our case this is not possible.  Indeed, from $K_{2n}$ one can construct the tensors $K_{0},...,K_{4n}$.  Thus if the source of explicit symmetry breaking has a `magnitude' $\mathcal{O}(\epsilon)$, then we expect to see lower multipoles at $\mathcal{O}(\epsilon^2)$.  This aspect remains an open question.  Nonetheless, the Gegenbauer Higgs constructions reveal that the minimality assumption commonly employed in the consideration of pNGB-like Higgs boson models has a significant impact on how we interpret fine-tuning in the context of those models.

\begin{figure}[t]
\centering
\includegraphics[height=3.2in]{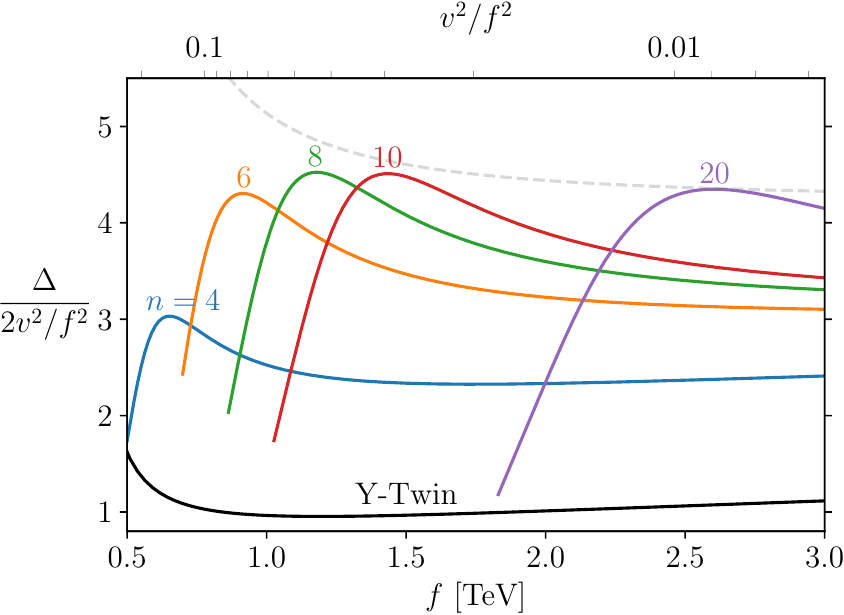} 
\caption{Na\"ive fine-tuning in a Gegenbauer Twin model, as compared to a standard Twin Higgs model, in this case denoted `Y-Twin'.}
\label{fig:geg}
\end{figure}

\subsection{Extra Dimensions}
I know what you're thinking...  What place do extra dimensions have in a chapter on Pion-like Higgs models.  It's a good question.  My first response is rather superficial, but holds some weight.  If one is considering physics on scales much smaller than the size of a pion, it makes no sense to talk about pions.  It is far wiser to discuss the quarks and gluons that make up the pion, and their interactions and dynamics.  Thus, one sense in which pions are scalars that have properties, including mass, which are insensitive to physics at the Planck scale is that above the QCD scale there is no pion at all!  In this way the only mass hierarchy to be explained is that between the pion mass and the QCD scale.  Similarly, if the Higgs is pion-like then in some sense the UV cutoff has been brought down to the TeV scale and one need not worry about the sensitivity of the Higgs mass to physics at the Planck scale.  It turns out that extra-dimensional models operate in a similar manner, bringing the cutoff of the EFT, including gravitational, down to much lower scales and consequentially reducing the hierarchy of scales between the EW scale and the true gravitational Planck scale.

My second response is less superficial, but more difficult to describe.  In the mid-1990's Maldacena rocked the theory community by giving compelling arguments that strongly coupled QFTs can be dual to gravitational theories in a larger number of dimensions \cite{Maldacena:1997re}.  By `dual' we mean that the two theories literally describe the same physics.  This is known, colloquially, as `holography'.  How far one can stretch this duality, and into which realms, is a difficult question to answer, however with many years of work it became clear that qualitative aspects can map between the twos sides of the duality for a great many types of strongly coupled theories.  In this respect it has become apparent that some 4D pNGB-like Higgs models, including those that we have discussed, may, in fact, be approximately dual to certain classes of 5D gravitational theories.  Thus in that sense it makes a great deal of sense to find a home for a discussion of extra-dimensions in this chapter.

To set the scene, in the late 1990's the theoretical physics community was electrified by an age-old question:  ``What if there are extra dimensions beyond the four we are familiar with''?  One reason, beyond the question of holography, this came to the fore was because it was realised that extra dimensions can solve the hierarchy problem, or at least turn it on it's head \cite{ArkaniHamed:1998rs,Antoniadis:1998ig}.  Essentially, the hierarchy problem is resolved if the cutoff of the SM EFT is actually at the weak scale.  If this is the case, and the true fundamental theory kicks in around a TeV, then we can understand why the Higgs mass is near to this scale, since this is the natural scale of the theory.  This part of the idea was thus not all that radical, however the really radical part was that even the cutoff of the gravity \emph{as we know it} is near the weak scale, rather than at $M_P \approx 2 \times 10^{18} \GeV$.  We will see how this works, through the example of the graviton, but first let me point you towards some excellent lectures available online \cite{Rattazzi:2003ea,Csaki:2004ay}.

The N-dimensional graviton is massless, thus it will in general have a momentum described by the null N-vector $P^M = (p,\underline{p}_3,\underline{p}_{E})$ where $\underline{p}_{E}$ is the spatial momentum in the extra dimensions.  Since the graviton is massless we have $P_M P^M = 0$, which we may rearrange as
\be
p^2 - |\underline{p}_3|^2 = |\underline{p}_{E}|^2 ~~.
\ee
Although simple, this equation is very revealing.  It tells us that a massless field in N-dimensions will have an apparent 4D mass given by its momentum in the extra dimension!
\be
m^2 =  |\underline{p}_{E}|^2 ~~.
\ee
We must still have our familiar massless graviton from 4D, which must then correspond to an N-dimensional graviton with vanishing extra-dimensional momentum $\underline{p}_{E} = \underline{0}$.  This also means that the wavefunction of the massless graviton in the extra dimension must be flat, since it carries no extra-dimensional momentum $\partial_m h_0 = 0$.

A massless graviton in N-dimensions has $N(N-3)/2$ degrees of freedom.  This means that from a 4D perspective we will expect to see not only the massless and massive spin-2 fields in 4D (2 and 5 degrees of freedom respectively), but also additional scalar, vector, and tensor fields all coming from the original N-dimensional metric.  For now, we only want to consider the 4D Planck constant, thus we need only consider the massless graviton we see in 4D.  Without loss of generality we may write the extra dimensional metric as a background metric accompanied by 4D metric fluctuations.  Let us consider extra-dimensional geometries in which, through a coordinate transformation, the metric can be taken to be `conformally flat',
\bea
ds^2 & = &  \tilde{g}^{MN} dx_M dx_N \\
& = &  e^{A(x^m)} g^{MN} dx_M dx_N \\
 & = &  e^{A(x^m)} \left( g_4^{\mu \nu} (x_\mu) dx_\mu dx_\nu + \sum_m dx_m^2 \right)
\eea
where Greek indices are for 4D coordinates and lowercase Latin indices are for extra dimensional coordinates.  The 4D fluctuations are taken independent of the extra dimensional coordinates, since the massless 4D graviton carries no extra-dimensional momentum.

The Einstein-Hilbert action for gravity in N-dimensions is given by
\be
S_{EH} = \int d^N x M_N^{N-2} \sqrt{-\tilde{g}} R(\tilde{g})
\label{eq:EH}
\ee
where $g$ is the determinant of the metric, and $R$ is the Ricci scalar.  Now, for $N>4$ we need to figure out what the effective 4D Planck constant will look like.    With a textbook bit of work, using standard properties of the Ricci scalar under Weyl transformations, and the fact that $\partial_m g^{\mu \nu} (x_\mu) = 0$, we may re-write the Einstein-Hilbert action as
\be
S_{EH} = \int d^N x M_N^{N-2} e^{\frac{N-2}{2} A(x^m)} \sqrt{-g_4} R_4(g_4)
\label{eq:weyl}
\ee
where $R_4(g_4)$ is the usual 4D Ricci scalar.  The usual 4D Einstein-Hilbert action is given by
\be
S_{EH} = \int d^N x M_P^{2} \sqrt{-g_4} R_4(g_4)
\ee
thus we may now identify the observed Planck's constant as
\be
M_P^2 = \int d^{N-4} x^m M_N^{N-2} e^{\frac{N-2}{2} A(x^m)} ~~.
\ee
Let us now consider some explicit examples.

\subsubsection{Flat extra dimensions}
If the extra dimensions are flat we have $A(x^m) = 0$.  Then, if the length of each extra dimension is $r_m$, we have
\be
M_P^2 = M_N^{N-2} \prod_m r_m ~~.
\ee
Let's take each extra dimension to be of the same size $r_0$, then, solving for $M_P \approx 2 \times 10^{18}$ GeV, we have that the required size of the extra dimensions are
\be
r_0 \approx 2 \times  \frac{10^{\frac{32}{N-4} - 19}}{5^{\frac{2}{N-4}}} \left( \frac{1 \text{ TeV}}{M_N} \right)^{\frac{N-2}{N-4}} \text{ m} ~~.
\ee
Clearly, for a single extra dimension we would need the extra dimension to have a size about as large as 5 astronomical units, roughly the distance from the Sun to Jupiter.  Since at distances below this scale gravitational physics would start to appear 5D, rather than 4D, the predictions of this theory would not match 4D Einstein's gravity.  In fact, gravitational physics on much smaller distance scales has already been probed, so this theory is ruled out.

However, for two extra dimensions, they only need to have a size in the mm level.  Gravitational physics on these distance scales is only just beginning to be probed, making this scenario very appealing for laboratory probes of gravity. 

Why does such a scenario solve the hierarchy problem?  The reason is that the cutoff of the theory is $M_N \sim $ TeV's, and not $2 \times 10^{18}$ GeV.  Essentially, the cutoff of field theory, where the full theory of quantum gravity must kick in, has been moved down to near the weak scale.  This means that, from the EFT perspective, there is no hierarchy problem, since the Higgs mass is indeed near the cutoff of the theory, exactly as expected.

There is, however, a delicate subtlety.  This comes down to the fact that we now have to understand why the extra dimension is so large \cite{ArkaniHamed:1998kx}.  After all, for the $N=6$ case the size of the extra dimension corresponds to an energy in the ballpark of $0.1$ meV.  Now there is a huge hierarchy between fundamental scales $M_N/(1/r_0)$!  This means that while the electroweak hierarchy problem is resolved, a new one pops up in its place regarding the volume of the extra dimensions.  There are, however, ways to get around this problem, as described in \cite{ArkaniHamed:1998kx}.

\subsubsection{Randall-Sundrum}
Let us make a jump from particle physics to cosmology.  In cosmology we have previously faced immense hierarchy problems, related to the flatness and homogeneity of the universe.  The flatness problem relates to the fact that the geometry of the Universe is very close to flat.  If the Universe expanded in a radiation-dominated manner since the big bang then the contribution of the curvature must have been initially very finely tuned to avoid the curvature dominating at late times.  Similarly, the horizon problem is also a fine-tuning problem in the sense that the initial conditions could have been very precisely fine-tuned to make it appear homogeneous now, however that is not what one would generally expect if the big bang only involved a radiation-dominated epoch.\footnote{I'll leave it to the cosmology talks to cover these topics in more detail!}

Of course, cosmologists have had tremendous success in solving these hierarchy problems through the theory of inflation, so let us revisit the details.  Einstein's equations in the presence of a cosmological constant are, in a general number of dimensions, given by
\be
G_{MN} = - \Lambda g_{MN}  ~~.
\ee
Let us consider two general metrics of the form
\be
ds^2 = - dt^2 + e^{\sqrt{\pm 2 \alpha/3} t} \sum_m dx_m^2 ~~,~~ ds^2 = e^{\sqrt{\pm 2 \alpha/3} x_M}\left( - dt^2 + \sum_{m\neq M} dx_m^2 \right) + dx_M^2
\ee
where the indices run over all spatial dimensions.  These metrics are clearly of the same form, yet in the first the time direction is `special', and in the latter a spatial direction is special.  Respectively they yield an Einstein tensor of the form
\be
G_{MN} = - \alpha g_{MN}  ~~,~~ G_{MN} =  \alpha g_{MN} ~~.
\ee
Thus, for $\Lambda > 0$, corresponding to a positive cosmological constant and de-Sitter geometry, we may choose $\alpha = \Lambda$ and we recover the usual solution for cosmological inflation.  As we move along the time dimension the proper distance between two space-time points grows exponentially.  This is highly non-trivial, as it can explain how spacetime events that appear to be causally connected only now, for example photons coming from opposite sides of the Universe, may in fact have been causally connected at earlier times.  This explains the horizon problem.  The flatness problem is similarly solved.  Thus inflation solves these hierarchy, or fine-tuning, problems very convincingly.  These hierarchies really do deal with hierarchies of scales, thus it is appealing to consider whether a similar mechanism could be used for the weak scale.

Now consider $\Lambda < 0$.  This is anti-de-Sitter geometry.  If there is an additional extra dimension, that we will parameterise with the coordinate $y$, then we can use the second metric with a scale factor $e^{\pm \sqrt{2 |\Lambda|/3} y}$.  Everything should follow in analogy with the inflationary case, however the scales will now become exponentially warped as we move along an extra spatial dimension.  This is, in fact, the proposal of Randall and Sundrum \cite{Randall:1999ee}.  Let us now see in detail how this works.

There are different ways in which one can frame this proposal, however the one which illuminates the mechanism most clearly is one in which all fundamental parameters are taken to be of order the Planck scale.  So let us take $M_P \sim M_5 \sim \alpha$.  The last parameter we will trade for $k = \alpha/3$, which can be easily inserted into the metric above.  Now let us imagine the Higgs boson is not a 5D field, but in fact lives on a 4D slice of the extra dimension, that we will locate at $y=y_0$.  Following the standard EFT rules we will write the Higgs mass at the same mass scale as the other parameters in the theory, thus the quadratic action for the Higgs living on this slice is
\be
\mathcal{L} = \int d^4 x dy \delta(y-y_0) \frac{\sqrt{-\tilde{g}}}{\sqrt{g_{55}}} \left( g_{\mu\nu} \partial_\mu H^\dagger \partial_\nu H^\dagger -\lambda ( |H|^2-f^2)^2 \right) ~~,
\ee
where, as indicated above, the decay constant is near the Planck scale $f \sim M_P$.  Let us now insert the explicit form of the metric and integrate over the delta function
\be
\mathcal{L} = \int d^4 x e^{-4 k y_0} \left( e^{2 k y_0} \eta^{\mu\nu} \partial_\mu H^\dagger \partial_\nu H^\dagger -\lambda ( |H|^2-f^2)^2 \right) ~~.
\ee
Finally, canonically normalising the Higgs field, we have 
\be
\mathcal{L} = \int d^4 x \left( \eta^{\mu\nu} \partial_\mu H^\dagger \partial_\nu H^\dagger -\lambda \left( |H|^2-f^2 e^{-2 k y_0} \right)^2 \right) ~~.
\ee
Remarkably, the natural scale for the Higgs vacuum expectation value is exponentially dependent on the position of the brane on which the Higgs field lives.  This is the essence of the Randall-Sundrum solution to the hierarchy problem, and we can arrive at a natural value for the weak scale with $y_0 \sim k^{-1} \log{v/M_P}$.  Since the brane position is only logarithmically dependent on the required separation of scales the hierarchy problem is truly solved in the sense that the radius of the extra dimension need not be hierarchically larger than the relevant length scales.

One might correctly object that we do not, in fact, live in an AdS universe.  This issue is, however, relatively straightforward to resolve.  To fully solve Einsteins equations one must also consider the boundaries of the extra dimension.  It turns out that one can place 4D slices at these boundaries with their own cosmological constant and if one has a finely-tuned value for the cosmological constant on these branes the final 4D Universe may in fact have a small cosmological constant.  This is a tuning, but it is none other than the fine tuning we must accept for the cosmological constant in the first place, thus it is not related to the electroweak hierarchy.

\subsubsection{Other geometries / Linear Dilaton theory}
There are many different geometries that may be interesting to study.  As an example, consider the following theory
\beq
{\mathcal S} = \int d^4 x\, dy\, \sqrt{-g} \, \frac{M_5^3}{2} e^{S} \left( {\mathcal R} +   g^{MN}\partial_M S
\, \partial_N S +4k^2 \right) ~,
\label{actionstring}
\eeq
wheer $S$ is a scalar field, usually referred to as the dilaton.  One can justify this action with a constant shift symmetry $S\to S + \alpha$, accompanied by a Weyl rescaling $g_{MN} \to e^{2 \alpha/3} g_{MN}$, broken only by the parameter $k$, which can thus be naturally small.  However, one should not become overly beguiled by such mixed Weyl-Scalar shift symmetries, as one can always perform a Weyl transformation into a frame, known as Einstein frame, where this is simply a shift symmetry acting on the scalar alone.  This transformation is
\beq
g_{MN} \to e^{-\frac{2S}{3}} g_{MN}
\label{tram}
\eeq
which turns the total action into
\bea
{\mathcal S} &=& \int d^4 x\, dy\, \sqrt{-g}   \frac{M_5^3}{2} \left( {\mathcal R} -\frac{1}{3} \,  g^{MN}\partial_M S
\, \partial_N S +e^{-\frac{2S}{3}}\, 4k^2 \right)  
~.
\label{eq:einstein}
\eea
Whether one works in Jordan frame or Einstein frame is irrelevant, the physics will be the same.  Let's stay in the Jordan frame.  The EOM for $S$ is
\be
\partial_y e^S \partial_y S - 4 k^2 e^S = 0 ~~.
\ee
This is solved for the spacetime-dependent background value $\langle S \rangle = \pm2 k y$.  Interestingly, the background metric in this frame is flat $g_{MN} = \eta_{MN}$.  Alternatively, one could have worked in the Einstein frame, and deduced the same result.

This setup also allows for a solution of the hierarchy problem, somewhere between the flat and RS cases, with very interesting phenomenology.  This scenario demonstrates an even richer realm of possibilities than before, as the behaviour of fields and their couplings is no longer solely determined by the metric, but by the interactions with the dilaton.  This can be seen in the Jordan frame, where the metric is entirely flat, whereas some bulk 5D operator, or brane localised operator, may couple differently to the dilaton
\be
\mathcal{L} = \sum_j c_j e^{\alpha_j S} \mathcal{O}_j
\ee
where $\mathcal{O}_j$ could contain SM fields, the coefficient $c_j$ depends on the microscopic structure of the UV theory, and $\alpha_j$ depends on the charge of the operator $\mathcal{O}_j$ under the dilaton shift symmetry.  This means that different quantities may be warped by different exponential factors, unlike in RS where one does not have this freedom.

\subsubsection{Mass spectra, wavefunctions, localisation, and all that.}
There is a tremendous amount of interesting phenomenology related to extra dimensional models.  This involves the spectra of additional resonances, and other notions such as localisation of fields, which is related to the wavefunctions of various modes in the extra dimension.  To develop some familiarity with these aspects let us consider a simplified scenario, which is a massless 5D scalar $\phi$ in a nontrivial background geometry.  Without loss of generality we may write the action as
\be
{\mathcal S} = -\frac 12 \int d^4 x\, \int_{-\pi R}^{\pi R} dy \, e^{A(y)} \left[ (\partial_\mu \phi)^2 +  (\partial_y \phi)^2   \right]  ~~.
\ee
where $A(y)$ is some general function.  Note that both Randall-Sundrum and Linear Dilaton models can be written in this form, and in general any 5D geometry will take this form as one may always perform a diffeomorphism to go to this `conformally flat' frame.  Interestingly, if the prefactor has come only from the metric, and not additional factors such as a background scalar profile, then it turns out that the mass spectrum and wavefunctions are the same for the massless bulk scalar as for the graviton.  Let us see what they are.

As the extra dimension is finite, extra dimensional momenta will be quantized, in just the same way as for a particle in a box in quantum mechanics.  Thus we may decompose the 5D field as an infinite tower of momentum eigenstates.  These eigenstates correspond to 4D mass eigenstates.  To do this we perform a Kaluza-Klein reduction into 4D fields
\beq
\phi(x,y) =\sum_{n=0}^{\infty} \frac{{\tilde \phi}_n(x)\, \psi_n(y)}{\sqrt{ \pi R}} ~~.
\eeq
The 5D field satisfies the equation of motion
\be
e^{A(y)} \partial_\mu \partial^\mu \phi+ \partial_y e^{A(y)} \partial_y \phi =0 ~~.
\ee
An on-shell 4D scalar satisfies the equation $ \partial_\mu \partial^\mu {\tilde \phi}_n(x) = m_n^2 {\tilde \phi}_n(x)$, thus we may rewrite this equation of motion, for each mode, as
\be
e^{A(y)} m_n^2 \psi_n(y)+ \partial_y e^{A(y)} \partial_y \psi_n(y) =0 ~~.
\ee
We must now consider the boundary conditions.  For a bulk scalar they can in general be complicated, however if there is a boundary mass term then they will typically be of the form $\partial_y \phi = m \phi$.  This comes from continuity of the equation of motion at the boundary, sometimes known as the `jump conditions'.  If the boundary mass term is vanishing we have $\partial_y \phi |_{y=0,\pi R}=0$, which is known as a Neumann boundary condition.  If the boundary mass term is infinite then we must have $\phi |_{y=0,\pi R}=0$, known as Dirichlet.  Note that when choosing boundary conditions the appropriate boundary potential must be there, in order to satisfy conditions known as junction conditions that follow from the discontinuity of the wavefunction over a boundary.  Anyway, lets keep life simple and choose to have no boundary potential, corresponding to Neumann boundary conditions.  This is the usual case for the graviton as well.

We see that for this general geometry we have a massless mode with a flat profile
\be
\psi_0(y) = \text{const} ~~,~~ m_0 = 0 ~~.
\ee
The other potential zero mode does not satisfy the boundary conditions.  This means, for example, that in any general 5D geometry the graviton wavefunction is flat.  However, as we will see, this does not imply that the graviton is not localised preferentially towards one end of the extra dimension.

To solve for the wavefunctions of the massive modes we must solve the eigenfunction equation
\be
m_n^2 \psi_n(y)+ A'(y) \psi'_n(y)+ \psi''_n(y) =0 ~~.
\ee
for whichever specific geometry we are interested in.\footnote{To simplify things, we may perform a field redefinition $\psi_n(y) = e^{-A(y)/2} \tilde{\psi}_n(y)$, such that the equation of motion becomes $(m_n^2-A'^2(y)/4-A''(y)/2) \psi_n(y)+ \psi''_n(y) = 0$.  This does not, of course, change the solutions, but may be useful in calculations.}

For a flat extra dimension this equation is very simple to solve, since $A'(y) = 0$.  Subject to the boundary conditions, this leads to the solutions
\bea
\psi_0 (y) &\propto&  \text{const}
\label{eigen1}\\
\psi_n (y) &\propto& \cos \frac{ny}{R} ~~,~~~ n \in \mathbb{N}
\label{eigen2}
\eea
with mass
\beq
m_{0}^2  =  0  ~~,~~~~~~~~
m_n^2  =  \frac{n^2}{R^2} ~~.
\label{uffamass}
\eeq

For the linear dilaton setup this equation is also simple to solve, since $A'(y) = -2 k$.  Subject to the boundary conditions, this leads to the solutions
\bea
\psi_0 (y) &\propto&  \text{const}
\label{eigen1}\\
\psi_n (y) &\propto&  e^{k |y|} \left(  \frac{k R}{n}\sin \frac{n |y|}{R}+\cos \frac{ny}{R} \right) ~~,~~~ n \in \mathbb{N}
\label{eigen2}
\eea
with mass
\beq
m_{0}^2  =  0  ~~,~~~~~~~~
m_n^2  =  k^2 + \frac{n^2}{R^2} ~~.
\label{uffamass}
\eeq

For Randall-Sundrum the solution is a little more complicated.  In this case we may write $A(y) = -3 \log |k y|$.\footnote{Note that in this basis we usually refer to the coordinate $z$, since $y$ is conventionally reserved for the non-conformally flat version of the metric $ds^2 = e^{2 k y} dx^2 + dy^2$, but for the sake of consistency of notation here we will stick with the current notation.}  The solution is now
\bea
\psi_0 (y) &\propto&  \text{const}
\label{eigen1}\\
\psi_n (y) &\propto & y^2 \left( J_2 (m_n y) - \frac{J_1 (m_n/k)}{Y_1 (m_n/k)} Y_2 (m_n y) \right) ... ~~,~~~ n \in \mathbb{N}
\label{eigen2}
\eea
with mass
\beq
m_{0}^2  =  0  ~~,~~~~~~~~
\text{Sol} (J_1 (m_n \pi R) Y_1 (m_n/k)-Y_1 (m_n \pi R) J_1 (m_n/k)) = 0 ~~.
\label{uffamass}
\eeq
Note that expressions that appear to be different exist in the literature, such as in \cite{Csaki:2004ay}, however one should realise that they are all the same, differing only by a wavefunction redefinition, or a change of coordinates.

To understand the localisation of a Kaluza-Klein modes we need to know where they `live' in the extra dimension.  To be sure of making physical statements, one must define a measure that is independent of changes of coordinates.  In other words, it must be a diffeomorphism-invariant quantity.  The obvious candidate is the field density
\be
\frac{dP_n (y)}{dy} = e^{A(y)} |\psi_n (y)|^2
\ee
since this originates from the diffeomorphism-invariant term $\sqrt{-g} d^5 x |\phi|^2$.  Thus we see that for a flat extra dimension the zero mode and excited modes are evenly distributed along the extra dimension.  For the linear dilaton case the excited modes are distributed in exactly the same way as a flat extra dimension, but the zero mode is exponentially distributed towards one end of the extra dimension.  For Randall-Sundrum it is conventional to make a change of coordinates to the metric $ds^2 = e^{2 k y} dx^2 + d\tilde{y}^2$, for which we would define $k y = e^{k \tilde{y}}$, where one sees that the zero mode is exponentially distributed and the excited modes are evenly distributed, but this time with Bessel functions, rather than sinusoids.  This is why we refer to gravity being exponentially localised in Randall-Sundrum, since the 4D graviton is the zero mode, which is indeed exponentially localised at one end of the extra dimension.

These are the basic tools for model building in extra dimensions, where it is possible to localise different fields for different purposes.  For example, this has been used to realise models for small neutrino masses, flavour hierarchies, and many other possibilities.  Let us now end our foray into the fifth dimension by stepping back into 4D, discretely.

\subsubsection{Dimensional Deconstruction}
While it is natural to associate the physics of extra dimensions with gravity, it is possible to do 5D model building without ever stepping foot into a fully-fledged 5D model.  This is based on the idea of `dimensional deconstruction' \cite{ArkaniHamed:2001ca}, which essentially borrows the theoretical technology of lattice QCD to do model building!  Let us start by considering $N+1$ scalar fields in 4D, with the usual kinetic terms
\be
\mathcal{L} = \int d^4 x \sum_j \frac{1}{2} \partial_\mu \phi_j \partial^\mu \phi_j ~~.
\ee
Now we will add `nearest-neighbour' mass terms between the scalars that are of a form more familiar from condensed matter physics than particle physics
\be
\mathcal{L} = \int d^4 x \sum_j \frac{1}{2} m^2 ( \phi_j - \phi_{j-1} )^2 ~~.
\ee
These interactions still respect a shift symmetry $\phi_j \to \phi_j + \text{const}$, thus although the mass terms involve every scalar, a massless mode must emerge once we diagonalise the mass terms since there is a degree of freedom protected by this shift symmetry.  Since the shift symmetry acts equally on all fields, when we go from the interaction basis $\phi_j$, to the mass basis $\tilde{\phi}_j$, through an orthogonal rotation, we will find that the massless mode has an equal overlap with each of the interaction basis fields.  The massive modes will have a spectrum that approaches $m_n \sim m n/N$, and the overlap between mass eigenstates and interaction eigenstates will be found to be sinusoidal.  This is of course very familiar from a flat extra dimension.

To see the connection, let's revisit some basics of lattice field theory.  Take a field living in $4+1$ dimensions $\phi(x,y)$, where $y$ is the extra dimension.  In a compact extra dimension this gives rise to an infinite number of 4D fields, as we have seen.  We can understand this by having a single 4D field living at every slice of the extra dimension.  Now we may discretise the extra dimension, turning it into a lattice.  Thus the position along the extra dimension becomes a discrete variable $y_j = j a$, where $a$ is the lattice spacing $a=L/N$, $L$ is the length of the dimension, and $N$ is the number of lattice sites.  Now we have $N$ 4D fields $\phi(x,y) \to \phi_j(x,)$ for each lattice site.  The final, crucial, ingredient is that we must have some way to deal with extra dimensional derivatives $\partial_y \phi(x,y) \to$?  The correct prescription is of course to simply use the definition of the derivative $\partial_y \phi(x,y)|_y \to (\phi_{j+1}(x)-\phi_{j}(x))/a$.  This is all the machinery we require to transform our extra dimension into a lattice.

Let's put this to practice for a bulk scalar in a flat extra dimension
\bea
{\mathcal S} & = & -\frac 12 \int d^4 x\, \int_{-\pi R}^{\pi R} dy \, \left[ (\partial_\mu \phi)^2 +  (\partial_y \phi)^2   \right] \\
 & \to & -\frac 12 \int d^4 x\,  \left[ \sum_{j=0}^N (\partial_\mu \phi_j)^2 + \sum_{j=0}^{N-1} \frac{1}{a^2} (\phi_{j+1}(x)-\phi_{j}(x))^2   \right] \\
\eea
This is simply the condensed-matter inspired action we wrote above!  Thus we see that, reversing the direction, the continuum limit $a\to 0$, $N a \to L$, is simply a massless scalar in a flat extra dimension.

\begin{figure}[t]
\centering
\includegraphics[height=1.5in]{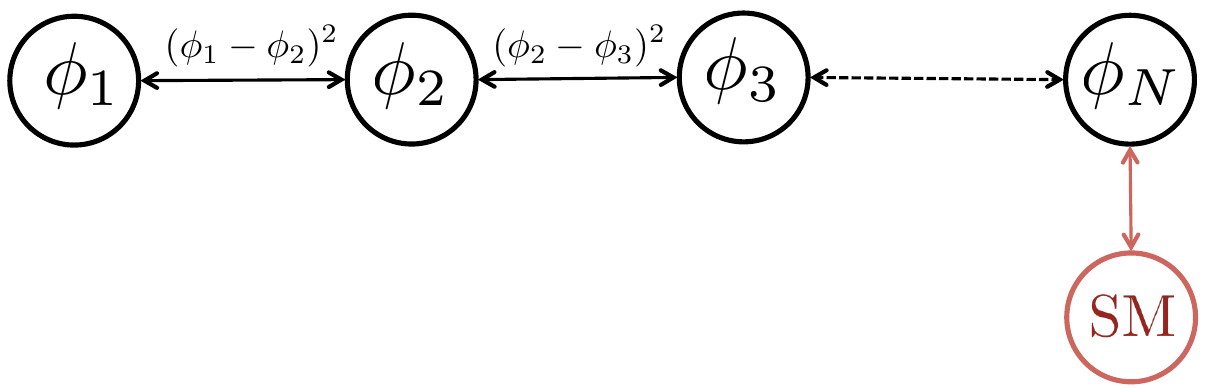}
\caption{A schematic of dimensional deconstruction with a multi-site model.  If some SM operator is coupled to the end of the chain it will inherit a suppressed coupling to the massless mode, scaling like $1/\sqrt{N}$, as well as a coupling to the massive fields.}
\label{fig:decsontruction}
\end{figure}

This may seem like a rather trivial set of steps, but it can be tremendously useful in model building.  The reason is that one can play with extra dimensional model building, and make use of particle locality in an extra dimension, by playing with models in `theory space', as sketched above.  This is shown schematically in \Fig{fig:decsontruction}.

While there is not time to go into it here, this notion of locality is very useful in terms of controlling radiative corrections.  For example, when one has a symmetry that is only broken when all interactions between the fields are considered, known as collective symmetry breaking, this means that loop corrections must involve all sites of the chain before they can transmit a symmetry-breaking spurion to some observable.  This then means that symmetry-breaking effects can be delayed to very high loop orders, which is particularly useful for composite Higgs models, where we already saw that gauge and Yukawa interactions break the shift symmetry of the Goldstone Higgs.  In dimensionally-deconstructed models the relevant dangerous corrections can then be delayed to higher loop order.

\subsection{Summary}
A Pion-like Higgs?  On the theory side the answer is not as straightforward as we might have hoped.  I have attempted to introduce you to a number of relevant paradigms which may play a genuine part in the origins and microscopic nature of the Higgs boson.  Goldstones, Twins, Warped Dimensions and Gegenbauers.  It may seem like a lot, but I have only scratched the surface.  Partial compositeness and the connection to flavour physics are obvious omissions.  Nonetheless, you are hopefully now well equipped, both conceptually and also hopefully technically, to dive into this fascinating question further if you wish.

Fortunately, on the experimental side things are relatively more rosy.  If there is one take-home message you bring from this chapter, I hope it is this:  In all Pion-like Higgs scenarios we expect modified Higgs couplings to show up at some level.  For this reason, if we truly want to understand the microscopic origins of the Higgs boson we simply must continue to study the Higgs boson with greater precision.  

\section{Beyond Symmetry?}
Before we move on to discuss cosmological approaches to the hierarchy problem it will be useful to briefly survey and discuss some potential alternatives which are incomplete, failed, or even contentless.  This section will thus have little structure, being more like a collection of theoretical odd socks hoping to find their pair and a purpose, but most likely headed for the trash after a prolonged period of denial.

\subsection{Just input parameters}
One trivial possibility is if the Higgs mass is not predicted in any way by the true fundamental theory.   In this case it simply becomes an input parameter and it does not make sense to question its value.  This is a possibility, however it would mean admitting defeat, and would essentially require that the reductionist paradigm has terminated, since this would mean that the Higgs mass, and presumably many (all?) other parameters, \emph{can never be explained}.  Thus, if one wishes the hierarchy problem away with this argument, the consequences for the rest of fundamental science must also be embraced.  I mention this possibility here just for the sake of being comprehensive, since it is a logical possibility, but with profoundly troubling consequences.  Buyer beware!

\subsection{Inseparable Scales}
The way in which we view the hierarchy problem is inherently Wilsonian.  When we are talking about EFTs in high energy physics we often refer to the picture as being `Wilsonian', because it was Ken Wilson who really put the entire EFT structure on a firm footing in quantum field theories, particularly including quantum corrections \cite{Wilson:1973jj}.\footnote{Ken Wilson was a truly remarkable human.  Not only did he run a mile in less than four minutes, but he also initiated the field of lattice quantum field theory.  See \cite{Jackiw:2013jba} for more on his life in physics.}  This Wilsonian picture of quantum field theories is, of course, built upon the foundations of quantum field theory and the Wilsonian picture amounts to being able to `integrate out' the physics on short distance scales (UV) to arrive at an effective description for long-distances (IR) which captures the effects of the short distance physics without recourse to having to account for it explicitly each time we calculate.  If the UV physics can not be factorised in this way then the programme would fail.  We refer to an instance like this as `UV/IR' mixing, since the IR physics depends crucially on what happens at short distances.

This would then seem a promising avenue for explaining the puzzle we now have concerning the Higgs mass, since UV/IR mixing would overthrow the Wilsonian reasoning which presents us with the problem in the first place.  It turns out, however, that no one has succeeded (yet) in putting forth a concrete theoretical realisation of UV/IR mixing which can apply to the weak scale.  The underlying reason is that the Wilsonian picture is built upon the foundations of quantum field theory and so, ultimately, to break the picture means compromising those foundations in some way or another.

Anyone who has worked through the first volume of The Quantum Theory of Fields by Weinberg will be (painfully) aware that QFT has various fundamental properties built in, such as unitarity, locality, causality, special relativity, cluster decomposition.  Unitarity is essentially the requirement that the sum over probabilities of all possible outcomes must equal unity.  Locality corresponds to the assumption that interactions occur at a spacetime point.  Prosaically, the action is an integral over all of spacetime and the Lagrangian has interactions at the same point, such as $\int d^4 x \phi^4(x)$, rather than $\int d^4 x d^4 y \phi^2(x) \phi^2(x+y)$, for instance.  Causality is the requirement that an effect can not occur before its cause.  This can occur, for example, if there are closed timelike curves, violations of the null energy condition, tachyons, ghosts, or violations of the Lorentz symmetry at high energies, and, importantly, a Hamiltonian with energy unbounded from below.  Now, if we were to mess with these ingredients then the Wilsonian picture can break down.  For example, if causality is violated then just as there may be correlations between space-like separated events, one might hope to have correlations, or cancellations, between physics in the far UV and the IR, since spacetime invariants are related to energy-momentum invariants.  If this were the case then could we understand the hierarchy problem as being solved/explained by such a correlation?

One might na\"ively expect something like a causality-violating theory to look very exotic, however it is straightforward to write down rather sane-looking Lorentz-invariant field theories which, upon closer inspection, reveal themselves to violate causality at the microscopic level.  One particularly simple example for a massless scalar field was exposed in \cite{Adams:2006sv}
\be
\mathcal{L} = \frac{1}{2} \partial_\mu \phi \partial^\mu \phi - \frac{1}{\Lambda^4} (\partial_\mu \phi \partial^\mu \phi)^2 ~~.
\ee
the negative sign in the second term is in fact the cause of the acausality, and allows for superluminal propagation.\footnote{Recently contested \cite{Kaplan:2024qtf}.}  Thus this theory is not actually an EFT in the Wilsonian sense, at all, although it sure looks like it at first glance!  Perhaps then something similar could be true of the Standard Model, where it is not actually an EFT, so the hierarchy problem, at least as we understand it, does not even exist.  

With regards to the hierarchy problem, this type of approach has a long history.  The best-known example is the Lee-Wick class of theories, in which the hierarchy problem may have been resolved through the presence of higher derivative operators which, upon closer inspection, can be understood as arising due to heavy ghost-like auxiliary fields \cite{Grinstein:2007mp}.  These extra fields lead to SM propagators that behave as
\be
\nabla (p^2) \sim \frac{1}{p^2-m^2} - \frac{1}{p^2-M^2}  ~~,
\ee
which at high energies cancel, just like in Pauli-Villars regularisation, whereas at energies $E\ll M$, look just like the usual SM propagators.  As a result these theories cancel quadratic UV-divergences as hoped, and the price paid for outflanking Wilsonian logic is a violation of causality by the propagators with extra `wrong-sign' residues.   However, this is only apparent at microscopic scales.

This is one example, however other attempts have been made which go beyond this, such as \cite{Craig:2019zbn} which is based on `non-commutative geometry'.  The non-commutative geometry follows from an $\mathcal{O}(1)$ violation of Lorentz invariance and thus, at face value, appears very much at odds with observations.  Nonetheless, I expect this line of investigation has not yet been fully exploited, but progress is hard and it takes a brave soul to start fiddling with the fundamentals.

\subsection{Swamp Monsters}
Closely related to the subject of UV/IR mixing is the notion of `The Swampland'.  The basic idea is that there exist perfectly self-consistent QFTs which are nonetheless perfectly inconsistent with a consistent theory of quantum gravity.  In other words, in the Venn diagram of QFTs and QGTs (quantum gravity theories) there is region of the former with no overlap with the latter.

One compelling argument for this concerns black hole decays.  Consider a global $\U(1)$ symmetry with associated global charges.  The Hawking radiation does not carry global charge, yet the black hole ultimately decays away to nothing.  In other words, quantum gravity violates the global symmetry, which essentially falls down a wormhole!  See \cite{Banks:2010zn} for enhanced lucidity on this point.  It turns out that there is a web of qualitative constraints like this concerning the interplay between quantum gravity (UV) and low energy QFT (IR) effects, with those QFTs which are inconsistent with a quantum gravity UV being known as lying in The Swampland \cite{Vafa:2005ui,Arkani-Hamed:2006emk,Adams:2006sv,Ooguri:2006in}.  The boundaries of the Swampland are marked by a now-vast interconnected web of conjectures of varying degrees of robustness.

This suggests a less direct form of UV/IR mixing than I had previously described, in which the lR dynamics can, happily, be captured by an EFT without recourse to the UV modes, however the structure of that EFT is constrained as a result of the UV.  Is it possible, then, that the reason the Higgs is so light cannot even be comprehended within QFT alone, being demanded instead by quantum gravity itself?  There have been many discussions and investigations in the direction of connecting IR oddities with the swampland, for instance see \cite{Ooguri:2016pdq,Ibanez:2017kvh,Ibanez:2017oqr,Hamada:2017yji,Lust:2017wrl,Gonzalo:2018tpb,Gonzalo:2018dxi,Craig:2018yvw,Craig:2019fdy}.  I cannot possibly review them all here, but I will attempt to describe one.

The Electric Weak Gravity Conjecture is that, in a $\U(1)$ gauge theory with gauge coupling $g$ coupled to gravity then there must exist a particle in the theory of charge $q$ of mass
\beq
m \lesssim g q M_P ~~.
\eeq
This follows from arguments that in the full quantum gravity theory there must exist one particle for which gravity is the weakest force, known as the Weak Gravity Conjecture.  The inequality is an immediate consequence from the comparison of Newtonian with Coulomb forces.   One way of arriving at the weak gravity conjecture involves remnants.  Consider the decay of a charged black hole, which continues all the way down to the Planck scale, at which one is ultimately faced with an extremal black hole of charge $Q$ and mass $Q g M_P$.  The only way for this black hole to be able to decay is if we have particles of charge $q<Q$ for it to decay into.  However, the number of particles ultimately produced will be $N=Q/q$.  Comparing the total mass before and after decay one finds
\beq
m \frac{Q}{q} \lesssim Q g M_P  ~~.
\eeq
Trivial rearrangement delivers the Weak Gravity Conjecture.\footnote{Note that the scaling of this result could have been deduced from dimensional analysis.}

How about the Higgs?  Well clearly this doesn't work for hypercharge, since the Higgs boson very trivially satisfies the bound.  Clearly to motivate a very very light field will require a very very weak coupling $g$.  It turns out that attempts along these broad lines have, indeed, been made \cite{Cheung:2014vva,Craig:2019fdy}.  However, such theories ultimately require new particles at or near to the EW scale, largely due to a confluence of interrelated weak gravity constraints, implying experimental signatures and no `desert' up to the Planck scale.  Similarly to the last subsection, this avenue warrants further investigation, however it is heartening to see that weak scale new physics seems to persist in any case.

\subsection{New Symmetries}
There is one topic I shouldn't really tell you about, for many reasons.  Chiefly I am under-qualified to explain it to you, but the fact that this topic has not yet shed any non-trivial light on the hierarchy problem also disqualifies it from these lecture notes.  Nonetheless, I'm compelled to bring it to you attention.  This topic would come under the banner of `Generalised Symmetries', but is more specifically concerned with either `Higher Form Symmetries' \cite{Gaiotto:2014kfa} or `Non-Invertible Symmetries'.   See e.g.\ \cite{Brennan:2023mmt, Bhardwaj:2023kri,Gomes:2023ahz,Luo:2023ive,Schafer-Nameki:2023jdn,Shao:2023gho} for lectures on both topics by qualified practitioners.

Suffice to say, there has been somewhat of a renaissance in our understanding of symmetries, generalising the basic Noether-current concepts you will be familiar with from undergraduate studies to greater vistas.  There is hope that this new broader picture may reframe our view of QFTs to an extent that it may offer an understanding of structural questions concerning the SM.  Indeed, in recent years a number of authors have made progress on that front \cite{Cordova:2018cvg,Wan:2019gqr,Davighi:2019rcd,Wang:2020xyo,Brennan:2020ehu,Hidaka:2020izy,Fan:2021ntg,Anber:2021upc,Wang:2021ayd,Wang:2021vki,Wang:2021hob,McNamara:2022lrw,Cordova:2022fhg,Cordova:2022qtz,Choi:2022rfe,Wang:2022eag,Yokokura:2022alv,Brennan:2023kpw,Cordova:2023her,Choi:2023pdp,vanBeest:2023dbu,Brennan:2023tae,Choi:2022fgx,Cordova:2023ent,Reece:2023iqn, Putrov:2023jqi, vanBeest:2023mbs,Aloni:2024jpb}.

Nothing substantive has yet emerged as concerns the EW hierarchy problem, but watch this space, whatever its co-dimension...

\subsection{Scale-Invariance}
The symmetry we encountered before which relates to particle masses was a shift symmetry for scalars.  What about scale-invariance as a symmetry?  This avenue seems all too obvious, since if the Higgs mass was the only spurion for scale symmetry breaking then it could be naturally small without any problem.  This warrants some thought.  Let us consider the scale symmetry for the pion example discussed above.  If one studies the charged pion interaction with the photon
\be
\mathcal{L}_{Kin} =  \frac{1}{2} (\partial_\mu \pi_0)^2 +| (\partial_\mu + i e A_\mu) \pi^+ |^2 ~~.
\ee
we see that the action respects a symmetry $x^\mu \to \alpha x^\mu$, $A^\mu \to \alpha^{-1} A^\mu$, $\pi^\pm \to \alpha^{-1} \pi^\pm$. This means that while the pion-photon interaction does break the pion shift symmetry, it also respects classical scale invariance.  This would then seem to imply that it alone cannot generate a mass term for the pion at the quantum level since this would break the scale symmetry.  Indeed, this is true, since $e$ carries no mass dimension.  One may be tempted to conclude then that no quantum corrections to the charged pion mass can arise.  However, this is clearly not true, both empirically and theoretically.  The reason is that this argument only works if there are no other dimensionful parameters in the theory.  However, in the SM there is the cutoff $\Lambda \sim $ GeV above the pion mass scale, and in combination with the coupling $e$ this generates the observed mass splitting.

Thus we see that having classical scale invariance cannot keep a scalar light if there are other high energy scales in the theory.  This means that one cannot simply state that in the SM the Higgs interactions respect scale invariance, thus do not actually generate a hierarchy problem.  This reasoning can only work if there are no other large dimensionful parameters in the entire theory of everything!  It is a tall order indeed to find a theory in which this logic can be used to evade the hierarchy problem.  Nonetheless, there are serious efforts to achieve precisely this goal (see e.g.\ \cite{Salvio:2014soa}).  The challenge is formidable, making this a very interesting problem.  Note that while gravity itself may not generate a Higgs mass correction $\delta m^2 \propto M_P^2$, since this is dimensionally forbidden, in combination with SM gauge or Yukawa couplings, there is no symmetry forbidding terms such as $\delta m^2 \propto g^2 M_P^2$.

\section{Hierarchies from Self-Organised Criticality?}\label{sec:dynamics}
The previous approaches to the hierarchy problem have utilised symmetry to try and explain a hierarchy between the Higgs mass and the UV scale.  These possibilities are, essentially, static in the sense that were some parameter to vary ever so slightly from its present value the hierarchy would persist.  However that is not the only possibility.  Suppose the fundamental parameters are, in fact, finely-tuned, but that fine-tuning arose naturally.  If that were to occur then the fine-tuning may have arisen dynamically in some sense, such that natural dynamics led to a special point in parameter space being preferred.

In what sense is the parameter point of the SM with a small Higgs mass, relative to UV scales, special?  It isn't special from the symmetry perspective since, as far as we're aware, no symmetry is enhanced when $m_H^2 \to 0$.  However it is special from the perspective of criticality.  Generally speaking, a critical point is a point between two different phases in some system.  Water held at a fixed temperature of $50^{\circ}$ Celcius is not critical; water held at boiling point is.  In the SM, if the microscopic scale of UV completion were far above the weak scale $\Lambda \gg v_{\text{EW}}$ then the SM is critical with respect to the weak scale, for the same reason.  As a result, in the context of a broader UV-completion the SM is at a critical point, which is special.

How, then, could dynamics pick out this critical point?  That has been the subject of considerable attention in the last ten or so years.  The first proposal to put forward a compelling possibility was the Relaxion, so let's start there.

\subsection{The Relaxion}
``The Relaxion'' was proposed in 2015 \cite{Graham:2015cka}.\footnote{A similar idea was considered much earlier for the cosmological constant problem \cite{Abbott:1984qf}, and alternative relaxation-based approaches to the hierarchy problem have also been explored \cite{Dvali:2003br,Dvali:2004tma} more recently.}  This approach uses symmetries in the underlying model, but the Higgs mass itself is not protected by a symmetry.  Instead, dynamical evolution of this Higgs mass in the early Universe halts at a point where it is tuned to be much smaller than the cutoff.

As described in \cite{Graham:2015cka}, if the Higgs is a fundamental scalar then the hierarchy problem relates to the fact that if we keep the theory fixed but change the Higgs mass, the point with a small Higgs mass is not a point of enhanced symmetry.  However, this may be a special point with regard to dynamics, since this is the point where the SM fields become light.

The structure of the theory is relatively simple to write down and we will, as always, rely on EFT arguments.  Let us consider the SM as an effective theory at the scale $M$, which is the cutoff of the theory.  Following the standard EFT rules we include all of the operators, including non-renormalizable ones, consistent with symmetries.  All dimensionful scales are taken to the cutoff $M$.  We add to this theory a scalar $\phi$ which is invariant under a continuous shift symmetry, $\phi \to \phi + \kappa$, where $\kappa$ is some constant.  This shift symmetry only allows for kinetic terms for $\phi$.  We then add a dimensionful spurion $g$ which breaks this shift symmetry.  As $g$ is the only source of shift symmetry breaking then a selection rule may be imposed, such that any potential terms for $\phi$ will enter in the combination $(g \phi / M^2)^n$.  Thus the theory is written
\be
\mathcal{L} = \mathcal{L}_{SM} - M^2 |H|^2 + g \phi |H|^2 + g M^2 \phi + g^2 \phi^2 + ...
\label{eq:relax}
\ee
where the ellipsis denote all of the other higher dimension terms and it should be understood that the coefficients of all the operators in \Eq{eq:relax} could vary by $\mathcal{O}(1)$ factors and the negative signs have been taken for ease of presentation.

The next step is to add an axion-like coupling of $\phi$ to the QCD gauge fields
\be
\frac{\phi}{32 \pi^2 f} G \widetilde{G} ~~.
\label{eq:ax}
\ee
This coupling is very special.  As $G \widetilde{G}$ is a total derivative, in perturbation theory \Eq{eq:ax} preserves the shift symmetry on $\phi$, thus it is consistent to include this operator without a factor of $g$ in the coupling.  Perturbatively this operator will not generate any potential for $\phi$, thus all of the shift-symmetry breaking terms involving $g$ remain radiatively stable and it is technically natural for them to be small.  However, non-perturbatively the full topological structure of the QCD vacuum breaks the shift symmetry $\phi \to \phi + \kappa$ down to a discrete shift symmetry $\phi \to \phi + 2 \pi f z$, where $z$ is an integer.  Thus the complete story behind the model is one of symmetries.  $\phi$ enjoys a shift symmetry which is broken to a discrete shift symmetry by QCD effects.  The discrete shift symmetry is then broken completely by $g$.

%Let us see how this works for a $\U(1)$ symmetry.  Take a complex scalar field $\phi$ with an action invariant under the transformation $\phi \to e^{i \theta} \phi$.  We may choose a Higgs-like potential for $\phi$, such that it obtains a vacuum expectation value $|\langle \phi \rangle| = f\sqrt{2}$.  We may parameterise the two degrees of freedom in $\phi$ however we wish.  One option is $\phi = (f+\phi_r + i \phi_i)/\sqrt{2}$, however another equally valid parameterisation is $\phi = (f+\rho) e^{i a/f}/\sqrt{2}$.

%The action is thus
%\bea
%\mathcal{L} & = & |\partial_\mu \phi|^2 - \lambda (|\phi|^2 - f^2/2)^2  \\
%& = & \frac{1}{2} \left(1+\frac{\rho}{f} \right)^2 (\partial_\mu a)^2+  \frac{1}{2} (\partial_\mu \rho)^2   - \frac{\lambda}{4} \left( (f+\rho)^2 - f^2 \right)^2  ~~.
%\eea
Let's see how this works.  At first pass the field $\phi$ is massless, and enjoys a shift symmetry $\phi \to \phi + f$.  This is the `nonlinear' realisation of a $\U(1)$ symmetry, and so we identify $\phi$ as a Goldstone boson.  Now let us return to the quarks and charge them under this symmetry, such that they cannot have a bare mass term, but can only have a Yukawa interaction with a complex scalar $\Phi$, of which $\phi/f$ is the phase, as enforced by the $\U(1)$ symmetry.  Once the scalar obtains a vev, spontaneously breaking the symmetry, then we can see that the action for the quarks becomes
\bea
\mathcal{L}& = &   i \overline{\psi} \gamma^\mu D_\mu  \psi + m_\psi e^{i \theta_q } \overline{\psi}  \psi +h.c. +\left(\theta+ \phi/f \right) \frac{g^2}{32 \pi^2} \epsilon_{\mu\nu\alpha\beta} G^a_{\mu\nu} G^a_{\alpha\beta} \\
& \to &  i \overline{\psi} \gamma^\mu D_\mu  \psi + m_\psi e^{i\left( \theta_q + \theta+ \phi/f \right)} \overline{\psi}  \psi + h.c. ~~,
\eea
where in the last line we performed an anomalous chiral rotation to move the QCD angle into the quark mass term, and the hermitian conjugate is just an alternative way of writing action without the $\gamma_5$ matrix.

The important point is that the Goldstone boson enters the action in just the same way as the bare CP-violating angles.  With only these terms this field would be the axion, and we will refer to this $\U(1)$ symmetry as $\U(1)_{PQ}$, after Roberto Peccei and Helen Quinn, who spotted that this global symmetry had very interesting implications for the strong-CP problem.  Since the axion has a shift symmetry, we may happily shift away the angles $\phi \to \phi-f(\theta_q + \theta)$ such that the action is simply
\be
\mathcal{L} = i \overline{\psi} \gamma^\mu D_\mu  \psi + m_\psi e^{i \phi/f} \overline{\psi}  \psi + h.c.
\ee
This is, of course, relating a shift of the axion field to a quark chiral field rotation!  The overall background value of the axion field $\langle \phi/f \rangle$ is now the total physical strong-CP phase.  For example, the neutron electric dipole moment is simply proportional to this value $n_{EDM} \propto \langle \phi/f \rangle$.  What should this value be?

Lets see what happens when the quarks condense and work now within the SM.  We will not include the neutral pion field, associated with the spontaneous breaking of the chiral $\SU(2)$ symmetry, however one should consult \cite{diCortona:2015ldu} for a clear and up-to-date treatment including the pions.  The result in the SM for the approximation $m_u = m_d = m_q$ is that $ m_q \langle \overline{\psi} \psi \rangle = f_\pi^2 m_\pi^2$, thus the action becomes
\bea
\mathcal{L}  & = &  e^{i \phi/f} \langle m_\psi \overline{\psi}  \psi \rangle + h.c. \\
& \to & f_\pi^2 m_\pi^2 e^{i \phi/f} + h.c.
\eea
Thus the potential generated for the axion, within QCD, is
\be
V(\phi) = - f_\pi^2 m_\pi^2 \cos \left(  \frac{\phi}{f} \right)  ~~.
\ee
Note that this is a very non-trivial result.  We started with a global symmetry which was spontaneously broken, leading to a massless Goldstone boson.  However, this symmetry was \emph{anomalous} at the quantum level, under QCD.  This means that although in perturbation theory no mass would ever be generated for the axion, there was no obstruction to generating a mass non-perturbatively, and this is precisely what has happened:  The $\U(1)_{PQ}$ symmetry was not a true quantum symmetry of the theory, and when QCD became strongly coupled non-perturbative effects become large.  Since these effects need not respect the global symmetry, they need not respect the shift symmetry of the axion, and they can, and do, generate a potential and a mass for the axion.

This becomes the crucial insight for the relaxion, since the $\phi$-potential generated by QCD effects depends on the light quark masses, which in turn depend on the Higgs vacuum expectation value and this will provide the dynamical back reaction.  In practice this potential is
\bea
V_{QCD} & \sim & f_\pi^2 m_\pi^2 \cos \phi/f  \\
& \propto & f_\pi^3 m_q \cos \phi/f  \\
& \propto & f_\pi^3 \lambda_{u,d} \langle |H| \rangle \cos \phi/f ~~.
\label{eq:axion}
\eea

Let us now consider the vacuum structure of the theory for two values of $\phi$, including also the effect of the $g$-terms.  If $M^2  - g \phi > 0$ then the effective Higgs mass-squared is positive.  QCD effects will break electroweak symmetry, and quark condensation will lead to a tadpole for the Higgs field, which will in turn lead to a very small vacuum expectation value for the Higgs.  Thus in this regime the axion potential of \Eq{eq:axion} exists but is extremely suppressed.  If $M^2  - g \phi < 0$ the effective Higgs mass-squared will be negative and the Higgs will obtain a vacuum expectation value, so the height of the axion potential will grow proportional to the vev.

\begin{figure}[t]
\centering
\includegraphics[height=2.0in]{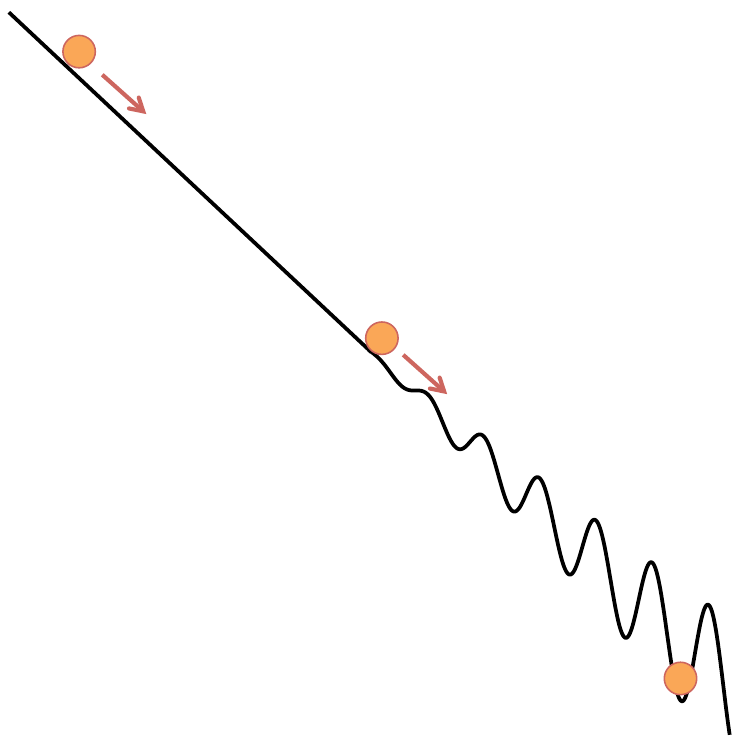}
\caption{Evolution of the relaxion field in the early Universe from a point where the effective Higgs mass-squared is postive (left), passing through zero (middle), and negative (right).}
\label{fig:relaxion}
\end{figure}

\subsubsection{Cosmological Evolution}
The general idea of the relaxion mechanism is sketched in \Fig{fig:relaxion}.  Imagine at the beginning of a period of inflation the relaxion field begins at values far from the minimum of the scalar potential.  We can, without loss of generality, take this to be at $\phi = 0$.  Due to its potential it will roll, with Hubble friction providing the necessary dissipation for this to occur in a controlled manner.  This Hubble friction can be understood from the equation of motion for a scalar in an inflating background
\be
\partial_t^2 \phi + 3 H \partial_t \phi \approx g M^2 + ... ~~,
\ee
where the ellipsis denotes higher order terms in $g$.  During inflation $H=$ const, and this term provides a constant source of friction, and for large $H$, one has a non-accelerating solution to the equations of motion $\phi \sim (g M^2/3 H) t$.  All the while the effective Higgs mass-squared is evolving.

Once the effective mass-squared passes through zero the Higgs will obtain a vacuum expectation value and the axion potential of \Eq{eq:axion} will turn on, growing linearly with the Higgs vev.  If the gradient of this potential becomes locally great enough to overcome the gradient of the $g$-induced relaxion potential, i.e.\
\be
\frac{f_\pi^3}{f} \lambda_{u,d} \langle |H| \rangle  > g M^2 ~~,
\ee
then the relaxion will stop rolling and become stuck.  Once it has become stuck the effective Higgs mass-squared has also stopped evolving.  If $g$ is taken to be appropriately small, then this evolution will cease at a point where the Higgs vev is small $\langle |H| \rangle \ll M$.  As $g$ is a parameter which can take values that are naturally small, and $g$ ends up determining the final Higgs vev, a naturally small value for the weak scale may be generated.

If it could be taken at face value, the picture painted above is quite a beautiful portrait involving SM and BSM symmetries and dynamics.  QCD plays a crucial role in determining the weak scale and solving the hierarchy problem.  Only an axion-like field, already motivated by the strong-CP problem, is added.   Inflation, which is already required in cosmology, provides the dissipation required for solving the hierarchy problem.  We even find an explanation for some other puzzles in the SM, such as why there are some quark masses determined by the weak scale which are nonetheless lighter than the QCD strong coupling scale.  However, as we will see, some puzzles remain to be understood, presenting a number of interesting areas to explore on the theoretical front.

\subsubsection{Parameter Constraints}
To determine the viability of the relaxion mechanism it is necessary to consider any constraints on the theory.  I will list them here.
\begin{itemize}
\item $\Delta \phi > M^2/g$:  For the relaxion to scan the entire $M^2$ of Higgs mass-squared it must traverse this distance in field space.
\item  $H_I > M^2/M_P$:  Inserting the previous $\Delta \phi$ into the potential we find that the vacuum energy must change by an amount $\Delta V \sim M^4$.  For the inflaton to dominate the vacuum energy during inflation we require $V_I > M^4$, which corresponds to the aforementioned constraint on the Hubble parameter during inflation.
\item  $H_I < \Lambda_{QCD}$:  For the non-perturbative QCD potential to form, the largest instantons, of size $l \sim 1/\Lambda_{QCD}$, must fit within the horizon.
\item $H_I < (g M^2)^{1/3}$:  Fluctuations in the relaxion field during inflation (due to finite Hubble) must not dominate over the classical evolution  if the theory is to predict a small weak scale.
\item $N_e \gtrsim H_I^2/g^2$:  Inflation must last long enough for the relaxion to roll over the required field range.
\item  $g M^2 f \sim \Lambda_{QCD}^4$:  It must be possible for a local minimum to form in the full relaxion potential whenever the Higgs vev is at the observed electroweak scale.
\end{itemize}
Combining these constraints it was found in \cite{Graham:2015cka} that the maximum allowed cutoff scale in the theory is
\be
M < \left( \frac{\Lambda^4 M^3_{Pl}}{f} \right)^{1/6} \sim 10^7 \text{ GeV} \times \left( \frac{10^9 \text{ GeV}}{f} \right)^{1/6} ~~.
\label{eq:cutoff}
\ee
It is compelling that such a large hierarchy can be realised within the relaxion framework.  Let us now saturate \Eq{eq:cutoff} and take $f=10^9$ GeV to explore the other parameters of the theory.  In this limit we find
\be
g \sim 10^{-26} \text{ GeV} ~~~,~ ~~  \Delta \phi \sim 10^{40} \text{ GeV} ~~~,~ ~~   5 \times 10^{-5} \text{ GeV} \lesssim H_I \lesssim 0.2 \text{ GeV}  ~~~, ~~~ N_e \gtrsim 10^{43} ~~.
\ee
All of these features are quite puzzling or unfamiliar.  As such they may represent interesting opportunities for continued theoretical investigation.  The parameter $g$ which explicitly breaks the shift symmetry is extremely small.  Recent work has shed some light on this question \cite{Ibanez:2015fcv}.  On a related note, the required field displacement is not only large, it is `super-duper Planckian' \cite{Kaplan:2015fuy}.  How such large field displacements can be accommodated by a story involving quantum gravity remains to be fully understood.

With regard to the inflationary aspects, the Hubble parameter is much smaller than is typical in inflationary models.  The number of e-foldings is huge (remember the scale factor grows during inflation by a factor $\sim e^{N_e}$).  Although not a problem in principle, it may be difficult to realise a natural inflationary model with the appropriate slow-roll parameters which reheats the Universe successfully and also accommodates the observed cosmological parameters.

A more tangible puzzle arises in the simplest QCD model presented above, as it is already excluded by experiment.  In the electroweak breaking vacuum the full relaxion potential will be minimized whenever
\be
\frac{\partial V_g}{\partial \phi} + \frac{\partial V_{QCD}}{\partial \phi} = 0  ~~,
\label{eq:min}
\ee
where $V_g$ is the scalar potential generated from the terms which explicitly break the shift symmetry, all originating from the parameter $g$, and $V_{QCD}$ is the axion-like potential coming from the non-perturbative QCD effects.  Since the relaxion is stopped by QCD effects before it reaches the minimum of $V_g$, the first term in \Eq{eq:min} is non zero.  This then implies that the second term in \Eq{eq:min} must also be non-zero.  By construction, $V_{QCD}$ is minimised whenever the effective strong-CP angle is zero, thus if it is not minimised the effective strong-CP angle must be non-zero.  In fact, it is typically expected to be close to maximal if the relaxion has stopped in one of the first minima that appears after the Higgs vev starts to grow.  This is in clear contradiction with experimental bounds on the strong-CP angle and so the model must be extended, and a number of options have been proposed.

\subsubsection{Summary}
The relaxion is not yet a complete story, so it is perhaps premature to include it in a lecture course.  However, it was the first major step towards a radically different perspective on the hierarchy problem, a perspective that may an important role in BSM theory for a long time to come, so it is included in these lectures.

\subsection{Self-Organised Localisation}
\begin{figure}[t]
\begin{center}
\includegraphics[width=0.95\columnwidth]{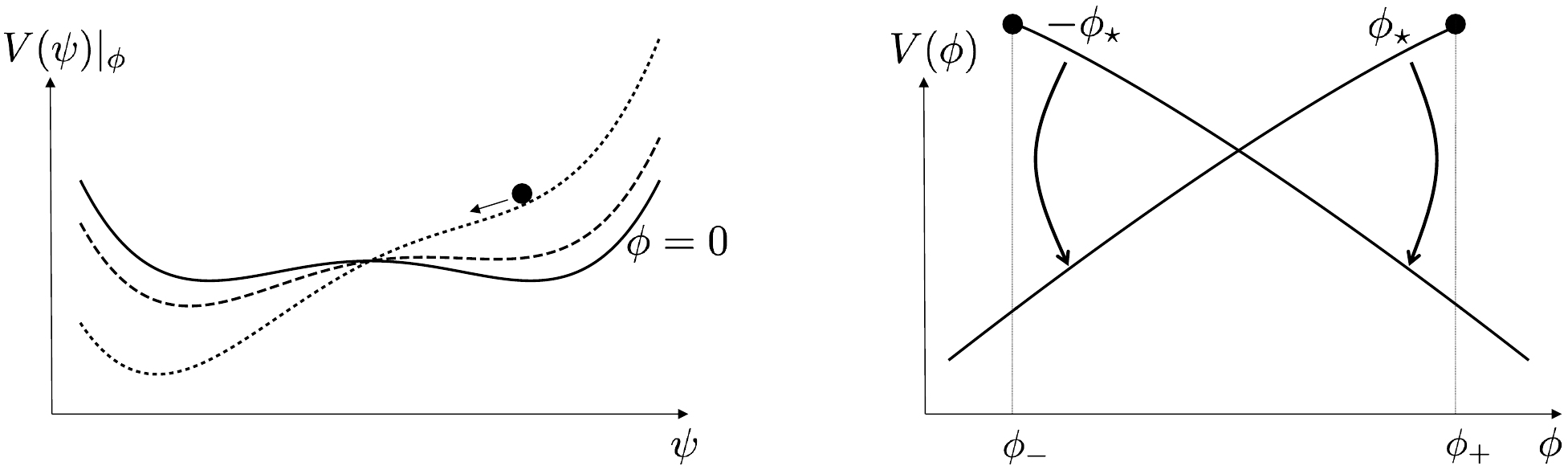} 
\end{center}
\caption{Vacuum structure in the Landau model, where $\phi$ plays an analogous role to an external magnetic field. Left panel: $V$ as a function of $\psi$ for different values of $\phi$. Right panel: $V$ as a function of $\phi$ with $\psi$ fixed at its two possible minima.}
\label{fig:1stvac}
\end{figure}

There is another aspect in which critical points are special, which is that they tend to exist at the top of the potential, as a function of some background field.  Let me be a bit more explicit.  Consider, as discussed at greater length in \cite{Giudice:2021viw}, the Landau model for a continuous, rather than discrete, spin variable (as in Ising model for a ferromagnet in the presence of an external magnetic field) with this potential
\beq
V=\frac{\lambda}{4} \left( \psi^2-\rho^2 \right)^2 + \kappa \phi \psi ~~,
\label{eq:Landau}
\eeq
where $\psi$ is the magnetisation and $\phi$ plays the role of the $Z_2$-breaking external magnetic field that we wish to vary.   The $\psi$ vacuum, as a function of $\phi$, has two configurations
\beq
\langle \psi \rangle_\pm =\pm \rho \,  C ( \phi / \phi_\pm ) ~~~~~{\rm for}~|\phi| < \phi_+
\eeq
\beq
\phi_\pm =\pm \frac{2\lambda \rho^3}{3\sqrt{3} \kappa} ~,~~~
C(x) = \left\{ 
\begin{array}{ll}
\frac{2}{\sqrt{3}} \, \cos \left[ \frac{\arccos (-x )}{3}\right] & {\rm for}~|x|<1  \vspace{0.2cm}\\
\frac{2}{\sqrt{3}} \, \cosh \left[  \frac{{\rm arcosh} (-x )}{3}\right] & {\rm for}~x < -1 
\end{array}
\right. ~.
\eeq
The two branches of the $\phi$ potential on these $\psi$ configurations are thus
\beq
V(\phi, \langle \psi \rangle_\pm )=
\frac{\lambda \rho^4}{4} \left[ 1+2 C^2 (\phi /\phi_\pm )-3C^4 (\phi /\phi_\pm )  \right] ~~~~~{\rm for}~|\phi| < \phi_+ ~,
\eeq
as shown in \Fig{fig:1stvac}. For $\phi$ large and negative, the true minimum of the $\psi$ potential is at $\psi_+$. As $\phi$ increases, the minimum is lifted and, at $\phi =0$, it becomes degenerate with the configuration $\psi_-$. However, $\psi_+$ persists as a local minimum in a `supercooled' phase beyond the critical point $\phi_c =0$ until it becomes classically unstable at $\phi = \phi_+$.

As shown in \Fig{fig:1stvac}, when $\phi$ increases beyond $\phi_+$, the metastable vacuum eventually disappears and $\psi$ rolls directly to the true vacuum. Depending on the dynamical timescales over which the external magnetic field is varied, $\psi$ may tunnel quantum-mechanically to the true vacuum long before $\phi_+$ has been reached, or the tunnelling process may be so slow that $\phi$ evolves along the supercooled branch all the way up to $\phi_+$  before proceeding to the true vacuum.  Nonetheless, we see that the highest point on the potential, as a function of $\phi$, is the point 
at which the higher `supercooled' vacuum for $\psi$ becomes an inflexion point, with the second derivative w.r.t. $\psi$ (i.e.\ the mass-squared) disappearing at that point!

This is a relatively common feature of quantum critical points, where `quantum' simply refers to the fact that it is some background field or parameter which is varying to go through the phase transition, rather than the temperature.  How, then, might this apply to particle physics?  Well first of all one would have to have some background scalar field, presumably light, which controls the parameters of the SM, most importantly the Higgs mass for our purposes.  Then one might hope to find that the critical point, which is the highest point on the potential, corresponds to a parameter choice with small Higgs mass.  The second requirement is that somehow we ought to find ourselves (our Universe, that is) at the top of the potential, not the bottom.

\subsubsection{Higgs Metastability}
\begin{figure}[t]
\begin{center}
\includegraphics[width=0.55\columnwidth]{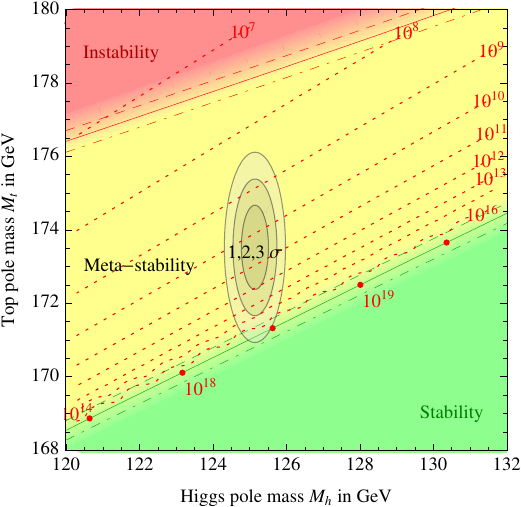} 
\end{center}
\caption{SM vacuum stability as a function of the Higgs and top quark mass, taken from \cite{Buttazzo:2013uya}.  The red contours show the value of the Higgs field, in natural units, at which the Higgs potential turns over and becomes lower than our present vacuum at small field values.}
\label{fig:meta}
\end{figure}

In \cite{Giudice:2021viw} it was argued that both requirements may be met in certain cosmological contexts.  To confront the first we will have to wrestle a little with Higgs metastability.  It has long been established that the SM vacuum we currently reside in could in fact be metastable \cite{Cabibbo:1979ay,Hung:1979dn,Lindner:1985uk,Sher:1988mj,Schrempp:1996fb,Altarelli:1994rb}.  In fact, given the status of present-day measurements of SM parameters such as the top and Higgs mass, it looks like our vacuum probably \emph{is} metastable \cite{Degrassi:2012ry,Buttazzo:2013uya,Bednyakov:2015sca,Andreassen:2017rzq}.  This can be understood by examining the quantum-corrected effective potential of the Higgs sector.  To a good approximation this is due to the RG evolution of the Higgs quartic interaction, where the evolution is not with respect to some momentum scale, instead being with respect to the Higgs field.  In other words, 
\be
V_h \approx \frac{\lambda(h)}{4}  h^4 - \frac{m_h^2}{2} h^2  ~~,
\ee
where
\bea
\frac{d\lambda(h) }{d \log h^2} & = & \beta_\lambda (h) \nonumber \\
& < & 0 ~~,
\eea
and I have not included the irrelevant logarithmic running of the Higgs mass since it is so small as to be irrelevant for this discussion.  Thus, ultimately, at large field values $h^2 \gg m_h^2/\lambda(v)$, the Higgs quartic coupling passes through zero and becomes negative; the SM Higgs potential actually turns over!  This is the reason our present vacuum may be unstable.  The running of the Higgs quartic is dominated by the contribution from the quartic itself and the top Yukawa, hence the Higgs and top quark mass are crucial parameters in establishing the nature of the Higgs vacuum at very large field values.  The present status of metastability is shown in \Fig{fig:meta}.

Now consider the Higgs mass-squared value.  In the SM it is negative and very very small as compared to the instability scale depicted in \Fig{fig:meta}.  However, imagine one increased its magnitude, becoming more and more negative until the point at which it is greater in magnitude than the instability scale.  In this case there would be no local minimum, metastable or otherwise, and the sole Higgs vacuum would be at large field values.  Thus we see that, due to vacuum metastability, the phase diagram of the SM, as we vary the Higgs mass, could look something like that of the ferromagnet in \Fig{fig:1stvac}, with the highest point being the critical point where one of the two vacua ceases to exist.

\begin{figure}[t]
\begin{center}
\includegraphics[width=0.55\columnwidth]{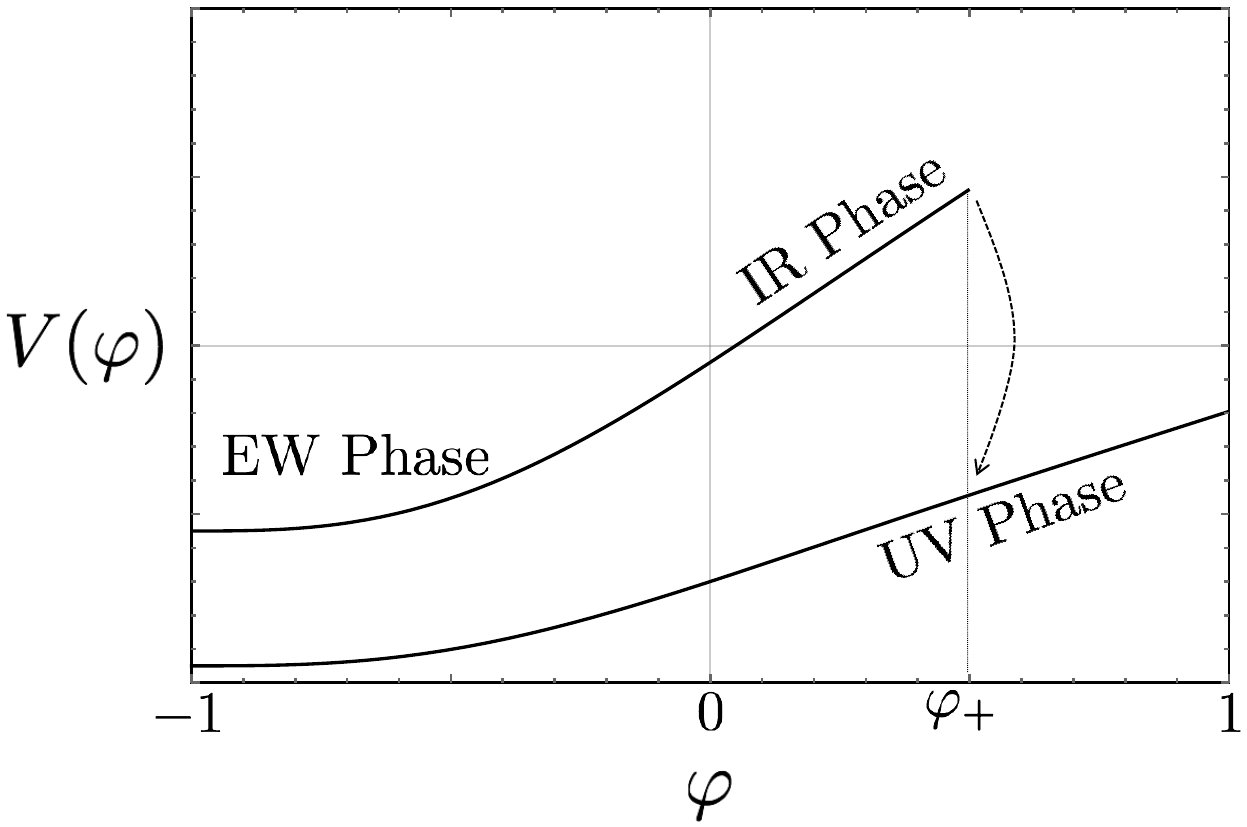} 
\end{center}
\caption{Vacuum structure of the theory in \Eq{eq:SOLH2}, where the Higgs mass is scanned by a scalar field up to, and beyond, the point at which the negative Higgs mass-squared surpasses the instability scale.}
\label{fig:SOLH2}
\end{figure}

Let's make this more concrete.  Suppose there is some very light dimensionless scalar field $\varphi= \phi/f$, where we have normalised relative to some field scale $f$, whose potential, including the Higgs sector, is of the form
\beq
V(\varphi) = M^2 f^2 F(\varphi) - \varphi M_H^2 |H|^2 +\frac{\lambda}{4} |H|^4 ~~,
\label{eq:SOLH2}
\eeq
where $F(\varphi)$ is some positive-gradient polynomial with $\mathcal{O}(1)$ coefficients and we have, without loss of generality, chosen the origin of field space to correspond to the point with vanishing Higgs mass.  Generically, the vacuum structure of this theory is of the form depicted in \Fig{fig:SOLH2}, where $\varphi_+$ is the field value at which the Higgs mass-squared becomes so large and negative that the SM Higgs vacuum no longer exists.  The upper branch corresponds to when the Higgs is in the SM-like vacuum at small field values, and the lower branch corresponds to when the Higgs rests in the high field-value vacuum.

Thus far, this very basic theory seems to have some success, in that the highest point on the potential for $\varphi$ naturally coincides with the SM instability scale.  This is interesting because the instability scale appears as the result of RG evolution, thus the instability scale can naturally be at scales exponentially below the ultimate mass scales associated with the UV cutoff of the theory, which we may denote $\Lambda$.  This is because the RG evolution of the quartic coupling is logarithmic.  Thus, if there were some reason that the Universe should find itself at the highest point on the potential then we would have a means for realising a natural hierarchy of mass scales.

\subsubsection{Inflation}
This is the juncture where we leave a well-marked road where, thanks to the power of EFT, we can be confident of what we're saying, and turn off onto a much more speculative gravel track.  This is possibly a moment for some broader perspective on model building.  It is a fundamentally speculative enterprise; we ought to be daring enough to speculate, but also conscious of the (relative) degree of speculation.  Read the last paragraph of `A Model of Leptons' by Weinberg \cite{Weinberg:1967tq} and you will find that daring theoretical leaps, employing speculative and, at times, unproven and potentially pathological theoretical structures, can occasionally be rewarded with truth.  However, daring leaps carry with them an inherent risk.  For the rest of this section the daring leap will be into the behaviour of light scalar fields during inflation.

During inflation spacetime is approximately de Sitter (dS).  As many of you will know, there are strong analogies between dS space and a thermal background, much as a black hole has a Hawking temperature \cite{Hawking:1975vcx}, dS has a Gibbons-Hawking temperature \cite{Gibbons:1976ue}.  Consider a light scalar field $\phi$, which is not necessarily the inflaton, in this background.  As with finite temperature, $\phi$ will experience fluctuations which depend on the dS horizon scale, which depends on the Hubble scale, i.e. $T\sim H$.  In fact, as Starobinsky \cite{Starobinsky:1985ibc,Starobinsky:1986fx} identified, the typical magnitude of a fluctuation during one Hubble time is of the order
\be
\delta^H \phi \sim \frac{H}{2 \pi} ~~.
\ee
On the other hand, during a Hubble time the field will have rolled a distance
\be
\delta^C \phi \sim -\frac{V'}{3 H^2}~~,
\ee
classically, due to slow-roll.  So, as time goes on the distribution of field values will have an average value that is evolving classically, but with an ever-increasing width due to the fluctuations.  Approximating this as a Gaussian we would thus expect
\be
P(\phi,t) \approx \sqrt{\frac{2 \pi}{H^3 t}} \exp^{- \frac{\left(\phi+ \frac{V' t}{3 H}\right)^2}{\frac{H^3}{2 \pi^2} t}} ~~,
\ee
assuming the initial field value was $\phi_I = 0$.  However, the patches that are further up the potential will also inflate more, since the inflationary rate is proportional to $\sqrt{V}$.  As a result, the \emph{volume}-weighted distribution, which is no longer a probability distribution, is expected to behave approximately as
\be
P_V(\phi,t) \approx \sqrt{\frac{2 \pi}{H^3 t}} \exp^{- \frac{\left(\phi+ \frac{V' t}{3 H}\right)^2}{\frac{H^3}{2 \pi^2} t}} \times \exp^{\frac{V' \phi}{2 H M_P^2} t}~~,
\ee
where I have dropped the overall $\exp{3 H t}$ term, and expanded to first order in $V'$.  Let's consider the gradient of this volume-weighted distribution.  Normalised to the positive distribution this gradient is
\be
\frac{P'_V(\phi,t)}{P_V(\phi,t)} \approx -\frac{4 \pi^2 (V'+3 H \phi/t)}{3 H^4} + \frac{t V'}{2 H M_P^2} ~~,
\ee
thus the position of the peak of the distribution is given by the field value where this gradient vanishes.  We thus see that the peak of the distribution initially begins evolving down the potential, however at a time $t \approx 4 M_P^2 \pi^2 / 3 H^2$ it turns around and begins climbing the potential!  This occurs at a potential value
\be
V_{\text{reverse}} \approx - \frac{2 (M_P \pi V')^2}{9 H^4} ~~,
\ee
thus, if the distribution has stayed within the realm of validity of the calculation whenever this turnaround occurs we can be relatively confident that the volume-weighted field distribution eventually climbs the potential, assuming the calculation is, itself, being correctly interpreted.  The only major approximation made was in Taylor expanding the potential, thus we require that $|V_{\text{reverse}}| \ll 3 M_P^2 H^2$.  Thus, if the condition
\be
|V'| \ll \sqrt{\frac{3}{2}} \frac{3 H^3}{\pi}
\ee
is satisfied then we would expect that the volume-weighted field distribution can actually climb up the scalar potential, ultimately peaking at the top.  See \cite{Rudelius:2019cfh,Giudice:2021viw} for more detailed discussions on this aspect.

Now I want to issue a big warning.  I would argue that the behaviour of very light scalar fields in dS is not yet understood.  There are many puzzles, paradoxes, concerns, eternal inflation, Boltzmann brains, youngness paradoxes, etc.  As a result, I would urge that you take the result of this calculation in the spirit it is intended; there are good reasons to believe that, if it even makes sense to do `statistics' on light fields in the multiverse, very light scalar fields may have multiversal distributions which are peaked at the top of their scalar potential.

\subsubsection{Summary}
We now appear to have all of the ingredients required to generate a natural mass-hierarchy for the Higgs, through the combination of a very light scalar field which scans the Higgs mass, inflation and the SM vacuum metastability.  There are many issues, not least that this doesn't actually work, since it predicts $m_h \sim \Lambda_\text{Instability}$.\footnote{Interestingly, one way to rectify this is to add vector-like fermions coupled to the Higgs which, curiously, is also the way to rectify the relaxion model too.}  But that isn't really the point, which is that it links a scalar mass to a scale which is generated by RG-evolution in a perturbative theory and, in any case, that aspect can be fixed up by making the SM more unstable!  The real, big, issues concern the cosmological aspects.  Because of these aspects one cannot view this approach as `complete'.  Rather, I present it to you to give you a sense of the sort of ingredients that could link cosmology and self-organised criticality for SM questions.

\subsection{NNaturalness}
The final set of ideas I would like to bring to your attention are contained in the `Nnaturalness' proposal \cite{Arkani-Hamed:2016rle}.  I have chosen to discuss this as it demonstrates a qualitatively different manner in which cosmology could play a role in explaining a mass hierarchy.  In this theory we suppose that our EFT, with cutoff $\Lambda$ contains a large number $N$ of SM-like sectors.  Note that they are essentially identical to the SM, but with a soft breaking of the resulting exchange symmetry that I will come to in due course.

What is the largest $\Lambda$ could be?  Normally we would expect this to be around the Planck scale, since that is the absolute last frontier for the SM, beyond which gravitational scattering amplitudes become strongly interacting and a UV-completion is called for.  However, at the quantum level all of these sectors play a role, hence gravitational interactions are renormalised and become strongly interacting at a scale $\Lambda^2 \sim g_\star^2 M_P^2 / N$, where $g_\star$ is some generic coupling, just to keep dimensions correct.  Thus the existence of these additional sectors actually brings the gravitational cutoff down, reducing the hierarchy with the EW scale.  You probably recognise that this is somewhat similar in spirit to the extra-dimensional approach to the EW-gravity hierarchy problem.  This aspect of the story is not all that new, in fact, having been emphasised long ago \cite{Dvali:2007hz,Dvali:2007wp}.  Note that it doesn't involve cosmology either.

\begin{figure}[t]
\begin{center}
\includegraphics[width=0.65\columnwidth]{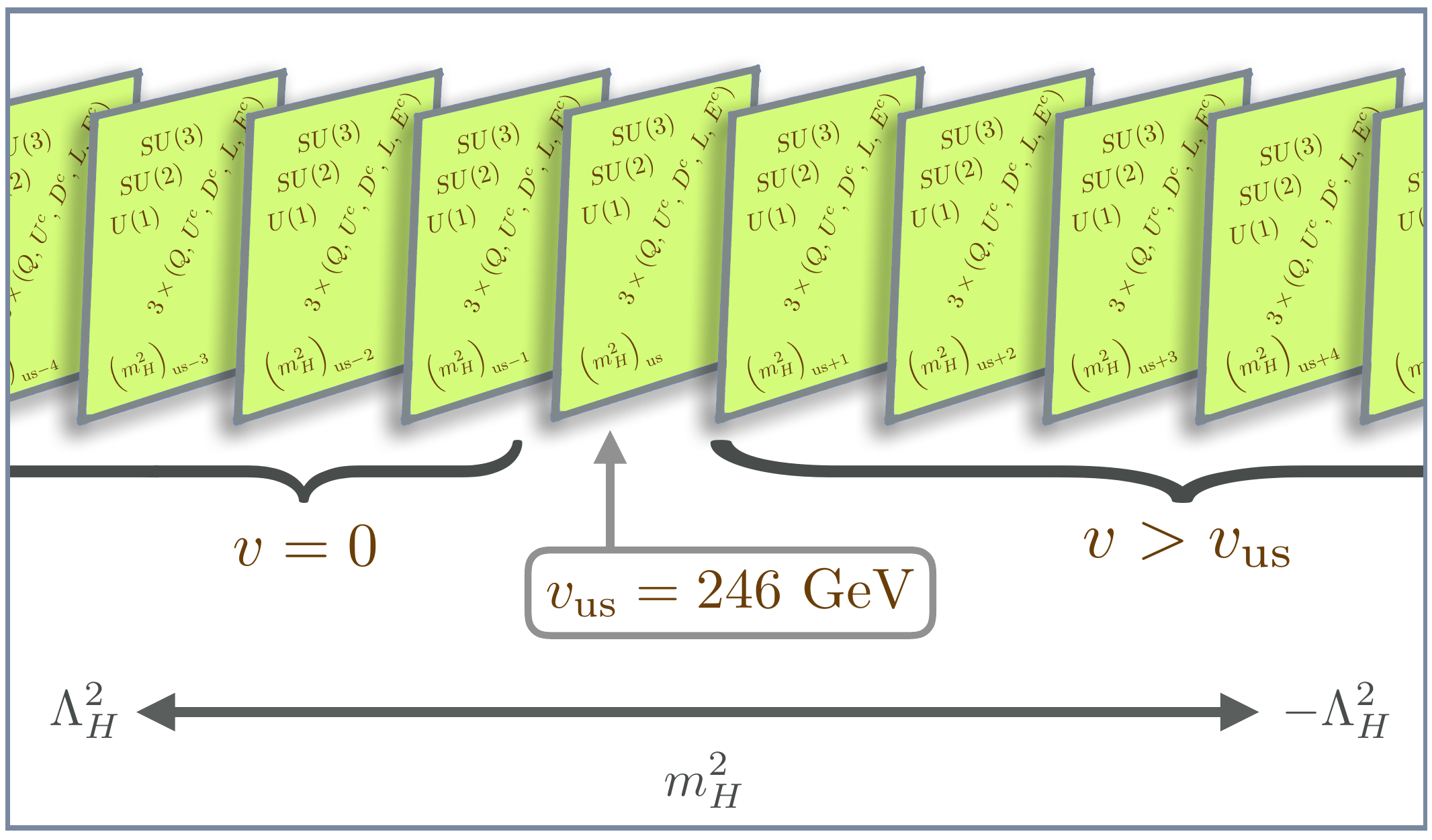} 
\end{center}
\caption{Depiction of the multiple sectors of an NNaturalness theory, taken from \cite{Arkani-Hamed:2016rle}.}
\label{fig:NNat}
\end{figure}

Now consider the various scalars in all these sectors.  We will assume the exchange symmetry will be softly broken by allowing for different Higgs masses in the different sectors. Typically we expect the Higgses to have mass-squared around the cutoff scale ($\sim \Lambda^2$).  However, there are many of them, so if the scalar masses are randomly distributed then occasionally accidents will happen, and some minority of the sectors will have accidentally light (fine-tuned) scalar masses, as depicted in \Fig{fig:NNat}.  Assuming a relatively flat distribution of bare masses the lightest scalar we would expect to encounter would have mass-squared $m_H^2 \sim \Lambda^2/N$, where I have suggestively identified this scalar, and sector, with the SM Higgs.  This is pretty neat, as the two effects combined can then explain a mass hierarchy
\be
\frac{m_H^2}{g_\star^2 M_P^2} \sim \frac{1}{N^2} ~~.
\ee
At this stage it might seem that the job is done, as far as explaining how a large hierarchy could come about naturally, we just need around $N \sim 10^{16}$ SM-like sectors to exist and a hierarchy to the Planck scale is naturally accommodated!  The problem, however, is that if all of these sectors got hot, like the SM, then they would totally over-close the Universe, looking nothing like the Universe we inhabit.  This is where a cute interplay between cosmology and naturalness can come into play.

In the standard inflationary picture for early Universe cosmology inflation is driven by some scalar field, the `Inflaton', rolling down a shallow potential.  The region of large vacuum energy density drives inflation, however in some patches the inflaton will roll right down to the bottom of the potential, begin to oscillate about the minimum of its potential, then decay.  This decay `reheats' the Universe to some high temperature $T_{\text{RH}} \gg 10$ MeV.  This last reheating step is essentially the moment the Big Bang went Bang.  In fact, it needn't be precisely the inflaton decay that causes reheating.  It could have been some other unstable particle that had a significant abundance at the end of inflation.  So we can generalise this picture and call the associated particle the `Reheaton'.

The big question then is how could this reheaton (hereafter $S$) preferentially reheat the lightest sector out of all the $N \sim 10^{16}$ of them, without any fine-tuning or gerrymandering funny business?  Consider the reheaton mass.  If it is a scalar then it can naturally be light; the simplest explanation being that it possesses an approximate shift symmetry $S\to S + $const, which is only broken by a small amount, proportional to $m_S^2/\Lambda^2$.  If it is a fermion then it can naturally be light due to an approximate chiral symmetry $S\to e^{i \theta} S$, again only broken by a small amount, this time proportional to the Majorana mass $m_S/\Lambda$.

Now we move on to the reheaton couplings.  Let's first assume that it couples more or less universally to all sectors.  At low energies the couplings which will dominate are the lowest dimension ones, thus $S^2 |H|^2$ for the scalar and $S H L$ for the fermion.  Note that both involve the Higgs.  Thus, for those sectors in which the Higgs-like fields are heavy the reheaton decays into those sectors will be strongly suppressed by powers of the reheaton mass over the Higgs mass.  Thus the reheaton will decay predominantly into the sector with the lightest Higgs!

In fact, one requires the reheaton masses to be close to the weak scale, otherwise the next highest up sector becomes too populated to be consistent with cosmological observations.  This is in some sense the least appealing feature: the reheaton mass and the lightest Higgs mass are uncorrelated parameters, yet for the model to work they must be relatively coincident.  This isn't a fine-tuning, but it is an odd coincidence.

In summary, through a coincidence of scales and a very large number of SM copies one can understand why the Higgs of the SM would be significantly, naturally, lighter than the cutoff of the theory \emph{and} why the sector with the lightest Higgs is the only one which is significantly reheated, giving rise to the observed cosmology.

\subsection{Summary}
That completes the ideas I would categorise as being `self-organised criticality', very broadly defined.  I have not been exhaustive and there are other and related ideas on the market (see e.g.\ \cite{Geller:2018xvz,Giudice:2019iwl,Strumia:2020bdy,Csaki:2020zqz,TitoDAgnolo:2021nhd,Khoury:2021zao,TitoDAgnolo:2021pjo}), however I intentionally hand-picked those which I think, combined, give good coverage of the range of elements that could be at play.

I must admit that I don't think any of the ideas I presented here is complete, particularly in a conceptual respect.  However, I believe their value is in showing that it is, in principle, possible or even plausible that the scale hierarchy we observe could have its origins in something very different than the symmetry-based scenarios such as SUSY or pNGB Higgs models.  This is very important as it shifts the narrative and widens the phenomenological goalposts.  In 2014 I would have been sceptical if I had been told that it's possible cosmology plays a role in explaining the EW hierarchy.  Nowadays it wouldn't surprise me at all if a complete and convincing cosmological narrative came along tomorrow.  Watch this space!

\section{Relativity, Extended?}\label{chap:SUSY}
Finally, we come to Supersymmetry (SUSY).  It is surely a topic you have heard of.  You have also presumably heard of the broad, but not unanimous, anticipation that SUSY would discovered at the LHC?  Well, it wasn't, and this has led to a lot of head scratching.  In the spirit of these lectures I plan to first describe the basics of the theoretical structure, which remains of interest.  If it hasn't been discovered so far, could it be discovered by the end of HL-LHC?  I will aim to answer as best I can.  But first, the theory...

In the last section we saw that a Higgs mass well below the cutoff can be explained if the Higgs is a Goldstone boson of a spontaneously broken global symmetry, but what about spacetime symmetries?  In 1967 Coleman and Mandula proved that it is impossible to combine the Poincar\'{e} and internal symmetries in any but a trivial way.  Intriguingly, this proof only applied to Lorentz scalar, i.e.\ bosonic, internal symmetries, and in 1975 Haag, Lopuszanski, and Sohnius showed that, in addition to internal and Poincar\'{e} symmetries, it is possible to extend the Poincar\'{e} symmetry to include spin-1/2 generators in a consistent quantum field theory.  This extension is known as supersymmetry.  See \cite{Martin:1997ns,Baer:2006rs,Weinberg:2000cr} for standard texts on SUSY.

Any continuous symmetry has generators, and as with global symmetries, the supersymmetry generators must commute with the Hamiltonian, but also convert fermionic states to bosonic states, and vice-versa.  We call the SUSY generators $Q_\alpha$ ($\alpha = 1,2$) and their complex conjugate $\overline{Q}_{\dot{\alpha}}$ ($\dot{\alpha}=\dot{1},\dot{2}$).  These are spinor quantities, and obey the commutation and anti-commutation relations

\begin{eqnarray}
[P^{\mu},Q_\alpha ]  & = &  [P^{\mu}, \overline{Q}_{\dot{\alpha}}] = 0      \\
\{ Q_\alpha, Q_\beta \}   & = & \{ \overline{Q}_{\dot{\alpha}}, \overline{Q}_{\dot{\beta}} \} = 0\\
\{ Q_\alpha, \overline{Q}_{\dot{\beta}} \}   & = & 2 \sigma^\mu_{\alpha \dot{\beta}} P_\mu
\label{SUSYanticomm}
\end{eqnarray}

where $P^\mu$ is the usual generator of translations, $\partial_\mu$.  I have shown only the commutation and anti-commutation relations for one set of supercharges, e.g.\ $\mathcal{N}=1$ SUSY, however it is straightforward to generalise these relations to more supercharges.  I will continue to focus on the case of $\mathcal{N}=1$ SUSY throughout this section.  As these generators change the spin of a state by a unit of $1/2$, one would expect that in a supersymmetric theory states come with some sort of `multiplet' structure, in which there is a state of spin $S$ and a state of spin $S+1/2$, where $S = 0,1/2$ for a renormalizable theory.  These multiplets are called `supermultiplets', and we will now consider how they are constructed.

In order to begin constructing such multiplets it is instructive to begin by considering the supersymmetry algebra as a graded Lie algebra.  By extending the analogy with space-time translations, we define the group element
\begin{equation}
G(x,\theta,\overline{\theta}) = e^{-i (x_\mu P^\mu - \theta Q - \overline{\theta} \overline{Q})} ~~,
\end{equation}
where $\theta$ and $\overline{\theta}$ are anti-commuting parameters.   Now, using Hausdorff's formula one can show that under a transformation with parameters $\{ \zeta, \overline{\zeta} \}$ we have the set of transformations
\begin{eqnarray}
x^\mu   & \rightarrow &  x^\mu + i \theta \sigma^\mu \overline{\zeta}  - i \zeta \sigma^\mu \overline{\theta}     \\
\theta   & \rightarrow & \theta + \zeta   \\
\overline{\theta} & \rightarrow & \overline{\theta}+\overline{\zeta} ~~.
\label{SUSYtrans}
\end{eqnarray}

This transformation in parameter space can be generated by the differential operators $Q$ and $\overline{Q}$

\begin{equation}
\zeta Q + \overline{\zeta} \overline{Q}   = \zeta^\alpha \left ( \frac{\partial}{\partial \theta^\alpha} - i \sigma_{\alpha \dot{\alpha}}^\mu \overline{\theta}^{\dot{\alpha}} \partial_\mu     \right ) + \overline{\zeta}_{\dot{\alpha}} \left ( \frac{\partial}{\partial \overline{\theta}_{\dot{\alpha}}} - i \theta^\alpha \sigma_{\alpha \dot{\beta}}^\mu \epsilon^{\dot{\beta} \dot{\alpha}} \partial_\mu     \right ) ~~.
\end{equation}

Again, by analogy with fields which are functions of space-time co-ordinates, we can define a superfield as a field which is a function of the co-ordinates $\{ x, \theta, \overline{\theta} \}$.  Henceforth we will write superfields in bold font, and their component fields in plain font.  As $\theta$ and $\overline{\theta}$ are Grassmann parameters, the Taylor expansion of a superfield in these co-ordinates terminates as, e.g.\ $\theta_1 \theta_1 =0$.  Thus, calling our superfield $\mb{F} (x, \theta, \overline{\theta})$, and expanding in the Grassmann parameters, we have

\begin{eqnarray}
\mb{F}(x,\theta,\overline{\theta})  &=& f(x) + \theta \phi(x) + \overline{\theta} \overline{\chi} (x) + \theta \theta m(x) + \overline{\theta} \overline{\theta} n(x) + \theta \sigma^\mu \overline{\theta} v_\mu (x) \nonumber \\
&& + \theta \theta \overline{\theta} \overline{\lambda} (x) + \overline{\theta} \overline{\theta} \theta \psi (x) + \theta \theta \overline{\theta} \overline{\theta} d(x) ~~,
\label{superfield}
\end{eqnarray}
which transforms under a SUSY transformation as
\begin{equation}
\delta_\zeta \mb{F}(x,\theta,\overline{\theta})\equiv (\zeta Q + \overline{\zeta} \overline{Q}) \mb{F} ~~.
\label{SUSYderiv}
\end{equation}
By comparing individual powers of $\theta$ after applying the SUSY transformation of eq.(\ref{SUSYderiv}) we can determine how the individual fields transform.  Also, as the Taylor expansion in $\theta$ terminates, we can see that the product of two, or more, superfields must itself be a superfield, where the individual component fields are products of component fields of the original `fundamental' superfields.

Now we have a linear representation of the SUSY algebra, however, this representation can be reduced.  We define a chiral superfield, $\mb{\Phi}$, by the constraint $\overline{D}_{\dot{\alpha}} \mb{\Phi} = 0$, where $\overline{D}_{\dot{\alpha}} = - \partial/\partial \overline{\theta}^{\dot{\alpha}} - i \theta^\alpha \sigma_{\alpha \dot{\alpha}}^\mu \partial_\mu$.  Thus our chiral superfield takes the form
\begin{eqnarray}
\mb{\Phi}(x,\theta)  &=& A(x)  + i \theta \sigma^\mu \overline{\theta} \partial_\mu (x) A(x)+ \frac{1}{4} \theta \theta \overline{\theta} \overline{\theta} \partial_\mu \partial^\mu A(x)   \nonumber \\
&& + \sqrt{2} \theta \psi(x) + \frac{i}{\sqrt{2}} \theta \theta \partial_\mu \psi \sigma^\mu \overline{\theta} + \theta \theta F(x)
\label{chiral}
\end{eqnarray}
and an anti-chiral superfield takes a similar form, following from $D_\alpha \mb{\Phi}^\dagger = 0$.

The components of the superfield in eq.(\ref{chiral}) transform under SUSY transformations as
\begin{eqnarray}
\delta_\zeta A   & = &  \sqrt{2} \zeta \psi     \\
\delta_\zeta \psi   & = & i \sqrt{2} \sigma^\mu \overline{\zeta} \partial_\mu A + \sqrt{2} \zeta F   \\
\delta_\zeta F & = & i \sqrt{2} \overline{\zeta}  \overline{\sigma}^\mu \partial_\mu \psi ~~.
\label{SUSYtranscomponents}
\end{eqnarray}
From these transformations we see that the $F$-term transforms into a total derivative.  If all fields vanish at infinity then the $F$-term of a chiral superfield thus forms a SUSY-invariant Lagrangian term.  It follows that the $F$-term of any product of chiral superfields is a SUSY-invariant Lagrangian term.  In addition, the $\theta^2 \overline{\theta}^2$ term in $\Phi^\dagger \Phi$ also transforms into a total derivative.  This term is then also a candidate for a SUSY-invariant term in the Lagrangian, and is given in component form as
\begin{eqnarray}
\mb{\Phi}^\dagger \mb{\Phi} |_{\overline{\theta}^2 \theta^2} & = & F^\ast F + \frac{1}{4} A^\ast \partial^2 A + \frac{1}{4} \partial^2 A^\ast A - \frac{1}{2} \partial_\mu A^\ast \partial^\mu A \nonumber\\
&& + \frac{i}{2} \partial_\mu \overline{\psi} \overline{\sigma}^\mu \psi - \frac{i}{2} \overline{\psi} \overline{\sigma}^\mu \partial_\mu \psi
\label{kinetic}
\end{eqnarray}
clearly giving the kinetic terms for the individual component fields.

Thus we are in a position to construct a SUSY-invariant theory with chiral superfields.  We can introduce one further ingredient which simplifies notation.  Defining $\int d \theta = 0$ and $\int \theta d \theta = 1$, then we can write our supersymmetric Lagrangian as
\begin{equation}
\mathcal{L} = \int d^2 \overline{\theta} d^2 \theta \mb{\Phi}^\dagger_i \mb{\Phi}_i + \left[ \int d^2 \theta (  f_{i} \mb{\Phi}_i+m_{ij} \mb{\Phi}_i \mb{\Phi}_j + \lambda_{ijk} \mb{\Phi}_i \mb{\Phi}_j \mb{\Phi}_k ) + \text{h.c} \right] ~~,
\label{eq:SUSY}
\end{equation}
where the first term is usually referred to as the \kahler potential, $K$, and the second term is the Superpotential, $W$.  The former picks up the $\theta^2 \overline{\theta}^2$ term in $K$, and the latter the $\theta^2$ term in $W$.

We can re-write this Lagrangian as
\begin{equation}
\mathcal{L} = \int d^2 \overline{\theta} d^2 \theta K (\mb{\Phi}^\dagger_i,\mb{\Phi}_i) + \int d^2 \theta W (\mb{\Phi}_i) + \int d^2 \overline{\theta} W^\ast (\mb{\Phi}^\ast_i) ~~,
\end{equation}
where $W$ is a function of chiral superfields only, and not their conjugates.  Because of this we say that the superpotential is a `holomorphic' function of the chiral superfields.  By defining $W^i = \frac{\partial W}{\partial A_i} |_{\theta=0}$, and $W^{i j}$ by analogy, then we find
\begin{equation}
\int d^2 \theta W (\mb{\Phi}_i) = W^i F_i -\frac{1}{2} W^{i j} \psi_i \psi_j - (\text{total derivative}) ~~.
\end{equation}

By inspecting the kinetic terms for the component fields in eq.(\ref{kinetic}) we can see that there are no derivative terms for the field $F$, and thus it does not propagate.  We can then simplify the supersymmetric Lagrangian by solving the Euler-Lagrange equation for $F$, i.e.\ solving $\partial \mathcal{L} / \partial F = 0$.  After performing this final step we find that $F_i = - W^{\ast i}$.  Using this, rearranging total derivative terms, and employing the equations of motion, our final supersymmetric Lagrangian, in component form, is
\begin{equation}
\mathcal{L} = \partial_\mu A^{\ast i} \partial^\mu A_i + i \overline{\psi}^i \overline{\sigma}^\mu \partial_\mu \psi_i - \frac{1}{2} \left( W^{i j} \psi_i \psi_j +  W^{\ast}_{i j} \overline{\psi}^i \overline{\psi}^j \right) - W^{\ast i} W_i ~~.
\end{equation}
This completes the construction of theories with $\mathcal{N}=1$ supersymmetry containing scalars and fermions.  It is quite remarkable that one can package fields in such a way that whatever you do, if the theory is written as in \Eq{eq:SUSY} the theory will inevitably be supersymmetric!

In order to include gauge interactions we must also construct theories with vector fields, which are contained in real vector superfields.  These arise by considering a superfield, $\mb{V}$, which is constrained for reality to satisfy $\mb{V}^\ast = \mb{V}$.  Such a superfield can be constructed from the general superfield in eq.(\ref{superfield}).  The general form for $\mb{V}$ contains numerous component fields, however it is possible to remove some of these fields by performing a suitable gauge transformation.  The supersymmetric generalisation of an Abelian gauge transformation acts on $\mb{V}$ as $\mb{V} \rightarrow \mb{V} + \mb{\Lambda} + \mb{\Lambda}^\ast$, where $\mb{\Lambda}$ is a chiral superfield.  It can be seen by comparing eq.(\ref{superfield}) and eq.(\ref{chiral}) that this will correspond to a gauge transformation of $v_\mu (x) \rightarrow v_\mu (x) + \partial_\mu (A (x) - A^\ast (x))$, as expected.

By choosing the `Wess-Zumino' gauge, in which the extra unwanted component fields are gauged away, we have
\begin{equation}
\mb{V} = - \theta \sigma^\mu \overline{\theta} v_\mu (x) + i \theta \theta \overline{\theta} \overline{\lambda} (x) -i \overline{\theta} \overline{\theta} \theta \lambda (x) + \frac{1}{2} \theta \theta \overline{\theta} \overline{\theta} D(x) ~~,
\label{vectorsuperfield}
\end{equation}
where $v_\mu (x)$ is a vector field, $\lambda (x)$ is its fermionic partner, the `gaugino', and $D(x)$ is a scalar field.  We see that $\mb{V}$ is the supersymmetric generalisation of the Yang-Mills potential $A^\mu$.  The transformation of the component fields under supersymmetry can again be calculated, and it is found that the field $D(x)$ transforms into a total derivative.  We will also need to construct a gauge-invariant field strength.  From eq.(\ref{vectorsuperfield}) we see that the lowest gauge-invariant components are $\lambda$ and $\overline{\lambda}$.  Hence we can construct a gauge-invariant chiral superfield $\mb{W}_\alpha = - \frac{1}{4} \overline{D} \overline{D} D_\alpha \mb{V}$, where chirality follows from $\overline{D}_{\dot{\beta}} \mb{W}_\alpha = 0$.

Now we can construct supersymmetric kinetic terms for the gauge fields as
\begin{equation}
\int d^2 \theta \frac{1}{4} \mb{W}^\alpha \mb{W}^\alpha + \int d^2 \overline{\theta} \frac{1}{4} \overline{\mb{W}}^{\dot{\alpha}} \overline{\mb{W}}_{\dot{\alpha}} = \frac{1}{2} D^2 -\frac{1}{4} F^{\mu \nu} F_{\mu \nu} - i \lambda \sigma^\mu \partial_\mu \overline{\lambda} ~~.
\label{gaugekinetic}
\end{equation}
This completes the construction of an Abelian SUSY gauge sector, however it now remains to include gauge interactions with matter fields.

The lowest component of a chiral superfield is a complex scalar field, which will transform under Abelian gauge transformations by multiplication of a space-time dependent phase.  It is clear, then, that in the language of superfields, a gauge transformation of a chiral superfield will proceed as $\mb{\Phi} \rightarrow e^{-i g \mb{\Lambda}} \mb{\Phi}$.  By considering the gauge transformation of a vector superfield we then see that the combination $\mb{\Phi}^\ast e^{g \mb{V}} \mb{\Phi}$ is gauge invariant, however $\mb{\Phi}^\ast \mb{\Phi}$ is not.  Therefore, to construct a supersymmetric theory with gauge interactions we use the gauge kinetic terms of eq.(\ref{gaugekinetic}), we impose that the superpotential $W$ is gauge invariant, and we adapt the \kahler potential terms to the form $\mb{\Phi}^\ast \mb{\Phi} \rightarrow \mb{\Phi}^\ast e^{g \mb{V}} \mb{\Phi}$

Finally, from eq.(\ref{gaugekinetic}) we see that we have a new non-propagating auxiliary field, $D$.  Once again we can solve for $\partial \mathcal{L} / \partial D = 0$ and find that $D = g \sum_i q_i A^{\ast i} A_i$.  After performing this final simplification, rearranging total derivative terms, and extending to the case where the chiral superfields transform under a non-abelian gauge symmetry, we have a supersymmetric gauge theory, with Lagrangian

\begin{eqnarray}
\mathcal{L} & = &  -\frac{1}{4} F^{a \mu \nu} F^a_{\mu \nu} + i \overline{\lambda}^a \sigma^\mu D_\mu \lambda^a + D_\mu A^{\ast i} D^\mu A_i + i \overline{\psi}^i \overline{\sigma}^\mu D_\mu \psi_i  \nonumber \\
& & + i \sqrt{2} g \sum_a (A^{\ast i} T^a \psi_i \lambda^a - \overline{\lambda}^a T^a A_i \overline{\psi}^i) - \frac{g^2}{2} \sum_a (\sum_i A^{\ast i} T^a A_i)^2  \nonumber \\
& & - \frac{1}{2} \left( W^{i j} \psi_i \psi_j +  W^{\ast}_{i j} \overline{\psi}^i \overline{\psi}^j \right) - W^{\ast i} W_i ~~,
\label{fullSUSYgauge}
\end{eqnarray}
where $D_\mu$ is the gauge-covariant derivative, and $T^a$ is a generator of the non-abelian gauge group.  This completes the construction of supersymmetric gauge theories.\footnote{Recall from \Sec{sec:EFTinQFT} that, due to $\hbar$ counting, since in a SUSY gauge theory the only dimensionful coupling is the gauge coupling then this is the only form the scalar quartic interactions could have taken, up to some overall numerical factor.}

\subsubsection{R-Symmetry}
It is possible, but not required, that a supersymmetric theory can also possess a global $U(1)$ symmetry under which $\theta$ transforms.  This symmetry is usually referred to as an $R$-symmetry, and it is special as it distinguishes between components of a supermultiplet.  If this symmetry is present, and $\theta$ has charge $q_R$ then the superpotential has charge $2 q_R$ and individual physical components of a chiral or vector superfield differ in charge by a unit of $q_R$.  Any other global symmetries act on the individual components of a chiral multiplet in the same way, and do not act on vector multiplets.

\subsubsection{Quadratic divergences}
One of the most attractive features of SUSY is the absence of quadratic divergences and hence the compelling link with the hierarchy problem. This can be explained quite simply.  In a supersymmetric theory in flat space the masses of fields in a SUSY multiplet are equal, by SUSY.  Fermion masses do not receive quadratic corrections to their mass thus, by tying the mass of the scalar to the mass of the fermion the scalar itself cannot receive quadratic corrections to it's mass.

There is another way of seeing this, where we also learn about something known as the `non-renormalization theorem'.  As shown before, the superpotential is a holomorphic function of the chiral superfields.  In addition, any relevant operators must arise from the superpotential in a renormalizable SUSY theory.  We can employ the usual spurion trick that has already come up many times and consider any parameters as SUSY-preserving vacuum expectation values (vevs) of some background chiral superfield.  We can then write our superpotential with the understanding that all parameters are actually vevs of fields, and assign global symmetry charges to these vevs to find the selection rules.  For example, we can consider a toy model with superpotential
\begin{equation}
W  =  \frac{m}{2} \mb{\phi}^2 + \frac{\lambda}{3} \mb{\phi}^3 ~~.
\label{nonrenorm}
\end{equation}
This theory has a global $U(1)_R$ symmetry and a global $U(1)$ symmetry, which are both broken by non-zero values for $m$ and $\lambda$.  We can still use the selection rules that arise as a result of these symmetries and write down all renormalizable, holomorphic, terms which behave well in the limits $m\rightarrow 0$ and $\lambda \rightarrow 0$.  Doing this we find that the only superpotential terms that are allowed are those already in eq.(\ref{nonrenorm}).  Thus if we consider renormalizing this theory down to some scale then no new terms can arise in the superpotential involving the cut-off.  This has been proven at a greater level of rigour for SUSY theories using supergraph techniques \cite{Amati:1988ft,Shifman:1986zi,Shifman:1991dz,Grisaru:1979wc}, and using the holomorphicity of the superpotential \cite{Seiberg:1993vc,Seiberg:1994bp}, and is in general referred to as the `Non-renormalization' theorem.

The \kahler potential gives the standard kinetic terms, which are still renormalized, giving rise to wavefunction renormalization.  Therefore terms in the superpotential are only renormalized through wavefunction renormalization.  Wavefunction renormalization is only logarithmic in the cut-off, hence no quadratic divergences occur in this theory.  Again, it can be shown, along these lines, that this is true in general for SUSY theories.

\subsection{Supersymmetry breaking}
As we have not observed any scalar particles with electric charge $-1$ and a mass of $511$ keV we must conclude that the Universe is not supersymmetric, i.e.\ supersymmetry is broken.  However, this does not mean that supersymmetric theories don't offer a resolution to the hierarchy problem:  If supersymmetry is restored at high energies then the hierarchy problem is relieved to the point that the only troublesome hierarchy is between the electroweak scale and the scale at which the theory becomes supersymmetric.

If we want a theory in which a symmetry is present at high energies, but apparently absent at low energies, we require that the symmetry is spontaneously broken somewhere along the way.  As supersymmetry is inherently tied to space-time symmetries we must be careful if we want to break supersymmetry spontaneously but not Lorentz symmetry.  From the last of the anti-commutation relations in eq.(\ref{SUSYanticomm}) we see that the vacuum energy, $P_0$, is given by
\begin{equation}
H = P_0 = \frac{1}{4} (\{ Q_1, \overline{Q}_{\dot{1}} \} + \{ Q_2, \overline{Q}_{\dot{2}} \}) ~~.
\end{equation}
As a result, in a globally supersymmetric theory, $\langle 0 | H | 0 \rangle \neq 0$ implies that $Q_\alpha | 0 \rangle \neq 0$ or $\overline{Q}_{\dot{\alpha}} | 0 \rangle \neq 0$, and supersymmetry is broken since the vacuum is not annihilated by the supercharge.  If we want to find a vacuum in which supersymmetry is spontaneously broken we must then find one with non-vanishing energy density.  If we want to maintain Lorentz symmetry then the only fields which obtain vacuum expectation values (VEVs) must be Lorentz scalars, hence the only candidate terms are from the scalar potential.  However the scalar potential comes from $V_{\text{scalar}} = \frac{1}{2} F^{\ast i} F_i + \frac{g^2}{2} D^a D^a$.  Thus we know that for a supersymmetric theory to spontaneously break supersymmetry requires a cosmologically stable vacuum in which $F_i \neq 0$ or $D^a \neq 0$.

By analogy with spontaneously broken global symmetries, which give rise to a massless Nambu-Goldstone boson, when global SUSY is spontaneously broken this leads to a massless Nambu-Goldstone fermion, named the `Goldstino'.  Why this is so can be seen quite simply for $F$-term breaking of SUSY.  At the minimum of the scalar potential we require that $dV/dA_i = 0$ and this implies that $W^{\ast}_i W^{i j} = 0$.  If there is $F$-term SUSY breaking then $\text{Abs} [W^{\ast}_i] \neq 0$, and hence $W_{i j}$ has a zero eigenvalue, with eigenvector $W^{\ast}_i$.  But the fermion mass matrix is given by $W_{i j}$, and, as a result, there must exist a massless fermion, which lives in the chiral multiplet that breaks SUSY.  A similar argument applies for $D$-term breaking, however in this case the goldstino is a gaugino of a vector multiplet.

The spontaneous breaking of supersymmetry leads to mass-splittings between component fields of a superfield.  It can be shown that in a theory with spontaneous SUSY breaking a mass-sum rule, $\text{Tr} [ M^2_{\text{scalars}} ] = \text{Tr} [ M^2_{\text{fermions}} ]$, where the scalars are real, is obeyed.  This rule implies that if SUSY is broken spontaneously in the visible sector we should have observed scalars as light as the lightest fermions.  As these scalars have not been observed then SUSY must be broken in another sector, and then this SUSY-breaking must be communicated to the visible sector, raising the masses of the unobserved superpartners.

This pattern of SUSY-breaking can be accounted for if we allow for some `spurion' superfield, $\mb{X}$, with non-zero $F$-term in the vacuum, i.e.\ $\langle \mb{X} \rangle = \theta^2 F_X$.  Alternatively one can consider a vector superfield with a non-zero $D$-term.  If some `messengers' which communicate between the SUSY-breaking sector and the visible sector have mass $M_M \gg M_{\text{weak}}$ we can integrate them out, including their effects by considering the effective field theory with higher dimension operators involving the field $\mb{X}$ and the visible sector fields.  Operators such as
\begin{equation}
K \supset \frac{\mb{X}^\dagger \mb{X}}{M_M^2} \mb{\Phi}^{\ast}_i \mb{\Phi}_i ~~, ~~ W \supset \frac{\mb{X}}{M_M} \mb{\Phi}_i \mb{\Phi}_j \mb{\Phi}_k ~~, ~~ W \supset \frac{\mb{X}}{M_M} \mb{W}^\alpha \mb{W}_\alpha ~~,
\end{equation}
lead to SUSY-breaking mass-terms for the scalars of a chiral superfield, $\tilde{m} = F_X/M_M$, trilinear scalar interactions, $|A_{i j k}| = F_X/M_M$, or mass terms for the gauginos in a vector superfield, $M_\lambda = F_X/M_M$.  All such terms break supersymmetry `softly', as they do not introduce new quadratic $UV$-divergences into the theory, and only lead to quadratic divergences up to the scale of the soft-terms.

The messenger superfields could be associated with some UV-completion, and would thus typically have $M_M \simeq M_P$, where $M_P$ is the Planck mass.  This scenario is usually referred to as `Gravity Mediation'.  Alternatively they could potentially have much lower mass, and communicate with the visible sector through gauge interactions.  In this case $M_M$ is not set, but the soft terms come dressed with a loop-factor involving gauge charges.

\subsubsection{Supergravity}
General relativity (GR) is the tremendously successful theory of gravity on macroscopic scales, and is hence desirable in any physical theory.  We can think of GR as a theory of gauged local Lorentz transformations, however, by going to a SUSY theory we have extended the Lorentz group to include fermionic generators.  Thus, if we gauge the Lorentz transformations we must also gauge local SUSY transformations in order to maintain SUSY.  In doing so we find a theory of \emph{local} supersymmetry.  This theory is called `Supergravity' (SUGRA).  It is sometimes touted as a surprising, and/or compelling, feature that gauging SUSY leads to GR, however this should really come as no surprise as we still have the Lorentz group as a subgroup of the general SUSY transformations, and one should then expect that gauging these transformations would lead to GR.

There are many interesting features of SUGRA, which is a subject of much study in its own right, however, for brevity, we will only comment on the features relevant to BSM.\footnote{An excellent textbook focussing specifically on SUGRA is \cite{Freedman:2012zz}.}  Perhaps the most interesting relevant feature of SUGRA is the requirement of a new spin-$3/2$ field, called the gravitino, which is partnered with the graviton.  This field has its own set of Planck-suppressed interactions with other SUSY fields.  An interesting analogy with local gauge theories arises when SUSY is spontaneously broken.  When a global symmetry is spontaneously broken we expect a massless Nambu-Goldstone boson, and if this symmetry is gauged we expect this boson to be `eaten' by the massless gauge boson, leading to a massive gauge boson.  Interestingly, when SUSY is spontaneously broken we have a massless fermion, the goldstino, however in a SUGRA theory this goldstino is `eaten' by the gravitino, leading to a massive gravitino.  

\subsection{The MSSM}\label{MSSM}
Now we are equipped to construct a supersymmetric theory of the known particles and interactions.  We will consider first the minimal model, a.k.a.\ the `Minimal Supersymmetric Standard Model' (MSSM).  In a supersymmetric version of the SM we will have to introduce superpartners for all of the known fields of the standard model.  It is conventional notation to denote a superpartner of a SM field with a tilde, i.e.\ a $\tilde{e}_L$ is the superpartner of the left-handed electron.  The fermions of the standard model are contained in chiral superfields, and thus we introduce `squarks' in addition to the quarks, and `sleptons' in addition to leptons.  Scalar partners of SM fermions are individually named with an `s' in front of the name of their fermion partner, i.e.\ sneutrino, selectron, sbottom, etc.  The gauge fields will have to live in a vector superfield and will thus have fermionic superpartners.  The partners of the gauge fields are termed `gauginos' and, in specific cases, are differentiated from their bosonic partners by the suffix `ino'.  Thus along with gluons we now have gluinos, with W-bosons winos, and with the hypercharge boson the bino.  After electroweak symmetry breaking we have charginos and two neutralinos from the electroweak gauge sector.

The simple extension of the SM to a SUSY theory enters difficulties when we consider the SM Higgs boson.  Because the Higgs is a scalar, in a SUSY theory it will have a fermionic partner, the higgsino.  This higgsino will have SM gauge charges and is a new fermion contributing to anomalies in the previously anomaly-free SM.  Thus in order to cancel this new contribution we must add an additional chiral superfield with the opposite gauge charges of the Higgs.  Hence a supersymmetric theory has two Higgs doublets, as opposed to one in the SM, and these doublets are `vector-like', as they have equal and opposite gauge charges.  It is often stated that, as the superpotential is holomorphic and terms such as $H_U^\dagger Q D^c$ are not allowed, then an extra Higgs doublet must be introduced in order to give down-type fermions mass.  However this is not strictly true, as we know that SUSY must be broken, and once SUSY is broken such arguments do not apply, whereas a gauge symmetry in QFT must be anomaly-free, regardless of SUSY.

\begin{table}[t]
\begin{center}
%% The following adds some white space between row-dividing lines and text %%
\def\str{\vrule height12pt width0pt depth7pt}
\begin{tabular}{| l | l | c | r | r |}
    \hline\str
    ~Field ~&~ Gauge rep. ~&~ R-parity ~&~ Supermultiplet ~ \\
    \hline\str
    ~~$\mb{Q}$ & ~~$(\rep{3},\rep{2}, \frac{1}{6})$ & $-1$ & Chiral \hspace{2em} \\
    \hline\str
    ~~$\mb{U^c}$ & ~~$(\conjrep{3},\rep{1}, -\frac 23)$ & $-1$ & Chiral \hspace{2em} \\
    \hline\str
    ~~$\mb{D^c}$ & ~~$(\conjrep{3},\rep{1}, \frac 13)$ & $-1$ & Chiral \hspace{2em} \\
    \hline\str
    ~~$\mb{L}$ & ~~$(\rep{1},\rep{2}, -\frac 12)$ & $-1$ & Chiral \hspace{2em} \\
    \hline\str
    ~~$\mb{E^c}$ & ~~$(\rep{1},\rep{1}, 1)$ & $-1$ & Chiral \hspace{2em} \\
    \hline\str
    ~~$\mb{H_u}$ & ~~$(\rep{1},\rep{2}, \frac 12)$ & $1$ & Chiral \hspace{2em} \\
    \hline\str
    ~~$\mb{H_d}$ & ~~$(\rep{1},\rep{2}, -\frac 12)$ & $1$ & Chiral \hspace{2em} \\
    \hline\str
    ~~$\mb{G}$ & ~~$(\rep{8},\rep{1}, 0)$ & $1$ & Vector \hspace{2em} \\
    \hline\str
    ~~$\mb{W}$ & ~~$(\rep{1},\rep{3}, 0)$ & $1$ & Vector \hspace{2em} \\
    \hline\str
    ~~$\mb{B}$ & ~~$(\rep{1},\rep{1}, 0)$ & $1$ & Vector \hspace{2em} \\
    \hline
\end{tabular}
\caption{The superfield content of the MSSM.}
\label{tab:MSSM}
\end{center}
\end{table}

The superfields of the MSSM are summarised in \Tab{tab:MSSM}.  The kinetic terms and gauge interactions for all fields are as in eq.(\ref{fullSUSYgauge}), and the superpotential for the MSSM is
\begin{equation}
W_{\text{MSSM}} = \mu \mb{H_u} \mb{H_d} + \lambda_u \mb{H_u} \mb{Q} \mb{U^c} + \lambda_d \mb{H_d} \mb{Q} \mb{D^c} + \lambda_e \mb{H_d} \mb{L} \mb{E^c}
\label{MSSMsuper}
\end{equation}
where the $\lambda$ are $3\times3$ Yukawa couplings and summation over flavour indices is implied.  Additional gauge-invariant, renormalizable, terms which violate baryon or lepton number are also allowed.  These are $\mb{L} \mb{L} \mb{E^c}$, $\mb{U^c} \mb{D^c} \mb{D^c}$, $\mb{L} \mb{Q} \mb{D^c}$ and $\mu_L \mb{L} \mb{H_u}$.  These terms can lead to rapid proton decay, amongst other forbidden processes, and thus should be suppressed.  To do this we impose an additional global symmetry by hand.  This symmetry is a discrete $\IZ_2$ symmetry which is a subgroup of R-symmetry, known as R-parity.  The R-parity charges of the MSSM superfields are shown in table \ref{tab:MSSM}, and the Grassmann parameter $\theta$ is also odd under this parity, hence the name `R'-parity.  As $\theta$ is charged under this parity superpartners within a supermultiplet have different charges.  Thus all SM fermions, gauge bosons, and both scalar Higgs doublets are even under this parity, whereas all superpartners such as gauginos, squarks, sleptons and higgsinos, are odd.  Hence R-parity distinguishes between the SM particles and those which we have added, with the exception of the extra Higgs doublet.

\subsubsection{Soft Masses}
The model as described so far is completely supersymmetric, however we have not observed any R-parity odd particles, and thus we must softly break the supersymmetry.  We say that the breaking is soft because we only include operators that preserve SUSY in the high energy limit.  To achieve this we add only terms that that are less and less relevant at high energies.  Such operators have mass dimension $D<4$, thus we can add explicit mass terms for fields that break SUSY at the scale $m$, safe in the knowledge that they do not spoil the cancellation of quadratic divergences at energies above $E\gg m$.  In practise these soft terms can be generated by some particular underlying model for SUSY breaking.

This is achieved at a phenomenological level by adding soft masses for all scalar fields and all gauginos.  We must also add to the scalar potential trilinear scalar interactions with the same structure as the trilinear terms in the superpotential in eq.(\ref{MSSMsuper}), as well as a `$B_\mu$' term $\mathcal{L} \supset B \mu H_u H_d$ which mixes the two Higgs fields.  All such soft-parameters are, in general, complex, and need not have the same flavour structure as the SM Yukawa couplings.  This completes the construction of the MSSM as a phenomenological model.

\subsubsection{Successes and motivation}
From the perspective of these lectures, the greatest success of the MSSM is that it addresses the hierarchy problem by removing quadratic divergences, thus stabilising the electroweak scale against corrections from unknown physics in the far UV.  There are however, additional hints that add to the appeal of the MSSM.  We briefly discuss these in no particular order.

\subsubsection{Dark Matter}
A particularly attractive feature of the MSSM arises as a result of protecting protons from decaying.  We saw that an extra global symmetry, R-parity, must be imposed in order to conserve baryon number and lepton number at the renormalizable level.  This extra symmetry largely distinguishes between SM particles and their partners, and has the consequence that the lightest of the superpartners cannot decay, and is thus cosmologically stable.  If this stable particle is charged, or coloured then this stability is disastrous, however if it is neutral then it may be a candidate for DM.  It turns out that there are plenty of neutral particles in the MSSM, four `neutralinos' which are each a mixture of the bino, zino, and two higgsinos, and there are also three neutral sneutrinos.  The correct abundance of all of these particles results from the thermal freeze-out mechanism, suggesting that they could be the DM.  DM direct detection experiments place stringent bounds on how strongly the DM can couple to nucleons, and this rules out the left-handed sneutrinos as DM candidates, however if the lightest neutralino is dominantly made up of higgsino, or bino, components then it can still make a good candidate for DM.  Thus, as a result of protecting the proton from decay, the MSSM contains a good candidate for DM.

\subsubsection{Radiative Electroweak Symmetry Breaking}
An additional, unexpected feature of the MSSM is that, for a large range of parameters, the mass of the up-type Higgs boson is driven negative by radiative corrections.  The result being that even if the electroweak gauge symmetry is unbroken in the theory at high energies, when one runs all of the parameters down to the weak scale the Higgs mass-squared becomes negative, and the electroweak gauge symmetry is spontaneously broken.  This is due to the large Yukawa coupling of the Higgs multiplet to the top multiplet.  Electroweak symmetry breaking in this manner is not always guaranteed, however it does seem to be a fairly generic feature of the MSSM.  Additionally, the Higgs seems to be special in this respect as for most parameter regions no other scalars are driven to develop a vev.

\subsubsection{Baryogenesis}
Another interesting hint lies in the problem of baryogenesis.  It is known that in order to generate an asymmetry between baryons and antibaryons in the early Universe the three conditions of baryon-number violation, CP-violation, and out of thermal equilibrium dynamics must be satisfied.  These are known as the `Sakharov conditions', after Andrei Sakharov, who first wrote them down.\footnote{Google him.}  It was once hoped that such conditions could be present during the electroweak phase transition, as there is CP-violation in the quark sector, baryon-number violation due to electroweak non-perturbative effects (sphalerons) and if the electroweak phase transition is strongly first-order enough then in the bubble walls, which separate the symmetric phase from the broken phase, there should exist out-of-thermal-equilibrium conditions.  Unfortunately, in the SM these conditions are not met to the extent that the observed asymmetry can be achieved.  However, going beyond the SM it is possible to meet these conditions, with the introduction of new sources for all three necessary conditions.  A plethora of models for baryogenesis exist, including the MSSM.

\subsubsection{Flavour and Neutrino Puzzles}
The MSSM itself does not provide an explanation for the puzzle of the hierarchies of quark and lepton Yukawas, nor for the relatively miniscule neutrino masses, however SUSY permits to have perturbatively stable mass hierarchies between fundamental scalars, thus if any solution to these puzzles requires new physics at high energies, SUSY provides a natural accommodation of a light Higgs mass despite these high energy scales.

\subsubsection{Gauge Coupling Unification}
\begin{figure}[tbp]\begin{center}
\includegraphics[width=0.6\textwidth]{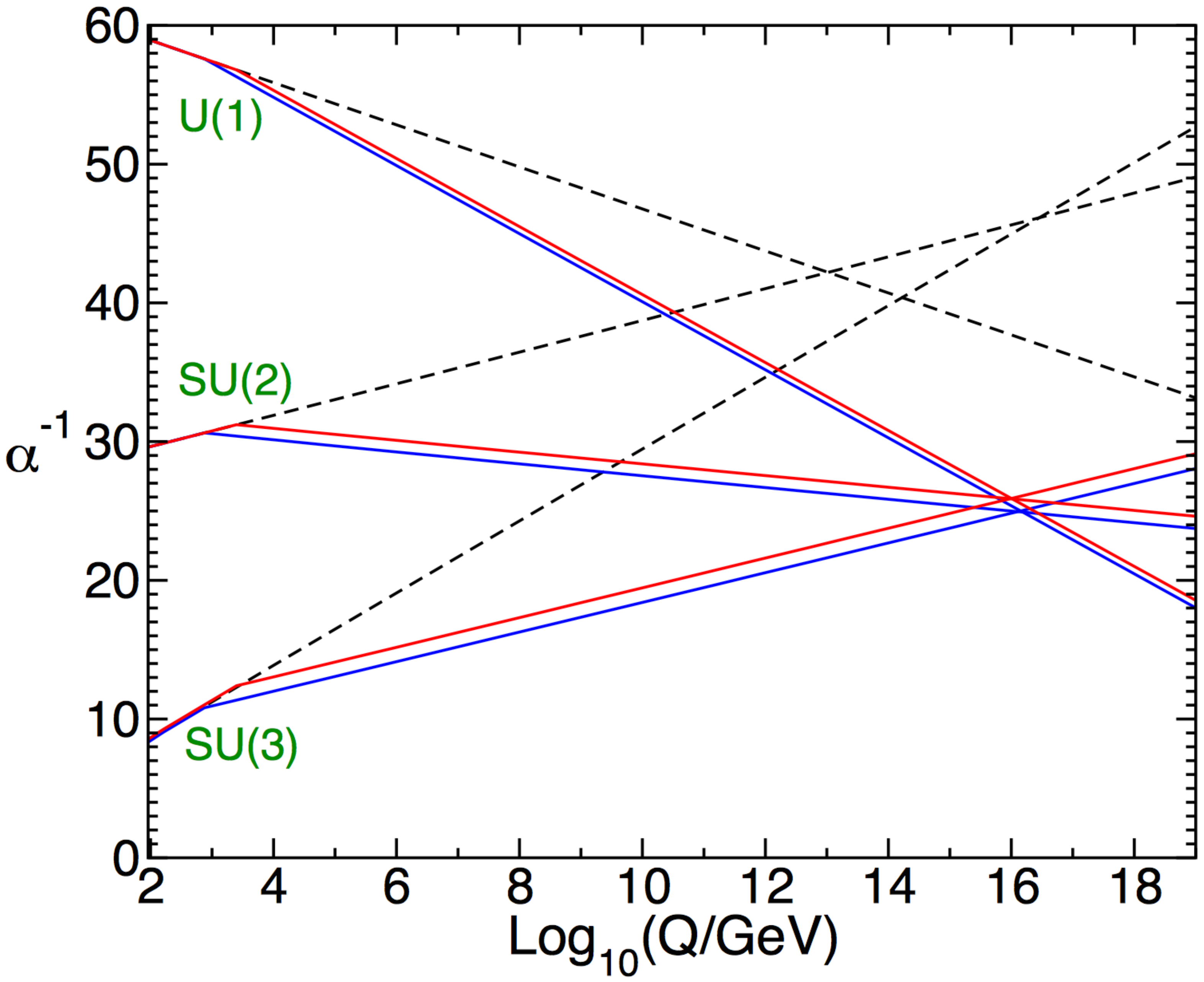}
\end{center}
\caption{Renormalization group evolution of gauge couplings up to high energies, taken from \cite{Martin:1997ns}.  The Standard Model gauge couplings are shown in dashed black and the gauge couplings with superpartners added, with masses in the range $0.75 \to 2.5$ TeV, are shown in red and blue.  Unification of the forces at high energies is clearly apparent in the supersymmetric case.}
\label{fig:unification}
\end{figure}

An unexpected surprise that arises whenever the Standard Model is supersymmetrized connects the behaviour of the Standard Model gauge couplings to a deep idea concerning the nature of the fundamental forces at extremely high energies.  With the superpartners added, it was found that upon evolving the $\text{U}(1)_{Y}$, $\text{SU}(2)_{W}$, and $\text{SU}(3)_{C}$ gauge couplings up to high energies they appeared to unify at energies close to $E\sim 10^{16}$ GeV \cite{Dimopoulos:1981yj,Dimopoulos:1981zb}!  This is shown in Fig.~\ref{fig:unification}.  Of course, that two lines will cross is almost guaranteed, however three lines crossing almost at a point is strongly suggestive of a deeper structure and a potential link between SUSY and unification.  This is especially compelling as SUSY is precisely the ingredient that would allow for a stable hierarchy between the unification scale and the weak scale!

%%%%%%%%%%%%%%%%%
%%%%%%%%%%%%%%%%%

Ever since the unification of the electroweak forces was discovered, it has been believed that further unification of all gauge forces, now including $\text{SU}(3)_{C}$, may occur at very high energies.  A variety of larger gauge groups into which they may unify have been proposed, however the simplest is arguably an $\text{SU}(5)$ gauge symmetry \cite{Georgi:1974sy}.\footnote{It is also possible that the gauge forces unify with gravity, in the context of String Theory, however we will not discuss this possibility here.}  It is deeply compelling that the Standard Model matter gauge representations neatly fall into multiplets of a larger symmetry, such as $\text{SU}(5)$, as this need not have been the case.  A key feature which must arise at the unification scale in such a theory is that the gauge couplings must themselves become equal.  Thus supersymmetric gauge coupling unification is strongly suggestive that supersymmetry may go hand-in-hand with the unification of the forces and, if discovered, the superpartners would provide a low energy echo of physics at extremely high energies.

When considering the role of the superpartners in supersymmetric unification one finds that some are more relevant than others.  The reason is that since the matter fermions of the Standard Model fill out complete unified representations, so must their partners, the squarks and the sleptons.  Thus although the masses of squarks and sleptons may change the scale at which unification occurs they do not significantly alter whether or not the couplings will unify, unless they are split by large mass differences themselves.  This means that the most important superpartners for gauge coupling unification are the fermions: the gauginos and the Higgsinos.

Studies of supersymmetric gauge coupling unification generally find that for successful unification it is necessary to have gauginos and higgsinos not too far from the weak scale.  If the gaugino and Higgsino mass parameters are taken equal, then unification requires $\mu,\widetilde{M}_{1/2} \lesssim \mathcal{O}(10 \text{ TeV})$ 
%at $1\sigma$, where $\sigma$ denotes the magnitude of estimated 
with some uncertainty due to unknown threshold corrections at the unification scale \cite{Arvanitaki:2012ps}.  The scalar soft masses, $\widetilde{m}_0$, may be arbitrarily heavy while preserving successful gauge coupling unification.  This realization led to the consideration of so-called `Split-Supersymmetry' theories \cite{Wells:2004di,ArkaniHamed:2004fb,Giudice:2004tc}, in which the main motivations for the mass spectrum are taken from gauge coupling unification and dark matter, as discussed previously.  

The fact that, in addition to the gauge forces, also the matter particles are unified in representations of the unified gauge symmetry group can imply relations between the Yukawa couplings of quarks and leptons at the unification scale \cite{Georgi:1974sy,Chanowitz:1977ye,Nanopoulos:1978hh,Georgi:1979df,Antusch:2009gu,Antusch:2013rxa}. To compare such predictions with the measured values of the fermion masses, one has to take into account the supersymmetric loop threshold corrections at the soft breaking mass scale \cite{Hempfling:1993kv,Hall:1993gn,Carena:1994bv,Blazek:1995nv,Antusch:2008tf,Antusch:2015nwi}, which depend on the masses of the superpartners. Including them in the analysis, and using the measured fermion masses and Higgs mass as constraints, unified theories are even capable of predicting the complete sparticle spectrum \cite{Antusch:2015nwi,Antusch:2016nak}.

%To summarize, as with dark matter, gauge coupling unification motivates Higgsino and gaugino soft masses in the approximate range $\mu,\widetilde{M}_{1/2} \lesssim \mathcal{O}(\text{10's TeV})$, once again suggesting that much of the parameter space motivated by this consideration should be within reach of a 100 TeV collider.

\subsubsection{The Higgs Mass}
As is common in physics, when new symmetries are introduced to a theory, the predictive power often increases.  Because supersymmetry is softly broken, many new parameters associated with this breaking are introduced and certain aspects of the increased predictivity are lost.  However, some predictability beyond the SM remains and the Higgs boson mass is a prime example.

In the Standard Model, when the theory is written in the unbroken electroweak phase there are only two fundamental parameters in the scalar potential, the doublet mass $m_H$, and the quartic coupling $\lambda$.  In the broken electroweak vacuum this translates to two fundamental parameters, the Higgs vacuum expectation value $v=246$ GeV, and the Higgs scalar mass $m_h$.  Once these two parameters are set, all other terms, such as the Higgs self-couplings, are determined.  Supersymmetric theories take this one step further as supersymmetry relates the Higgs scalar potential quartic coupling to the electroweak gauge couplings in a fixed manner.  The story is complicated a little relative to the Standard Model by the two Higgs doublets required in supersymmetric theories, however since the quartic couplings in the scalar potential are no longer free parameters, once the vacuum expectation value is set $v = \sqrt{v_u^2 + v_d^2} = 246$ GeV, the Higgs mass is now also predicted by the theory.  At tree level, this prediction is
\begin{equation}
m_h = M_Z |\cos 2 \beta| ~~.
\end{equation}
Clearly for any value of $\beta$ this prediction is at odds with the observed value of $m_h \approx 125$ GeV and thus for consistency additional contributions to the Higgs doublet quartic terms are required.  Within the MSSM the only potential source is from radiative corrections at higher orders in perturbation theory.  The dominant corrections arise from loops of particles with the greatest coupling to the Higgs, the stop squarks \cite{Ellis:1990nz,Ellis:1991zd}.  If the soft mass splitting between the top-quark and stop squarks is large enough then radiative corrections which are sensitive to this supersymmetry breaking may spoil the supersymmetric prediction for the Higgs quartic couplings and allow for contributions that may bring the Higgs boson mass within the observed window.

%%%%%%%%%%%%%%%%%
\begin{figure}[tbp]\begin{center}
\includegraphics[width=0.4\textwidth]{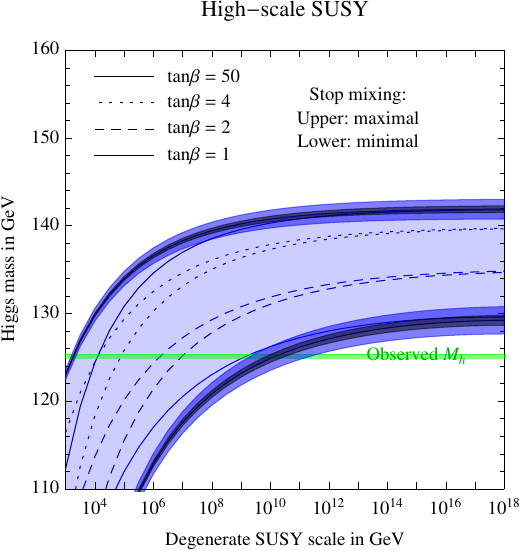} \qquad \includegraphics[width=0.4\textwidth]{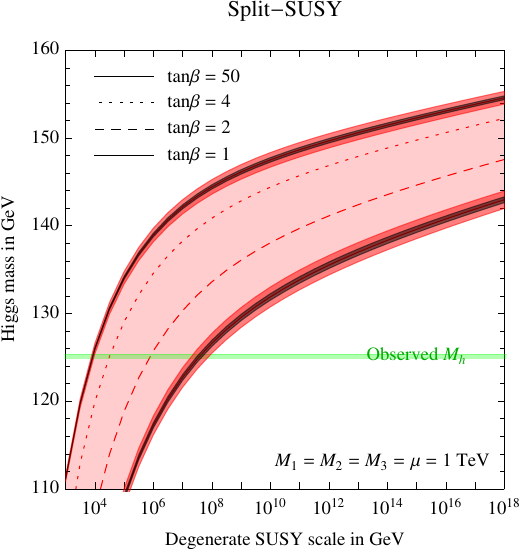}
\end{center}
\caption{Higgs mass predictions as a function of the supersymmetry breaking soft mass scale and the Higgs sector parameter $\tan \beta$, taken from \cite{Bagnaschi:2014rsa}.  In the High-Scale scenario all soft masses $\mu,\widetilde{M}_{1/2},\widetilde{m}_0$ are varied together, whereas in the Split SUSY scenario $\mu,\widetilde{M}_{1/2}$ are kept at 1 TeV and only the scalar soft masses $\widetilde{m}_0$ are varied.}
\label{fig:MSSMhiggsmass}
\end{figure}%
%%%%%%%%%%%%%%%%%

In Fig.~\ref{fig:MSSMhiggsmass} we show the expected soft mass parameter scales which reproduce the observed Higgs mass.  Clearly, within the MSSM the observed Higgs mass may be reproduced for scalar masses in the range $1 \text{ TeV} \lesssim \widetilde{m}_0 \lesssim 10^8 \text{ TeV}$.\footnote{In fact, if the soft scalar trilinear term $\tilde{A}_t$ is chosen so as to maximise the shift in the Higgs mass, the lightest stop squark could be as light as $\sim 500$ GeV \cite{Hall:2011aa}.}  Furthermore, if we consider the range $ \tan \beta > 4$, then scalar masses below $\mathcal{O}(\text{10's TeV})$ are required.  This is the first upper bound we have encountered for the scalar soft masses, resulting directly from the Higgs mass measurements.  Theoretically, this has given rise to a reduction in the allowed parameter space of supersymmetric theories and in the context of so-called Split SUSY, where previously scalar masses could take almost any value, the Higgs mass measurements have led to the so-called `Mini-Split' scenario \cite{Arvanitaki:2012ps,ArkaniHamed:2012gw}, where there is an upper bound on the value of the scalar soft masses.

There are variants of the MSSM in which the Higgs mass may also be raised above the MSSM tree-level prediction by utilizing additional effects deriving from couplings to new fields.  If the coupling is to new fields in the superpotential then such theories are typically variants of the NMSSM, in which the Higgs doublets couple to an additional gauge singlet.  Alternatively, the corrections may arise from coupling to new gauge fields, due to additional contributions to the quartic scalar potential predicted by supersymmetric gauge interactions.  Importantly, in these scenarios the additional enhancements of the Higgs mass only serve to reduce the required value of the radiative corrections, and hence the required value of the scalar soft masses.  Thus the required scalar soft mass values shown in Fig.~\ref{fig:MSSMhiggsmass} serve as an approximate upper limit for theories beyond the MSSM.

To summarize, the measurement of the Higgs mass has now provided information that is key to understanding the expected mass ranges of superpartners, particularly for the stop squarks.  Although scalar masses may be as large as $\widetilde{m}_0 \sim 10^8$ TeV, for a broad range of parameter space, if it is the case that $\tan \beta > 4$ this upper bound is reduced significantly to $\widetilde{m}_0 \lesssim \mathcal{O}(10\text{'s TeV})$.

\subsection{Summary:  SUSY now and the Hierarchy Problem}
With SUSY broken at a scale $\tilde{m}$, which represents the soft mass scale, the Higgs mass is no longer protected from quantum corrections.  Thus supersymmetry is effective in protecting the Higgs mass all the way down from a high mass scale to the supersymmetry breaking scale $M_{\text{New}} \to \tilde{m}$, however from the soft mass scale down to the weak scale, $\tilde{m} \to m_h$ supersymmetry is no longer present.  This means that for a natural theory without tuning we must expect $\tilde{m} \sim m_h$, and conversely if $\tilde{m} \gg m_h$ there must be some fine tuning to realize the weak scale below the soft mass scale.  These qualitative arguments may be made quantitative.  A well motivated measure for the degree of tuning in the weak scale with respect to a given fundamental parameter in the theory, $a$, is \cite{Ellis:1986yg,Barbieri:1987fn}
\begin{equation}
\Delta [a] = \frac{\partial \log M_Z^2}{\partial \log a^2}  ~~.
\end{equation}
By minimising the weak scale potential at large $\tan \beta$ we find
\begin{equation}
M_Z^2 = -2 (m^2_{H_u} + |\mu|^2)  ~~,
\end{equation}
where $m^2_{H_u}$ is the soft mass for the up-type Higgs which includes all radiative corrections.  Let us consider the tree-level contribution from the $\mu$-term, along with the one-loop contributions from stop squarks and the winos, and the two-loop contribution from gluinos, which are given by
\begin{eqnarray}
\delta m^2_{H_u} (\tilde{t}) &=& -\frac{3 y_t^2}{4 \pi^2} m^2_{\tilde{t}} \log (\Lambda/m_{\tilde{t}}) \\
 \delta m^2_{H_u} (\widetilde{W}) &=&-\frac{3 g^2}{8 \pi^2} (m^2_{\widetilde{W}} +m^2_{\tilde{h}})\log (\Lambda/m_{\widetilde{W}}) \\
  \delta m^2_{\tilde{t}} &=& \frac{2 g_s^2}{3 \pi^2} m_{\tilde{g}}^2 \log (\Lambda/m_{\tilde{g}}) ~~,
\end{eqnarray}
where $\Lambda$ is a UV-cutoff at which the full UV-completion of supersymmetry kicks in, and the last term may be inserted into the first to obtain an estimate of the tuning from gluinos.  Conservatively taking $\Lambda=10$ TeV we arrive at the following expectations for a theory which is only tuned at the $10\%$ level \cite{Craig:2013cxa}:
\begin{equation}
\mu \lesssim 200 \text{ GeV} ~~,~~ m_{\tilde{t}} \lesssim 400 \text{ GeV} ~~,~~ m_{\widetilde{W}} \lesssim 1 \text{ TeV} ~~,~~ m_{\tilde{g}} \lesssim 800 \text{ GeV} ~~,~~ 
\end{equation}
This picture is clearly at odds with the stop mass values required to achieve the observed Higgs mass in the MSSM, shown in Fig.~\ref{fig:MSSMhiggsmass}.  However it may be that non-minimal structure beyond the MSSM lifts the Higgs mass without requiring large stop masses, thus this constraint is not too significant.  More importantly, current constraints on the Higgs boson couplings, which would typically be modified if the stop squarks were light, already place stringent constraints on light stop scenarios.

\begin{figure}[tbp]\begin{center}
\includegraphics[height=0.35\textwidth]{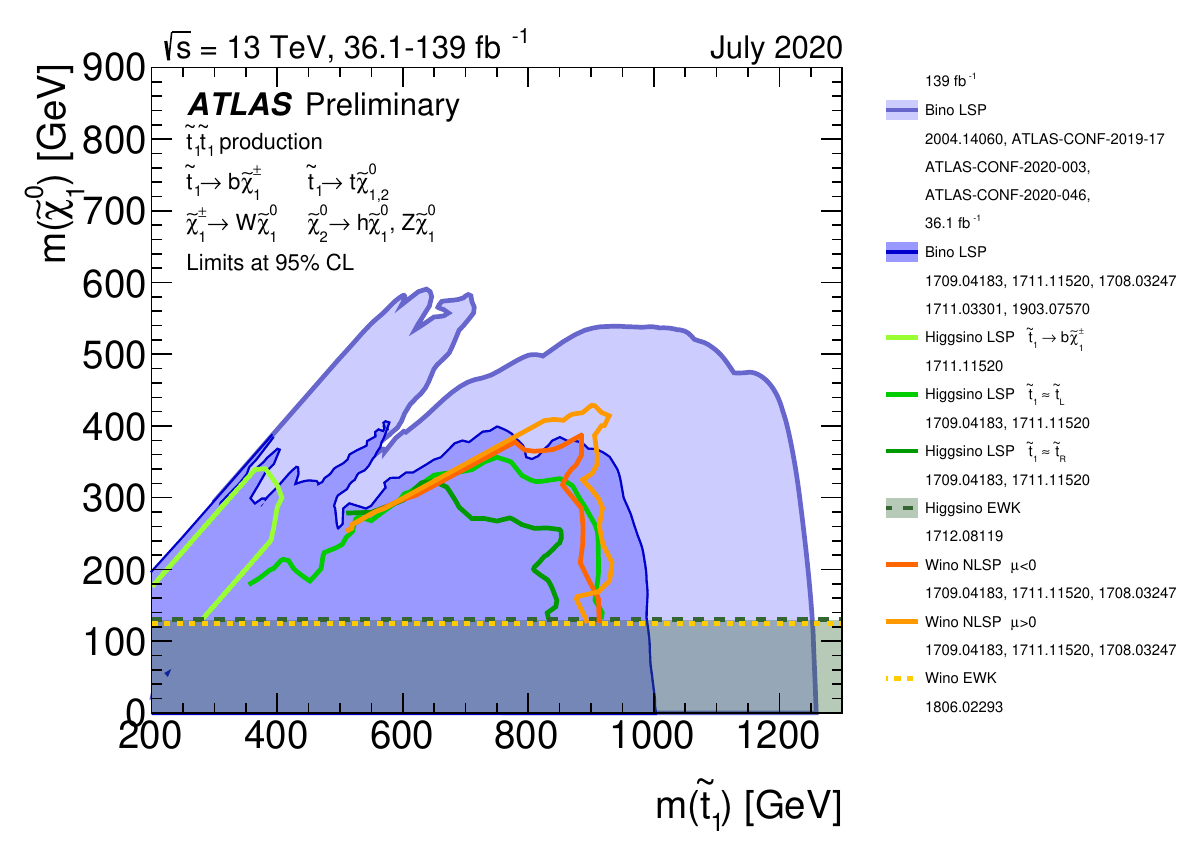} \qquad \includegraphics[height=0.35\textwidth]{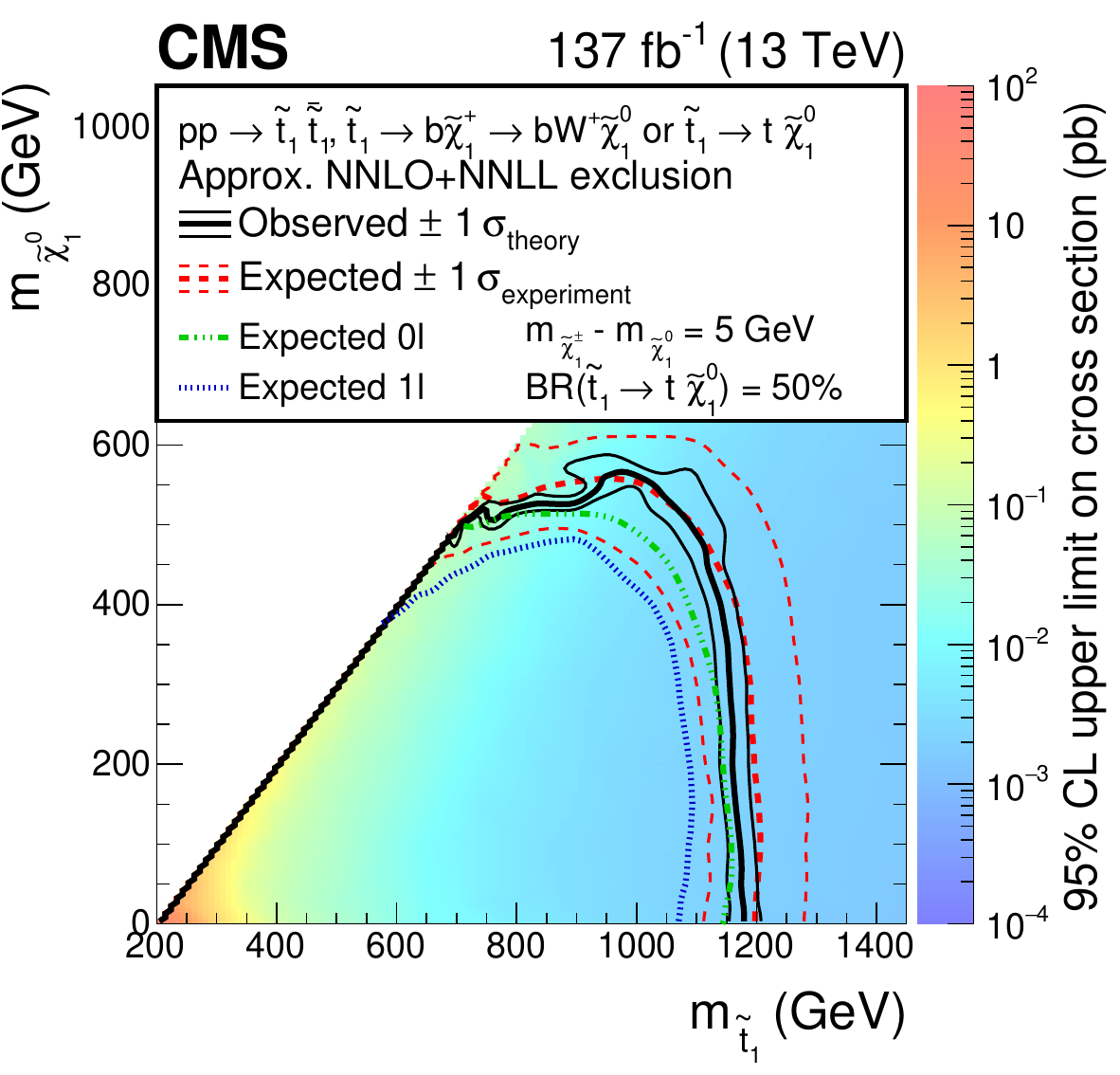}
\end{center}
\caption{Current experimental limits on a simplified model with stop squarks and a neutralino.  These limits are already placing significant pressure on SUSY naturalness for this class of models.}
\label{fig:stoplims}
\end{figure}

\begin{figure}[tbp]\begin{center}
\includegraphics[width=0.4\textwidth]{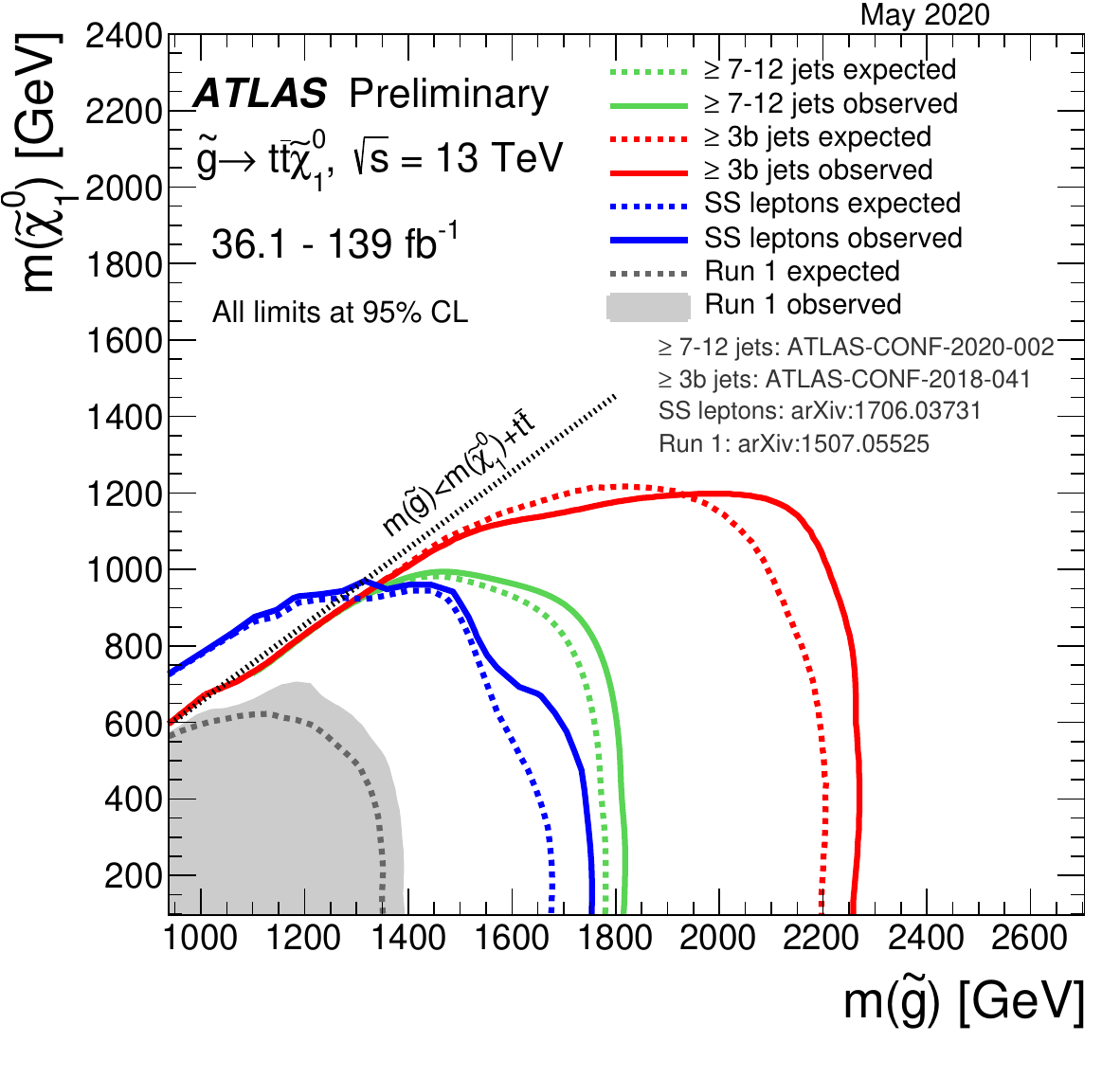} 
\end{center}
\caption{Experimental limits on a simplified model with gluinos and a neutralino.}
\label{fig:gluinolims}
\end{figure}

In many (but not all) concrete scenarios it is expected that the first two generation squarks should not be significantly heavier than the stop squarks and, as the production cross section is enhanced due to valence quarks in the initial state, constraints on first two generation squarks are very strong, indirectly placing strong constraints on the naturalness of many supersymmetric theories.  Most relevant, however, are the direct searches for stops and gluinos, which already show that significant portions of this parameter space are in tension with LHC 13 TeV data, as seen in \Fig{fig:stoplims} and \Fig{fig:gluinolims}.

Generally speaking, the stops have to be heavier than $\sim 1.2$ TeV, suggesting a fine-tuning of
\bea
\frac{1}{\Delta} & \lesssim &  \left( \frac{400}{1200} \right)^2 \times 10 \% \nonumber \\
& \lesssim &  1.1 \% ~~.
\eea
And for the gluinos
\bea
\frac{1}{\Delta} & \lesssim &  \left( \frac{800}{2200} \right)^2 \times 10 \% \nonumber \\
& \lesssim &  1.3 \% ~~.
\eea
So, both from the stop and gluino sectors it appears we are approaching $1\%$ fine-tuning for typical SUSY models.  How will things change by the end of HL-LHC?  Well, a na\"ive \href{http://cern.ch/collider-reach}{Collider Reach} estimate suggests the reach for stops could extend to $\sim1.8$ TeV, and for gluinos to $\sim3.2$ TeV.  This is essentially confirmed by dedicated studies \cite{CidVidal:2018eel}, which project a reach of $1.7$ TeV for stop exclusion and $3.2$ TeV for gluinos.  Note that to estimate the change in mass reach one cannot simply perform a $\sqrt{\mathcal{L}}$ rescaling, since the cross section and PDF dependence on the mass is all-important.  In any case, if these limits were borne out we would face a fine-tuning at the level of
\bea
\frac{1}{\Delta} & \lesssim &  \left( \frac{400}{1700} \right)^2 \times 10 \% \nonumber \\
& \lesssim &  0.6 \% ~~.
\eea
And for the gluinos
\bea
\frac{1}{\Delta} & \lesssim &  \left( \frac{800}{3200} \right)^2 \times 10 \% \nonumber \\
& \lesssim &  0.6 \% ~~.
\eea

What is more interesting, I think, is that the projected discovery reach is $1.2$ TeV for stop exclusion and $2.8$ TeV for gluinos \cite{CidVidal:2018eel}.  Thus it seems unlikely at this stage that from now until the end of HL-LHC we will discover any vanilla incarnation of stops, however for gluinos there is still the possibility for discovery.  This may be surprising to some, but is a reflection of the power of all that extra integrated luminosity at the HL-LHC.

%%%%%%%%%%%%%%%%%
%%%%%%%%%%%%%%%%%

It remains to ask where the strong current constraints leave the supersymmetric solution to the hierarchy problem?  It could be that the weak scale is meso-tuned, as in Mini-Split supersymmetry, and the \ae sthetic motivations for supersymmetry as a new spacetime symmetry are justified, whereas the naturalness arguments were misguided, to at least some degree, since supersymmetry does solve the big hierarchy problem and we are instead left with a relatively small tuning of the weak scale up to energies as high as $\mathcal{O} (10^8)$ TeV.  This scenario is in some sense quite successful.  A fundamental Higgs boson of mass $m_h \lesssim 135$ GeV is predicted, gauge coupling unification and successful dark matter candidates are realized, all at the cost of accepting some meso-tuning.  Although not necessarily guaranteed, the gauginos should be below mass scales of $\sim \mathcal{O}(\text{few TeV})$, mostly driven by the dark matter requirement.

Another possibility is that the Mini-Split spectrum is realized in nature, with all of the above successes, however the theory is not actually tuned due to a hidden dynamical mechanism which renders the hierarchy from the weak scale to the soft mass scale natural \cite{Batell:2015fma}.  This can be achieved by employing the cosmological relaxation mechanism of \cite{Graham:2015cka} in a supersymmetric context.  In this case both the \ae sthetic arguments for supersymmetry and the naturalness arguments for the weak scale were well founded, however the two may have manifested in an entirely unexpected manner, with a cocktail of symmetries and dynamics protecting the naturalness of the weak scale up to the highest energies.  As before, the gauginos should be below mass scales of $\sim \mathcal{O}(\text{10's TeV})$, however this expectation comes from the fact that a loop factor suppression between scalars and gauginos is expected in this model and in addition the scalars cannot be arbitrarily heavy due to the finite cutoff of the cosmological relaxation mechanism.

%%%%%%%%%%%%%%%%%
%%%%%%%%%%%%%%%%%

Alternatively, a re-evaluation of the fine-tuning in the infrared may be required if a spectrum with heavy squarks is made natural due to correlations between soft mass UV-boundary conditions and the infrared value of the Higgs mass, as in `Focus Point' supersymmetry \cite{Feng:1999zg} or in the recently proposed `Radiatively-Driven' natural Supersymmetry \cite{Baer:2012up}.  In these cases gauginos, Higgsinos, and most likely also stop and sbottom squarks are expected to still be in the sub-$10$ TeV range.  The first two generation squarks may be somewhat heavier.
 
Finally, it is still possible, although increasingly unlikely, that the weak scale is relatively natural due to supersymmetry, however the sparticles have evaded detection until now.  If this is the case it is likely the stop squarks are still relatively light, in the range of a few 100's of GeV, and the Higgs mass is raised by an additional tree-level term.  For the stop squarks to evade detection there are a number of possible scenarios.  We will discuss just a few here.  One is an example of a so-called `compressed' spectrum (see e.g.\ \cite{LeCompte:2011cn}), where the mass splitting between the stop and the stable neutralino is so small that the tell-tale missing energy signature carried away by the neutralino is diminished to the point of being unobservable.  Another possibility is `Stealth Supersymmetry' \cite{Fan:2011yu,Fan:2012jf}, where again the missing energy signatures are diminished, however in this case from sparticle decays passing through a hidden sector.  Yet another possibility is for R-parity violating decays of the superpartners \cite{Barbier:2004ez}, since in this case missing-energy signatures are removed and the collider searches must contend with larger backgrounds (see e.g.\ \cite{Csaki:2011ge} for models and collider phenomenology).  For a natural spectrum the first two generations of squarks must also have evaded detection.  One possibility is to raise their mass above experimental bounds, which is compatible with naturalness if they stay within an order of magnitude or so of the gluinos and stops \cite{Papucci:2011wy,Dimopoulos:1995mi,Cohen:1996vb}.  Dirac gauginos also offer opportunities for suppressing collider signatures, at no cost to the naturalness of the theory \cite{Fox:2002bu,Kribs:2012gx}, as Dirac gauginos may naturally be heavier than their Majorana counterparts.  This scenario allows not only for the suppression of gluino signatures at the LHC, but also suppresses the t-channel gluino exchange production of the first two generation squarks.

In summary, if we wish for supersymmetry to provide a comprehensive explanation for the origin of the EW scale, then the full cohort of sparticles should lie below $\mathcal{O}(\text{few TeV})$.  Otherwise we are forced into considering at least some fine tuning of the weak scale or alternatively the introduction of an additional mechanism, beyond supersymmetry, to enable a natural weak scale.

\section{Wrapping Up}
We are at the end of these notes.  I hope they have been of use, or may be in future.  I had a blast preparing them.  Some material is new, written for this course, whereas some other material is adapted from previous lectures such as at \href{https://www.ggi.infn.it/showevent.pl?id=464}{GGI}, \href{https://www.trisep.ca/2018/}{TRISEP}, \href{https://sites.google.com/view/busstepp2017/home}{BUSSTEPP 2017}, \href{https://sites.google.com/view/busstepp2018/home}{2018}, \href{https://sites.google.com/view/busstepp2019/home}{2019} and an \href{https://indico.ictp.it/event/8690/}{ICTP Summer School}.

I'm very grateful to have had opportunity to teach at TASI.  The major source of this gratitude is really that questions of the hierarchy problem and the origin of the electroweak scale have been given some prominence, through these dedicated lectures, as a topic of relevance to young theorists like yourselves.  This is important.  Of all the questions we seek to answer (it is our job to answer questions) the ones concerning the Higgs are the only ones that point to a particular energy scale and family of experimental phenomena.  For instance, it is extremely important that we search for dark matter, however we shouldn't deceive ourselves as to the possibility that it may remain forever beyond our experimental reach.  We may need to fish in many different ponds before darkness bites, if at all.  The experimental frontier of the hierarchy problem is much more immediate.

A recent fashion in our field has been to interpret the non-observation of Supersymmetry at the LHC as suggesting that nothing new is going on near the weak scale and, subsequently, to conclude that the search for Higgs naturalness is a wild goose chase.  While it's true that this conclusion may ultimately be correct I have to say I find it premature, and its basis rather myopic and unscientific.  As you have seen, supersymmetry is a very special symmetry.  My first hesitation in tying supersymmetry so closely to the broader question of naturalness follows from the fact that its discovery would have marked a turning point in reductionist science, being the first moment in which a natural description of nature that remains valid (effective) over an exponential range of scales had been discovered.  Why would we be the ones, in all mammalian history, to have picked those winning scientific lottery numbers?  Any expectation that would place our time, our LHC, so much at the centre of scientific development makes me uneasy; it's a sort of Copernican anxiety.   Secondly, thanks to having so much symmetry, the experimental signatures of supersymmetry are relatively specific, and should not be taken as representative of all things that could be going on at the weak scale.  Surely there are more things in heaven and earth than jets+MET?  I'm being facetious, of course supersymmetry has many different experimental signatures, but they do not cover \emph{all} possible new physics signatures, so why abandon the broader question just because we haven't found supersymmetry?

The lesson I draw from the lack of observed new particles at the LHC, and from the era before the LHC, is that we ought to avoid hubris at all costs:  Nature needn't care about what we \emph{want} to discover, about what we find \ae sthetically appealing, about our love of `minimality'.  It is very possible that we haven't yet succeeded in exploring the full landscape of theoretical possibilities and their associated phenomena because we have been anchored at fixed points by unnecessary human-imposed desires.  There could be local minima in the theory landscape, distant from the theories we know well, still hidden from our theoretical eyes.  The only way to find them will be to explore far and wide, taking leaps into the unknown.

Consider Columbus.  According to \href{https://en.wikipedia.org/wiki/Christopher_Columbus}{Wikipedia}, not only did he expect that by sailing west from Europe would he find Asia, but he even had estimated it would be about $4400$ km west of the Canary Islands.  People had for a long time been aware that the Earth is a sphere, but they didn't have very precise measurements or estimates of its coverage.  So, at the time, the overall picture of a spherical Earth was correct, the expectation that if one sailed westwards for a distance of $\mathcal{O}(4000~\text{km})$ one would hit land of some sort was correct to within a factor of $2$.  \emph{However}, they had to sail a bit further than was na\"ively expected (about $\mathcal{O}(6000~\text{km})$ as the crow flies, and they ultimately landed somewhere completely unexpected, unenvisaged.  It's a tired analogy, but I suggest you take a quiet moment while hiking to ponder the parallels between this and our own theoretical picture and experimental exploration of what lies above the electroweak scale...

Time to climb down from the soap box.   Thanks for your patience.

\section{Acknowledgements}
I have so many people to thank.  Regarding naturalness and the weak scale, conversations with Nathaniel Craig, Tim Cohen, Gian Giudice, John March-Russell and Riccardo Rattazzi have, for me, been formative.  In addition, the ideas and insights of Steven Weinberg, the dimensionalists Nima Arkani-Hamed, Savas Dimopoulos, Gia Dvali, Lisa Randall, and Raman Sundrum, the relaxionists Peter Graham, David Kaplan and Surjeet Rajendran, and the neutralists Gustavo Burdman, Zackaria Chacko, Hock-Seng Goh and Roni Harnik, have been instrumental in shaping the field and my own view of things through to today.

%\section{Homework}
%Give them a signal, a number of events and a significance.  They have to work out the potential model, the reach at HL-LHC, and the broader framework within which it might sit.

\appendix

\bibliographystyle{JHEP}
\bibliography{refs}

\providecommand{\href}[2]{#2}\begingroup\raggedright\begin{thebibliography}{100}

\bibitem{Das:1967it}
T.~Das, G.~S. Guralnik, V.~S. Mathur, F.~E. Low, and J.~E. Young, {\it
  {Electromagnetic mass difference of pions}},  {\em Phys. Rev. Lett.} {\bf 18}
  (1967) 759--761.

\bibitem{Nambu:1960tm}
Y.~Nambu, {\it {Quasiparticles and Gauge Invariance in the Theory of
  Superconductivity}},  {\em Phys. Rev.} {\bf 117} (1960) 648--663.

\bibitem{Goldstone:1961eq}
J.~Goldstone, {\it {Field Theories with Superconductor Solutions}},  {\em Nuovo
  Cim.} {\bf 19} (1961) 154--164.

\bibitem{Contino:2010rs}
R.~Contino, {\it {The Higgs as a Composite Nambu-Goldstone Boson}},  in {\em
  {Theoretical Advanced Study Institute in Elementary Particle Physics}:
  {Physics of the Large and the Small}}, pp.~235--306, 2011.
\newblock \href{http://xxx.lanl.gov/abs/1005.4269}{{\tt arXiv:1005.4269}}.

\bibitem{Schmaltz:2005ky}
M.~Schmaltz and D.~Tucker-Smith, {\it {Little Higgs review}},  {\em Ann. Rev.
  Nucl. Part. Sci.} {\bf 55} (2005) 229--270,
  [\href{http://xxx.lanl.gov/abs/hep-ph/0502182}{{\tt hep-ph/0502182}}].

\bibitem{Coleman:1969sm}
S.~R. Coleman, J.~Wess, and B.~Zumino, {\it {Structure of phenomenological
  Lagrangians. 1.}},  {\em Phys. Rev.} {\bf 177} (1969) 2239--2247.

\bibitem{Callan:1969sn}
C.~G. Callan, Jr., S.~R. Coleman, J.~Wess, and B.~Zumino, {\it {Structure of
  phenomenological Lagrangians. 2.}},  {\em Phys. Rev.} {\bf 177} (1969)
  2247--2250.

\bibitem{DeSimone:2012fs}
A.~De~Simone, O.~Matsedonskyi, R.~Rattazzi, and A.~Wulzer, {\it {A First Top
  Partner Hunter's Guide}},  {\em JHEP} {\bf 04} (2013) 004,
  [\href{http://xxx.lanl.gov/abs/1211.5663}{{\tt arXiv:1211.5663}}].

\bibitem{Panico:2015jxa}
G.~Panico and A.~Wulzer, {\em {The Composite Nambu-Goldstone Higgs}}, vol.~913.
\newblock Springer, 2016.

\bibitem{Durieux:2021riy}
G.~Durieux, M.~McCullough, and E.~Salvioni, {\it {Gegenbauer Goldstones}},
  {\em JHEP} {\bf 01} (2022) 076,
  [\href{http://xxx.lanl.gov/abs/2110.06941}{{\tt arXiv:2110.06941}}].

\bibitem{Durieux:2022sgm}
G.~Durieux, M.~McCullough, and E.~Salvioni, {\it {Gegenbauer\textquoteright{}s
  Twin}},  {\em JHEP} {\bf 05} (2022) 140,
  [\href{http://xxx.lanl.gov/abs/2202.01228}{{\tt arXiv:2202.01228}}].

\bibitem{ATLAS:2022vkf}
{\bf ATLAS} Collaboration, G.~Aad {\em et.~al.}, {\it {A detailed map of Higgs
  boson interactions by the ATLAS experiment ten years after the discovery}},
  {\em Nature} {\bf 607} (2022), no.~7917 52--59,
  [\href{http://xxx.lanl.gov/abs/2207.00092}{{\tt arXiv:2207.00092}}].
  [Erratum: Nature 612, E24 (2022)].

\bibitem{CMS:2022dwd}
{\bf CMS} Collaboration, A.~Tumasyan {\em et.~al.}, {\it {A portrait of the
  Higgs boson by the CMS experiment ten years after the discovery.}},  {\em
  Nature} {\bf 607} (2022), no.~7917 60--68,
  [\href{http://xxx.lanl.gov/abs/2207.00043}{{\tt arXiv:2207.00043}}].
  [Erratum: Nature 623, (2023)].

\bibitem{deBlas:2019rxi}
J.~de~Blas {\em et.~al.}, {\it {Higgs Boson Studies at Future Particle
  Colliders}},  {\em JHEP} {\bf 01} (2020) 139,
  [\href{http://xxx.lanl.gov/abs/1905.03764}{{\tt arXiv:1905.03764}}].

\bibitem{CMS:2022krd}
{\bf CMS} Collaboration, A.~Tumasyan {\em et.~al.}, {\it {Search for new
  physics in the lepton plus missing transverse momentum final state in
  proton-proton collisions at $\sqrt{s} =$ 13 TeV}},  {\em JHEP} {\bf 07}
  (2022) 067, [\href{http://xxx.lanl.gov/abs/2202.06075}{{\tt
  arXiv:2202.06075}}].

\bibitem{Thamm:2015zwa}
A.~Thamm, R.~Torre, and A.~Wulzer, {\it {Future tests of Higgs compositeness:
  direct vs indirect}},  {\em JHEP} {\bf 07} (2015) 100,
  [\href{http://xxx.lanl.gov/abs/1502.01701}{{\tt arXiv:1502.01701}}].

\bibitem{CMS:2022fck}
{\bf CMS} Collaboration, A.~Tumasyan {\em et.~al.}, {\it {Search for pair
  production of vector-like quarks in leptonic final states in proton-proton
  collisions at $ \sqrt{s} $ = 13 TeV}},  {\em JHEP} {\bf 07} (2023) 020,
  [\href{http://xxx.lanl.gov/abs/2209.07327}{{\tt arXiv:2209.07327}}].

\bibitem{CMS:2022jeb}
{\bf CMS} Collaboration, {\it {Search for a vector-like quark T decaying to bW,
  tZ, tH in the single lepton final state at the HL-LHC}}, .

\bibitem{CidVidal:2018eel}
X.~Cid~Vidal {\em et.~al.}, {\it {Report from Working Group 3}: {Beyond the
  Standard Model physics at the HL-LHC and HE-LHC}},  {\em CERN Yellow Rep.
  Monogr.} {\bf 7} (2019) 585--865,
  [\href{http://xxx.lanl.gov/abs/1812.07831}{{\tt arXiv:1812.07831}}].

\bibitem{Chacko:2005pe}
Z.~Chacko, H.-S. Goh, and R.~Harnik, {\it {The Twin Higgs: Natural electroweak
  breaking from mirror symmetry}},  {\em Phys. Rev. Lett.} {\bf 96} (2006)
  231802, [\href{http://xxx.lanl.gov/abs/hep-ph/0506256}{{\tt
  hep-ph/0506256}}].

\bibitem{Craig:2015pha}
N.~Craig, A.~Katz, M.~Strassler, and R.~Sundrum, {\it {Naturalness in the Dark
  at the LHC}},  {\em JHEP} {\bf 07} (2015) 105,
  [\href{http://xxx.lanl.gov/abs/1501.05310}{{\tt arXiv:1501.05310}}].

\bibitem{Strassler:2006im}
M.~J. Strassler and K.~M. Zurek, {\it {Echoes of a hidden valley at hadron
  colliders}},  {\em Phys. Lett. B} {\bf 651} (2007) 374--379,
  [\href{http://xxx.lanl.gov/abs/hep-ph/0604261}{{\tt hep-ph/0604261}}].

\bibitem{Strassler:2006ri}
M.~J. Strassler and K.~M. Zurek, {\it {Discovering the Higgs through
  highly-displaced vertices}},  {\em Phys. Lett. B} {\bf 661} (2008) 263--267,
  [\href{http://xxx.lanl.gov/abs/hep-ph/0605193}{{\tt hep-ph/0605193}}].

\bibitem{Han:2007ae}
T.~Han, Z.~Si, K.~M. Zurek, and M.~J. Strassler, {\it {Phenomenology of hidden
  valleys at hadron colliders}},  {\em JHEP} {\bf 07} (2008) 008,
  [\href{http://xxx.lanl.gov/abs/0712.2041}{{\tt arXiv:0712.2041}}].

\bibitem{Curtin:2015fna}
D.~Curtin and C.~B. Verhaaren, {\it {Discovering Uncolored Naturalness in
  Exotic Higgs Decays}},  {\em JHEP} {\bf 12} (2015) 072,
  [\href{http://xxx.lanl.gov/abs/1506.06141}{{\tt arXiv:1506.06141}}].

\bibitem{Maldacena:1997re}
J.~M. Maldacena, {\it {The Large N limit of superconformal field theories and
  supergravity}},  {\em Adv. Theor. Math. Phys.} {\bf 2} (1998) 231--252,
  [\href{http://xxx.lanl.gov/abs/hep-th/9711200}{{\tt hep-th/9711200}}].

\bibitem{ArkaniHamed:1998rs}
N.~Arkani-Hamed, S.~Dimopoulos, and G.~R. Dvali, {\it {The Hierarchy problem
  and new dimensions at a millimeter}},  {\em Phys. Lett. B} {\bf 429} (1998)
  263--272, [\href{http://xxx.lanl.gov/abs/hep-ph/9803315}{{\tt
  hep-ph/9803315}}].

\bibitem{Antoniadis:1998ig}
I.~Antoniadis, N.~Arkani-Hamed, S.~Dimopoulos, and G.~R. Dvali, {\it {New
  dimensions at a millimeter to a Fermi and superstrings at a TeV}},  {\em
  Phys. Lett. B} {\bf 436} (1998) 257--263,
  [\href{http://xxx.lanl.gov/abs/hep-ph/9804398}{{\tt hep-ph/9804398}}].

\bibitem{Rattazzi:2003ea}
R.~Rattazzi, {\it {Cargese lectures on extra-dimensions}},  in {\em {Cargese
  School of Particle Physics and Cosmology: the Interface}}, pp.~461--517, 8,
  2003.
\newblock \href{http://xxx.lanl.gov/abs/hep-ph/0607055}{{\tt hep-ph/0607055}}.

\bibitem{Csaki:2004ay}
C.~Csaki, {\it {TASI lectures on extra dimensions and branes}},  in {\em
  {Theoretical Advanced Study Institute in Elementary Particle Physics (TASI
  2002): Particle Physics and Cosmology: The Quest for Physics Beyond the
  Standard Model(s)}}, pp.~605--698, 4, 2004.
\newblock \href{http://xxx.lanl.gov/abs/hep-ph/0404096}{{\tt hep-ph/0404096}}.

\bibitem{ArkaniHamed:1998kx}
N.~Arkani-Hamed, S.~Dimopoulos, and J.~March-Russell, {\it {Stabilization of
  submillimeter dimensions: The New guise of the hierarchy problem}},  {\em
  Phys. Rev. D} {\bf 63} (2001) 064020,
  [\href{http://xxx.lanl.gov/abs/hep-th/9809124}{{\tt hep-th/9809124}}].

\bibitem{Randall:1999ee}
L.~Randall and R.~Sundrum, {\it {A Large mass hierarchy from a small extra
  dimension}},  {\em Phys. Rev. Lett.} {\bf 83} (1999) 3370--3373,
  [\href{http://xxx.lanl.gov/abs/hep-ph/9905221}{{\tt hep-ph/9905221}}].

\bibitem{ArkaniHamed:2001ca}
N.~Arkani-Hamed, A.~G. Cohen, and H.~Georgi, {\it {(De)constructing
  dimensions}},  {\em Phys. Rev. Lett.} {\bf 86} (2001) 4757--4761,
  [\href{http://xxx.lanl.gov/abs/hep-th/0104005}{{\tt hep-th/0104005}}].

\bibitem{Wilson:1973jj}
K.~G. Wilson and J.~B. Kogut, {\it {The Renormalization group and the epsilon
  expansion}},  {\em Phys. Rept.} {\bf 12} (1974) 75--199.

\bibitem{Jackiw:2013jba}
R.~Jackiw, {\it {Ken Wilson -- The Early Years}},  {\em Int. J. Mod. Phys. A}
  {\bf 29} (2014) 1430008, [\href{http://xxx.lanl.gov/abs/1312.6634}{{\tt
  arXiv:1312.6634}}].

\bibitem{Adams:2006sv}
A.~Adams, N.~Arkani-Hamed, S.~Dubovsky, A.~Nicolis, and R.~Rattazzi, {\it
  {Causality, analyticity and an IR obstruction to UV completion}},  {\em JHEP}
  {\bf 10} (2006) 014, [\href{http://xxx.lanl.gov/abs/hep-th/0602178}{{\tt
  hep-th/0602178}}].

\bibitem{Kaplan:2024qtf}
D.~E. Kaplan, S.~Rajendran, and F.~Serra, {\it {Wrong Signs are Alright}},
  \href{http://xxx.lanl.gov/abs/2406.06681}{{\tt arXiv:2406.06681}}.

\bibitem{Grinstein:2007mp}
B.~Grinstein, D.~O'Connell, and M.~B. Wise, {\it {The Lee-Wick standard
  model}},  {\em Phys. Rev. D} {\bf 77} (2008) 025012,
  [\href{http://xxx.lanl.gov/abs/0704.1845}{{\tt arXiv:0704.1845}}].

\bibitem{Craig:2019zbn}
N.~Craig and S.~Koren, {\it {IR Dynamics from UV Divergences: UV/IR Mixing,
  NCFT, and the Hierarchy Problem}},  {\em JHEP} {\bf 03} (2020) 037,
  [\href{http://xxx.lanl.gov/abs/1909.01365}{{\tt arXiv:1909.01365}}].

\bibitem{Banks:2010zn}
T.~Banks and N.~Seiberg, {\it {Symmetries and Strings in Field Theory and
  Gravity}},  {\em Phys. Rev. D} {\bf 83} (2011) 084019,
  [\href{http://xxx.lanl.gov/abs/1011.5120}{{\tt arXiv:1011.5120}}].

\bibitem{Vafa:2005ui}
C.~Vafa, {\it {The String landscape and the swampland}},
  \href{http://xxx.lanl.gov/abs/hep-th/0509212}{{\tt hep-th/0509212}}.

\bibitem{Arkani-Hamed:2006emk}
N.~Arkani-Hamed, L.~Motl, A.~Nicolis, and C.~Vafa, {\it {The String landscape,
  black holes and gravity as the weakest force}},  {\em JHEP} {\bf 06} (2007)
  060, [\href{http://xxx.lanl.gov/abs/hep-th/0601001}{{\tt hep-th/0601001}}].

\bibitem{Ooguri:2006in}
H.~Ooguri and C.~Vafa, {\it {On the Geometry of the String Landscape and the
  Swampland}},  {\em Nucl. Phys. B} {\bf 766} (2007) 21--33,
  [\href{http://xxx.lanl.gov/abs/hep-th/0605264}{{\tt hep-th/0605264}}].

\bibitem{Ooguri:2016pdq}
H.~Ooguri and C.~Vafa, {\it {Non-supersymmetric AdS and the Swampland}},  {\em
  Adv. Theor. Math. Phys.} {\bf 21} (2017) 1787--1801,
  [\href{http://xxx.lanl.gov/abs/1610.01533}{{\tt arXiv:1610.01533}}].

\bibitem{Ibanez:2017kvh}
L.~E. Ibanez, V.~Martin-Lozano, and I.~Valenzuela, {\it {Constraining Neutrino
  Masses, the Cosmological Constant and BSM Physics from the Weak Gravity
  Conjecture}},  {\em JHEP} {\bf 11} (2017) 066,
  [\href{http://xxx.lanl.gov/abs/1706.05392}{{\tt arXiv:1706.05392}}].

\bibitem{Ibanez:2017oqr}
L.~E. Ibanez, V.~Martin-Lozano, and I.~Valenzuela, {\it {Constraining the EW
  Hierarchy from the Weak Gravity Conjecture}},
  \href{http://xxx.lanl.gov/abs/1707.05811}{{\tt arXiv:1707.05811}}.

\bibitem{Hamada:2017yji}
Y.~Hamada and G.~Shiu, {\it {Weak Gravity Conjecture, Multiple Point Principle
  and the Standard Model Landscape}},  {\em JHEP} {\bf 11} (2017) 043,
  [\href{http://xxx.lanl.gov/abs/1707.06326}{{\tt arXiv:1707.06326}}].

\bibitem{Lust:2017wrl}
D.~Lust and E.~Palti, {\it {Scalar Fields, Hierarchical UV/IR Mixing and The
  Weak Gravity Conjecture}},  {\em JHEP} {\bf 02} (2018) 040,
  [\href{http://xxx.lanl.gov/abs/1709.01790}{{\tt arXiv:1709.01790}}].

\bibitem{Gonzalo:2018tpb}
E.~Gonzalo, A.~Herr\'aez, and L.~E. Ib\'a\~nez, {\it {AdS-phobia, the WGC, the
  Standard Model and Supersymmetry}},  {\em JHEP} {\bf 06} (2018) 051,
  [\href{http://xxx.lanl.gov/abs/1803.08455}{{\tt arXiv:1803.08455}}].

\bibitem{Gonzalo:2018dxi}
E.~Gonzalo and L.~E. Ib\'a\~nez, {\it {The Fundamental Need for a SM Higgs and
  the Weak Gravity Conjecture}},  {\em Phys. Lett. B} {\bf 786} (2018)
  272--277, [\href{http://xxx.lanl.gov/abs/1806.09647}{{\tt
  arXiv:1806.09647}}].

\bibitem{Craig:2018yvw}
N.~Craig, I.~Garcia~Garcia, and S.~Koren, {\it {Discrete Gauge Symmetries and
  the Weak Gravity Conjecture}},  {\em JHEP} {\bf 05} (2019) 140,
  [\href{http://xxx.lanl.gov/abs/1812.08181}{{\tt arXiv:1812.08181}}].

\bibitem{Craig:2019fdy}
N.~Craig, I.~Garcia~Garcia, and S.~Koren, {\it {The Weak Scale from Weak
  Gravity}},  {\em JHEP} {\bf 09} (2019) 081,
  [\href{http://xxx.lanl.gov/abs/1904.08426}{{\tt arXiv:1904.08426}}].

\bibitem{Cheung:2014vva}
C.~Cheung and G.~N. Remmen, {\it {Naturalness and the Weak Gravity
  Conjecture}},  {\em Phys. Rev. Lett.} {\bf 113} (2014) 051601,
  [\href{http://xxx.lanl.gov/abs/1402.2287}{{\tt arXiv:1402.2287}}].

\bibitem{Gaiotto:2014kfa}
D.~Gaiotto, A.~Kapustin, N.~Seiberg, and B.~Willett, {\it {Generalized Global
  Symmetries}},  {\em JHEP} {\bf 02} (2015) 172,
  [\href{http://xxx.lanl.gov/abs/1412.5148}{{\tt arXiv:1412.5148}}].

\bibitem{Brennan:2023mmt}
T.~D. Brennan and S.~Hong, {\it {Introduction to Generalized Global Symmetries
  in QFT and Particle Physics}},
  \href{http://xxx.lanl.gov/abs/2306.00912}{{\tt arXiv:2306.00912}}.

\bibitem{Bhardwaj:2023kri}
L.~Bhardwaj, L.~E. Bottini, L.~Fraser-Taliente, L.~Gladden, D.~S.~W. Gould,
  A.~Platschorre, and H.~Tillim, {\it {Lectures on generalized symmetries}},
  {\em Phys. Rept.} {\bf 1051} (2024) 1--87,
  [\href{http://xxx.lanl.gov/abs/2307.07547}{{\tt arXiv:2307.07547}}].

\bibitem{Gomes:2023ahz}
P.~R.~S. Gomes, {\it {An introduction to higher-form symmetries}},  {\em
  SciPost Phys. Lect. Notes} {\bf 74} (2023) 1,
  [\href{http://xxx.lanl.gov/abs/2303.01817}{{\tt arXiv:2303.01817}}].

\bibitem{Luo:2023ive}
R.~Luo, Q.-R. Wang, and Y.-N. Wang, {\it {Lecture notes on generalized
  symmetries and applications}},  {\em Phys. Rept.} {\bf 1065} (2024) 1--43,
  [\href{http://xxx.lanl.gov/abs/2307.09215}{{\tt arXiv:2307.09215}}].

\bibitem{Schafer-Nameki:2023jdn}
S.~Schafer-Nameki, {\it {ICTP lectures on (non-)invertible generalized
  symmetries}},  {\em Phys. Rept.} {\bf 1063} (2024) 1--55,
  [\href{http://xxx.lanl.gov/abs/2305.18296}{{\tt arXiv:2305.18296}}].

\bibitem{Shao:2023gho}
S.-H. Shao, {\it {What's Done Cannot Be Undone: TASI Lectures on Non-Invertible
  Symmetries}},  \href{http://xxx.lanl.gov/abs/2308.00747}{{\tt
  arXiv:2308.00747}}.

\bibitem{Cordova:2018cvg}
C.~C\'ordova, T.~T. Dumitrescu, and K.~Intriligator, {\it {Exploring 2-Group
  Global Symmetries}},  {\em JHEP} {\bf 02} (2019) 184,
  [\href{http://xxx.lanl.gov/abs/1802.04790}{{\tt arXiv:1802.04790}}].

\bibitem{Wan:2019gqr}
Z.~Wan and J.~Wang, {\it {Beyond Standard Models and Grand Unifications:
  Anomalies, Topological Terms, and Dynamical Constraints via Cobordisms}},
  {\em JHEP} {\bf 07} (2020) 062,
  [\href{http://xxx.lanl.gov/abs/1910.14668}{{\tt arXiv:1910.14668}}].

\bibitem{Davighi:2019rcd}
J.~Davighi, B.~Gripaios, and N.~Lohitsiri, {\it {Global anomalies in the
  Standard Model(s) and Beyond}},  {\em JHEP} {\bf 07} (2020) 232,
  [\href{http://xxx.lanl.gov/abs/1910.11277}{{\tt arXiv:1910.11277}}].

\bibitem{Wang:2020xyo}
J.~Wang, {\it {Anomaly and Cobordism Constraints Beyond the Standard Model:
  Topological Force}},  \href{http://xxx.lanl.gov/abs/2006.16996}{{\tt
  arXiv:2006.16996}}.

\bibitem{Brennan:2020ehu}
T.~D. Brennan and C.~Cordova, {\it {Axions, higher-groups, and emergent
  symmetry}},  {\em JHEP} {\bf 02} (2022) 145,
  [\href{http://xxx.lanl.gov/abs/2011.09600}{{\tt arXiv:2011.09600}}].

\bibitem{Hidaka:2020izy}
Y.~Hidaka, M.~Nitta, and R.~Yokokura, {\it {Global 3-group symmetry and 't
  Hooft anomalies in axion electrodynamics}},  {\em JHEP} {\bf 01} (2021) 173,
  [\href{http://xxx.lanl.gov/abs/2009.14368}{{\tt arXiv:2009.14368}}].

\bibitem{Fan:2021ntg}
J.~Fan, K.~Fraser, M.~Reece, and J.~Stout, {\it {Axion Mass from Magnetic
  Monopole Loops}},  {\em Phys. Rev. Lett.} {\bf 127} (2021), no.~13 131602,
  [\href{http://xxx.lanl.gov/abs/2105.09950}{{\tt arXiv:2105.09950}}].

\bibitem{Anber:2021upc}
M.~M. Anber and E.~Poppitz, {\it {Nonperturbative effects in the Standard Model
  with gauged 1-form symmetry}},  {\em JHEP} {\bf 12} (2021) 055,
  [\href{http://xxx.lanl.gov/abs/2110.02981}{{\tt arXiv:2110.02981}}].

\bibitem{Wang:2021ayd}
J.~Wang, Z.~Wan, and Y.-Z. You, {\it {Cobordism and deformation class of the
  standard model}},  {\em Phys. Rev. D} {\bf 106} (2022), no.~4 L041701,
  [\href{http://xxx.lanl.gov/abs/2112.14765}{{\tt arXiv:2112.14765}}].

\bibitem{Wang:2021vki}
J.~Wang and Y.-Z. You, {\it {Gauge Enhanced Quantum Criticality Between Grand
  Unifications: Categorical Higher Symmetry Retraction}},
  \href{http://xxx.lanl.gov/abs/2111.10369}{{\tt arXiv:2111.10369}}.

\bibitem{Wang:2021hob}
J.~Wang and Y.-Z. You, {\it {Gauge enhanced quantum criticality beyond the
  standard model}},  {\em Phys. Rev. D} {\bf 106} (2022), no.~2 025013,
  [\href{http://xxx.lanl.gov/abs/2106.16248}{{\tt arXiv:2106.16248}}].

\bibitem{McNamara:2022lrw}
J.~McNamara and M.~Reece, {\it {Reflections on Parity Breaking}},
  \href{http://xxx.lanl.gov/abs/2212.00039}{{\tt arXiv:2212.00039}}.

\bibitem{Cordova:2022fhg}
C.~Cordova, S.~Hong, S.~Koren, and K.~Ohmori, {\it {Neutrino Masses from
  Generalized Symmetry Breaking}},
  \href{http://xxx.lanl.gov/abs/2211.07639}{{\tt arXiv:2211.07639}}.

\bibitem{Cordova:2022qtz}
C.~Cordova and S.~Koren, {\it {Higher Flavor Symmetries in the Standard
  Model}},  {\em Annalen Phys.} {\bf 535} (2023), no.~8 2300031,
  [\href{http://xxx.lanl.gov/abs/2212.13193}{{\tt arXiv:2212.13193}}].

\bibitem{Choi:2022rfe}
Y.~Choi, H.~T. Lam, and S.-H. Shao, {\it {Noninvertible Time-Reversal
  Symmetry}},  {\em Phys. Rev. Lett.} {\bf 130} (2023), no.~13 131602,
  [\href{http://xxx.lanl.gov/abs/2208.04331}{{\tt arXiv:2208.04331}}].

\bibitem{Wang:2022eag}
J.~Wang, Z.~Wan, and Y.-Z. You, {\it {Proton stability: From the standard model
  to beyond grand unification}},  {\em Phys. Rev. D} {\bf 106} (2022), no.~2
  025016, [\href{http://xxx.lanl.gov/abs/2204.08393}{{\tt arXiv:2204.08393}}].

\bibitem{Yokokura:2022alv}
R.~Yokokura, {\it {Non-invertible symmetries in axion electrodynamics}},
  \href{http://xxx.lanl.gov/abs/2212.05001}{{\tt arXiv:2212.05001}}.

\bibitem{Brennan:2023kpw}
T.~D. Brennan, S.~Hong, and L.-T. Wang, {\it {Coupling a Cosmic String to a
  TQFT}},  {\em JHEP} {\bf 03} (2024) 145,
  [\href{http://xxx.lanl.gov/abs/2302.00777}{{\tt arXiv:2302.00777}}].

\bibitem{Cordova:2023her}
C.~Cordova, S.~Hong, and L.-T. Wang, {\it {Axion domain walls, small
  instantons, and non-invertible symmetry breaking}},  {\em JHEP} {\bf 05}
  (2024) 325, [\href{http://xxx.lanl.gov/abs/2309.05636}{{\tt
  arXiv:2309.05636}}].

\bibitem{Choi:2023pdp}
Y.~Choi, M.~Forslund, H.~T. Lam, and S.-H. Shao, {\it {Quantization of
  Axion-Gauge Couplings and Noninvertible Higher Symmetries}},  {\em Phys. Rev.
  Lett.} {\bf 132} (2024), no.~12 121601,
  [\href{http://xxx.lanl.gov/abs/2309.03937}{{\tt arXiv:2309.03937}}].

\bibitem{vanBeest:2023dbu}
M.~van Beest, P.~Boyle~Smith, D.~Delmastro, Z.~Komargodski, and D.~Tong, {\it
  {Monopoles, Scattering, and Generalized Symmetries}},
  \href{http://xxx.lanl.gov/abs/2306.07318}{{\tt arXiv:2306.07318}}.

\bibitem{Brennan:2023tae}
T.~D. Brennan, {\it {A New Solution to the Callan Rubakov Effect}},
  \href{http://xxx.lanl.gov/abs/2309.00680}{{\tt arXiv:2309.00680}}.

\bibitem{Choi:2022fgx}
Y.~Choi, H.~T. Lam, and S.-H. Shao, {\it {Non-invertible Gauss law and
  axions}},  {\em JHEP} {\bf 09} (2023) 067,
  [\href{http://xxx.lanl.gov/abs/2212.04499}{{\tt arXiv:2212.04499}}].

\bibitem{Cordova:2023ent}
C.~Cordova and K.~Ohmori, {\it {Quantum duality in electromagnetism and the
  fine structure constant}},  {\em Phys. Rev. D} {\bf 109} (2024), no.~10
  105019, [\href{http://xxx.lanl.gov/abs/2307.12927}{{\tt arXiv:2307.12927}}].

\bibitem{Reece:2023iqn}
M.~Reece, {\it {Axion-gauge coupling quantization with a twist}},  {\em JHEP}
  {\bf 10} (2023) 116, [\href{http://xxx.lanl.gov/abs/2309.03939}{{\tt
  arXiv:2309.03939}}].

\bibitem{Putrov:2023jqi}
P.~Putrov and J.~Wang, {\it {Categorical Symmetry of the Standard Model from
  Gravitational Anomaly}},  \href{http://xxx.lanl.gov/abs/2302.14862}{{\tt
  arXiv:2302.14862}}.

\bibitem{vanBeest:2023mbs}
M.~van Beest, P.~Boyle~Smith, D.~Delmastro, R.~Mouland, and D.~Tong, {\it
  {Fermion-Monopole Scattering in the Standard Model}},
  \href{http://xxx.lanl.gov/abs/2312.17746}{{\tt arXiv:2312.17746}}.

\bibitem{Aloni:2024jpb}
D.~Aloni, E.~Garc\'\i{}a-Valdecasas, M.~Reece, and M.~Suzuki, {\it
  {Spontaneously Broken $(-1)$-Form U(1) Symmetries}},
  \href{http://xxx.lanl.gov/abs/2402.00117}{{\tt arXiv:2402.00117}}.

\bibitem{Salvio:2014soa}
A.~Salvio and A.~Strumia, {\it {Agravity}},  {\em JHEP} {\bf 06} (2014) 080,
  [\href{http://xxx.lanl.gov/abs/1403.4226}{{\tt arXiv:1403.4226}}].

\bibitem{Graham:2015cka}
P.~W. Graham, D.~E. Kaplan, and S.~Rajendran, {\it {Cosmological Relaxation of
  the Electroweak Scale}},  {\em Phys. Rev. Lett.} {\bf 115} (2015), no.~22
  221801, [\href{http://xxx.lanl.gov/abs/1504.07551}{{\tt arXiv:1504.07551}}].

\bibitem{Abbott:1984qf}
L.~F. Abbott, {\it {A Mechanism for Reducing the Value of the Cosmological
  Constant}},  {\em Phys. Lett. B} {\bf 150} (1985) 427--430.

\bibitem{Dvali:2003br}
G.~Dvali and A.~Vilenkin, {\it {Cosmic attractors and gauge hierarchy}},  {\em
  Phys. Rev. D} {\bf 70} (2004) 063501,
  [\href{http://xxx.lanl.gov/abs/hep-th/0304043}{{\tt hep-th/0304043}}].

\bibitem{Dvali:2004tma}
G.~Dvali, {\it {Large hierarchies from attractor vacua}},  {\em Phys. Rev. D}
  {\bf 74} (2006) 025018, [\href{http://xxx.lanl.gov/abs/hep-th/0410286}{{\tt
  hep-th/0410286}}].

\bibitem{diCortona:2015ldu}
G.~Grilli~di Cortona, E.~Hardy, J.~Pardo~Vega, and G.~Villadoro, {\it {The QCD
  axion, precisely}},  {\em JHEP} {\bf 01} (2016) 034,
  [\href{http://xxx.lanl.gov/abs/1511.02867}{{\tt arXiv:1511.02867}}].

\bibitem{Ibanez:2015fcv}
L.~E. Ibanez, M.~Montero, A.~Uranga, and I.~Valenzuela, {\it {Relaxion
  Monodromy and the Weak Gravity Conjecture}},  {\em JHEP} {\bf 04} (2016) 020,
  [\href{http://xxx.lanl.gov/abs/1512.00025}{{\tt arXiv:1512.00025}}].

\bibitem{Kaplan:2015fuy}
D.~E. Kaplan and R.~Rattazzi, {\it {Large field excursions and approximate
  discrete symmetries from a clockwork axion}},  {\em Phys. Rev. D} {\bf 93}
  (2016), no.~8 085007, [\href{http://xxx.lanl.gov/abs/1511.01827}{{\tt
  arXiv:1511.01827}}].

\bibitem{Giudice:2021viw}
G.~F. Giudice, M.~McCullough, and T.~You, {\it {Self-organised localisation}},
  {\em JHEP} {\bf 10} (2021) 093,
  [\href{http://xxx.lanl.gov/abs/2105.08617}{{\tt arXiv:2105.08617}}].

\bibitem{Cabibbo:1979ay}
N.~Cabibbo, L.~Maiani, G.~Parisi, and R.~Petronzio, {\it {Bounds on the
  Fermions and Higgs Boson Masses in Grand Unified Theories}},  {\em Nucl.
  Phys. B} {\bf 158} (1979) 295--305.

\bibitem{Hung:1979dn}
P.~Q. Hung, {\it {Vacuum Instability and New Constraints on Fermion Masses}},
  {\em Phys. Rev. Lett.} {\bf 42} (1979) 873.

\bibitem{Lindner:1985uk}
M.~Lindner, {\it {Implications of Triviality for the Standard Model}},  {\em Z.
  Phys. C} {\bf 31} (1986) 295.

\bibitem{Sher:1988mj}
M.~Sher, {\it {Electroweak Higgs Potentials and Vacuum Stability}},  {\em Phys.
  Rept.} {\bf 179} (1989) 273--418.

\bibitem{Schrempp:1996fb}
B.~Schrempp and M.~Wimmer, {\it {Top quark and Higgs boson masses: Interplay
  between infrared and ultraviolet physics}},  {\em Prog. Part. Nucl. Phys.}
  {\bf 37} (1996) 1--90, [\href{http://xxx.lanl.gov/abs/hep-ph/9606386}{{\tt
  hep-ph/9606386}}].

\bibitem{Altarelli:1994rb}
G.~Altarelli and G.~Isidori, {\it {Lower limit on the Higgs mass in the
  standard model: An Update}},  {\em Phys. Lett. B} {\bf 337} (1994) 141--144.

\bibitem{Degrassi:2012ry}
G.~Degrassi, S.~Di~Vita, J.~Elias-Miro, J.~R. Espinosa, G.~F. Giudice,
  G.~Isidori, and A.~Strumia, {\it {Higgs mass and vacuum stability in the
  Standard Model at NNLO}},  {\em JHEP} {\bf 08} (2012) 098,
  [\href{http://xxx.lanl.gov/abs/1205.6497}{{\tt arXiv:1205.6497}}].

\bibitem{Buttazzo:2013uya}
D.~Buttazzo, G.~Degrassi, P.~P. Giardino, G.~F. Giudice, F.~Sala, A.~Salvio,
  and A.~Strumia, {\it {Investigating the near-criticality of the Higgs
  boson}},  {\em JHEP} {\bf 12} (2013) 089,
  [\href{http://xxx.lanl.gov/abs/1307.3536}{{\tt arXiv:1307.3536}}].

\bibitem{Bednyakov:2015sca}
A.~V. Bednyakov, B.~A. Kniehl, A.~F. Pikelner, and O.~L. Veretin, {\it
  {Stability of the Electroweak Vacuum: Gauge Independence and Advanced
  Precision}},  {\em Phys. Rev. Lett.} {\bf 115} (2015), no.~20 201802,
  [\href{http://xxx.lanl.gov/abs/1507.08833}{{\tt arXiv:1507.08833}}].

\bibitem{Andreassen:2017rzq}
A.~Andreassen, W.~Frost, and M.~D. Schwartz, {\it {Scale Invariant Instantons
  and the Complete Lifetime of the Standard Model}},  {\em Phys. Rev. D} {\bf
  97} (2018), no.~5 056006, [\href{http://xxx.lanl.gov/abs/1707.08124}{{\tt
  arXiv:1707.08124}}].

\bibitem{Weinberg:1967tq}
S.~Weinberg, {\it {A Model of Leptons}},  {\em Phys. Rev. Lett.} {\bf 19}
  (1967) 1264--1266.

\bibitem{Hawking:1975vcx}
S.~W. Hawking, {\it {Particle Creation by Black Holes}},  {\em Commun. Math.
  Phys.} {\bf 43} (1975) 199--220. [Erratum: Commun.Math.Phys. 46, 206 (1976)].

\bibitem{Gibbons:1976ue}
G.~W. Gibbons and S.~W. Hawking, {\it {Action Integrals and Partition Functions
  in Quantum Gravity}},  {\em Phys. Rev. D} {\bf 15} (1977) 2752--2756.

\bibitem{Starobinsky:1985ibc}
A.~A. Starobinsky, {\it {Multicomponent de Sitter (Inflationary) Stages and the
  Generation of Perturbations}},  {\em JETP Lett.} {\bf 42} (1985) 152--155.

\bibitem{Starobinsky:1986fx}
A.~A. Starobinsky, {\it {STOCHASTIC DE SITTER (INFLATIONARY) STAGE IN THE EARLY
  UNIVERSE}},  {\em Lect. Notes Phys.} {\bf 246} (1986) 107--126.

\bibitem{Rudelius:2019cfh}
T.~Rudelius, {\it {Conditions for (No) Eternal Inflation}},  {\em JCAP} {\bf
  08} (2019) 009, [\href{http://xxx.lanl.gov/abs/1905.05198}{{\tt
  arXiv:1905.05198}}].

\bibitem{Arkani-Hamed:2016rle}
N.~Arkani-Hamed, T.~Cohen, R.~T. D'Agnolo, A.~Hook, H.~D. Kim, and D.~Pinner,
  {\it {Solving the Hierarchy Problem at Reheating with a Large Number of
  Degrees of Freedom}},  {\em Phys. Rev. Lett.} {\bf 117} (2016), no.~25
  251801, [\href{http://xxx.lanl.gov/abs/1607.06821}{{\tt arXiv:1607.06821}}].

\bibitem{Dvali:2007hz}
G.~Dvali, {\it {Black Holes and Large N Species Solution to the Hierarchy
  Problem}},  {\em Fortsch. Phys.} {\bf 58} (2010) 528--536,
  [\href{http://xxx.lanl.gov/abs/0706.2050}{{\tt arXiv:0706.2050}}].

\bibitem{Dvali:2007wp}
G.~Dvali and M.~Redi, {\it {Black Hole Bound on the Number of Species and
  Quantum Gravity at LHC}},  {\em Phys. Rev. D} {\bf 77} (2008) 045027,
  [\href{http://xxx.lanl.gov/abs/0710.4344}{{\tt arXiv:0710.4344}}].

\bibitem{Geller:2018xvz}
M.~Geller, Y.~Hochberg, and E.~Kuflik, {\it {Inflating to the Weak Scale}},
  {\em Phys. Rev. Lett.} {\bf 122} (2019), no.~19 191802,
  [\href{http://xxx.lanl.gov/abs/1809.07338}{{\tt arXiv:1809.07338}}].

\bibitem{Giudice:2019iwl}
G.~F. Giudice, A.~Kehagias, and A.~Riotto, {\it {The Selfish Higgs}},  {\em
  JHEP} {\bf 10} (2019) 199, [\href{http://xxx.lanl.gov/abs/1907.05370}{{\tt
  arXiv:1907.05370}}].

\bibitem{Strumia:2020bdy}
A.~Strumia and D.~Teresi, {\it {Relaxing the Higgs mass and its vacuum energy
  by living at the top of the potential}},  {\em Phys. Rev. D} {\bf 101}
  (2020), no.~11 115002, [\href{http://xxx.lanl.gov/abs/2002.02463}{{\tt
  arXiv:2002.02463}}].

\bibitem{Csaki:2020zqz}
C.~Cs\'aki, R.~T. D'Agnolo, M.~Geller, and A.~Ismail, {\it {Crunching Dilaton,
  Hidden Naturalness}},  {\em Phys. Rev. Lett.} {\bf 126} (2021) 091801,
  [\href{http://xxx.lanl.gov/abs/2007.14396}{{\tt arXiv:2007.14396}}].

\bibitem{TitoDAgnolo:2021nhd}
R.~Tito~D'Agnolo and D.~Teresi, {\it {Sliding Naturalness: New Solution to the
  Strong-$CP$ and Electroweak-Hierarchy Problems}},  {\em Phys. Rev. Lett.}
  {\bf 128} (2022), no.~2 021803,
  [\href{http://xxx.lanl.gov/abs/2106.04591}{{\tt arXiv:2106.04591}}].

\bibitem{Khoury:2021zao}
J.~Khoury and T.~Steingasser, {\it {Gauge hierarchy from electroweak vacuum
  metastability}},  {\em Phys. Rev. D} {\bf 105} (2022), no.~5 055031,
  [\href{http://xxx.lanl.gov/abs/2108.09315}{{\tt arXiv:2108.09315}}].

\bibitem{TitoDAgnolo:2021pjo}
R.~Tito~D'Agnolo and D.~Teresi, {\it {Sliding naturalness: cosmological
  selection of the weak scale}},  {\em JHEP} {\bf 02} (2022) 023,
  [\href{http://xxx.lanl.gov/abs/2109.13249}{{\tt arXiv:2109.13249}}].

\bibitem{Martin:1997ns}
S.~P. Martin, {\it {A Supersymmetry primer}},  {\em Adv. Ser. Direct. High
  Energy Phys.} {\bf 18} (1998) 1--98,
  [\href{http://xxx.lanl.gov/abs/hep-ph/9709356}{{\tt hep-ph/9709356}}].

\bibitem{Baer:2006rs}
H.~Baer and X.~Tata, {\em {Weak scale supersymmetry: From superfields to
  scattering events}}.
\newblock Cambridge University Press, 5, 2006.

\bibitem{Weinberg:2000cr}
S.~Weinberg, {\em {The quantum theory of fields. Vol. 3: Supersymmetry}}.
\newblock Cambridge University Press, 6, 2013.

\bibitem{Amati:1988ft}
D.~Amati, K.~Konishi, Y.~Meurice, G.~C. Rossi, and G.~Veneziano, {\it
  {Nonperturbative Aspects in Supersymmetric Gauge Theories}},  {\em Phys.
  Rept.} {\bf 162} (1988) 169--248.

\bibitem{Shifman:1986zi}
M.~A. Shifman and A.~I. Vainshtein, {\it {Solution of the Anomaly Puzzle in
  SUSY Gauge Theories and the Wilson Operator Expansion}},  {\em Nucl. Phys. B}
  {\bf 277} (1986) 456.

\bibitem{Shifman:1991dz}
M.~A. Shifman and A.~I. Vainshtein, {\it {On holomorphic dependence and
  infrared effects in supersymmetric gauge theories}},  {\em Nucl. Phys. B}
  {\bf 359} (1991) 571--580.

\bibitem{Grisaru:1979wc}
M.~T. Grisaru, W.~Siegel, and M.~Rocek, {\it {Improved Methods for
  Supergraphs}},  {\em Nucl. Phys. B} {\bf 159} (1979) 429.

\bibitem{Seiberg:1993vc}
N.~Seiberg, {\it {Naturalness versus supersymmetric nonrenormalization
  theorems}},  {\em Phys. Lett. B} {\bf 318} (1993) 469--475,
  [\href{http://xxx.lanl.gov/abs/hep-ph/9309335}{{\tt hep-ph/9309335}}].

\bibitem{Seiberg:1994bp}
N.~Seiberg, {\it {The Power of holomorphy: Exact results in 4-D SUSY field
  theories}},  in {\em {Particles, Strings, and Cosmology (PASCOS 94)}},
  pp.~0357--369, 5, 1994.
\newblock \href{http://xxx.lanl.gov/abs/hep-th/9408013}{{\tt hep-th/9408013}}.

\bibitem{Freedman:2012zz}
D.~Z. Freedman and A.~Van~Proeyen, {\em {Supergravity}}.
\newblock Cambridge Univ. Press, Cambridge, UK, 5, 2012.

\bibitem{Dimopoulos:1981yj}
S.~Dimopoulos, S.~Raby, and F.~Wilczek, {\it {Supersymmetry and the Scale of
  Unification}},  {\em Phys. Rev. D} {\bf 24} (1981) 1681--1683.

\bibitem{Dimopoulos:1981zb}
S.~Dimopoulos and H.~Georgi, {\it {Softly Broken Supersymmetry and SU(5)}},
  {\em Nucl. Phys. B} {\bf 193} (1981) 150--162.

\bibitem{Georgi:1974sy}
H.~Georgi and S.~L. Glashow, {\it {Unity of All Elementary Particle Forces}},
  {\em Phys. Rev. Lett.} {\bf 32} (1974) 438--441.

\bibitem{Arvanitaki:2012ps}
A.~Arvanitaki, N.~Craig, S.~Dimopoulos, and G.~Villadoro, {\it {Mini-Split}},
  {\em JHEP} {\bf 02} (2013) 126,
  [\href{http://xxx.lanl.gov/abs/1210.0555}{{\tt arXiv:1210.0555}}].

\bibitem{Wells:2004di}
J.~D. Wells, {\it {PeV-scale supersymmetry}},  {\em Phys. Rev. D} {\bf 71}
  (2005) 015013, [\href{http://xxx.lanl.gov/abs/hep-ph/0411041}{{\tt
  hep-ph/0411041}}].

\bibitem{ArkaniHamed:2004fb}
N.~Arkani-Hamed and S.~Dimopoulos, {\it {Supersymmetric unification without low
  energy supersymmetry and signatures for fine-tuning at the LHC}},  {\em JHEP}
  {\bf 06} (2005) 073, [\href{http://xxx.lanl.gov/abs/hep-th/0405159}{{\tt
  hep-th/0405159}}].

\bibitem{Giudice:2004tc}
G.~F. Giudice and A.~Romanino, {\it {Split supersymmetry}},  {\em Nucl. Phys.
  B} {\bf 699} (2004) 65--89,
  [\href{http://xxx.lanl.gov/abs/hep-ph/0406088}{{\tt hep-ph/0406088}}].
  [Erratum: Nucl.Phys.B 706, 487--487 (2005)].

\bibitem{Chanowitz:1977ye}
M.~S. Chanowitz, J.~R. Ellis, and M.~K. Gaillard, {\it {The Price of Natural
  Flavor Conservation in Neutral Weak Interactions}},  {\em Nucl. Phys. B} {\bf
  128} (1977) 506--536.

\bibitem{Nanopoulos:1978hh}
D.~V. Nanopoulos and D.~A. Ross, {\it {Limits on the Number of Flavors in Grand
  Unified Theories from Higher Order Corrections to Fermion Masses}},  {\em
  Nucl. Phys. B} {\bf 157} (1979) 273--284.

\bibitem{Georgi:1979df}
H.~Georgi and C.~Jarlskog, {\it {A New Lepton - Quark Mass Relation in a
  Unified Theory}},  {\em Phys. Lett. B} {\bf 86} (1979) 297--300.

\bibitem{Antusch:2009gu}
S.~Antusch and M.~Spinrath, {\it {New GUT predictions for quark and lepton mass
  ratios confronted with phenomenology}},  {\em Phys. Rev. D} {\bf 79} (2009)
  095004, [\href{http://xxx.lanl.gov/abs/0902.4644}{{\tt arXiv:0902.4644}}].

\bibitem{Antusch:2013rxa}
S.~Antusch, S.~F. King, and M.~Spinrath, {\it {GUT predictions for quark-lepton
  Yukawa coupling ratios with messenger masses from non-singlets}},  {\em Phys.
  Rev. D} {\bf 89} (2014), no.~5 055027,
  [\href{http://xxx.lanl.gov/abs/1311.0877}{{\tt arXiv:1311.0877}}].

\bibitem{Hempfling:1993kv}
R.~Hempfling, {\it {Yukawa coupling unification with supersymmetric threshold
  corrections}},  {\em Phys. Rev. D} {\bf 49} (1994) 6168--6172.

\bibitem{Hall:1993gn}
L.~J. Hall, R.~Rattazzi, and U.~Sarid, {\it {The Top quark mass in
  supersymmetric SO(10) unification}},  {\em Phys. Rev. D} {\bf 50} (1994)
  7048--7065, [\href{http://xxx.lanl.gov/abs/hep-ph/9306309}{{\tt
  hep-ph/9306309}}].

\bibitem{Carena:1994bv}
M.~Carena, M.~Olechowski, S.~Pokorski, and C.~E.~M. Wagner, {\it {Electroweak
  symmetry breaking and bottom - top Yukawa unification}},  {\em Nucl. Phys. B}
  {\bf 426} (1994) 269--300,
  [\href{http://xxx.lanl.gov/abs/hep-ph/9402253}{{\tt hep-ph/9402253}}].

\bibitem{Blazek:1995nv}
T.~Blazek, S.~Raby, and S.~Pokorski, {\it {Finite supersymmetric threshold
  corrections to CKM matrix elements in the large tan Beta regime}},  {\em
  Phys. Rev. D} {\bf 52} (1995) 4151--4158,
  [\href{http://xxx.lanl.gov/abs/hep-ph/9504364}{{\tt hep-ph/9504364}}].

\bibitem{Antusch:2008tf}
S.~Antusch and M.~Spinrath, {\it {Quark and lepton masses at the GUT scale
  including SUSY threshold corrections}},  {\em Phys. Rev. D} {\bf 78} (2008)
  075020, [\href{http://xxx.lanl.gov/abs/0804.0717}{{\tt arXiv:0804.0717}}].

\bibitem{Antusch:2015nwi}
S.~Antusch and C.~Sluka, {\it {Predicting the Sparticle Spectrum from GUTs via
  SUSY Threshold Corrections with SusyTC}},  {\em JHEP} {\bf 07} (2016) 108,
  [\href{http://xxx.lanl.gov/abs/1512.06727}{{\tt arXiv:1512.06727}}].

\bibitem{Antusch:2016nak}
S.~Antusch and C.~Sluka, {\it {Testable SUSY Spectra from GUTs at a 100 TeV pp
  Collider}},  {\em Int. J. Mod. Phys. A} {\bf 31} (2016), no.~33 1644011,
  [\href{http://xxx.lanl.gov/abs/1604.00212}{{\tt arXiv:1604.00212}}].

\bibitem{Ellis:1990nz}
J.~R. Ellis, G.~Ridolfi, and F.~Zwirner, {\it {Radiative corrections to the
  masses of supersymmetric Higgs bosons}},  {\em Phys. Lett. B} {\bf 257}
  (1991) 83--91.

\bibitem{Ellis:1991zd}
J.~R. Ellis, G.~Ridolfi, and F.~Zwirner, {\it {On radiative corrections to
  supersymmetric Higgs boson masses and their implications for LEP searches}},
  {\em Phys. Lett. B} {\bf 262} (1991) 477--484.

\bibitem{Bagnaschi:2014rsa}
E.~Bagnaschi, G.~F. Giudice, P.~Slavich, and A.~Strumia, {\it {Higgs Mass and
  Unnatural Supersymmetry}},  {\em JHEP} {\bf 09} (2014) 092,
  [\href{http://xxx.lanl.gov/abs/1407.4081}{{\tt arXiv:1407.4081}}].

\bibitem{Hall:2011aa}
L.~J. Hall, D.~Pinner, and J.~T. Ruderman, {\it {A Natural SUSY Higgs Near 126
  GeV}},  {\em JHEP} {\bf 04} (2012) 131,
  [\href{http://xxx.lanl.gov/abs/1112.2703}{{\tt arXiv:1112.2703}}].

\bibitem{ArkaniHamed:2012gw}
N.~Arkani-Hamed, A.~Gupta, D.~E. Kaplan, N.~Weiner, and T.~Zorawski, {\it
  {Simply Unnatural Supersymmetry}},
  \href{http://xxx.lanl.gov/abs/1212.6971}{{\tt arXiv:1212.6971}}.

\bibitem{Ellis:1986yg}
J.~R. Ellis, K.~Enqvist, D.~V. Nanopoulos, and F.~Zwirner, {\it {Observables in
  Low-Energy Superstring Models}},  {\em Mod. Phys. Lett. A} {\bf 1} (1986) 57.

\bibitem{Barbieri:1987fn}
R.~Barbieri and G.~F. Giudice, {\it {Upper Bounds on Supersymmetric Particle
  Masses}},  {\em Nucl. Phys. B} {\bf 306} (1988) 63--76.

\bibitem{Craig:2013cxa}
N.~Craig, {\it {The State of Supersymmetry after Run I of the LHC}},  in {\em
  {Beyond the Standard Model after the first run of the LHC}}, 9, 2013.
\newblock \href{http://xxx.lanl.gov/abs/1309.0528}{{\tt arXiv:1309.0528}}.

\bibitem{Batell:2015fma}
B.~Batell, G.~F. Giudice, and M.~McCullough, {\it {Natural Heavy
  Supersymmetry}},  {\em JHEP} {\bf 12} (2015) 162,
  [\href{http://xxx.lanl.gov/abs/1509.00834}{{\tt arXiv:1509.00834}}].

\bibitem{Feng:1999zg}
J.~L. Feng, K.~T. Matchev, and T.~Moroi, {\it {Focus points and naturalness in
  supersymmetry}},  {\em Phys. Rev. D} {\bf 61} (2000) 075005,
  [\href{http://xxx.lanl.gov/abs/hep-ph/9909334}{{\tt hep-ph/9909334}}].

\bibitem{Baer:2012up}
H.~Baer, V.~Barger, P.~Huang, A.~Mustafayev, and X.~Tata, {\it {Radiative
  natural SUSY with a 125 GeV Higgs boson}},  {\em Phys. Rev. Lett.} {\bf 109}
  (2012) 161802, [\href{http://xxx.lanl.gov/abs/1207.3343}{{\tt
  arXiv:1207.3343}}].

\bibitem{LeCompte:2011cn}
T.~J. LeCompte and S.~P. Martin, {\it {Large Hadron Collider reach for
  supersymmetric models with compressed mass spectra}},  {\em Phys. Rev. D}
  {\bf 84} (2011) 015004, [\href{http://xxx.lanl.gov/abs/1105.4304}{{\tt
  arXiv:1105.4304}}].

\bibitem{Fan:2011yu}
J.~Fan, M.~Reece, and J.~T. Ruderman, {\it {Stealth Supersymmetry}},  {\em
  JHEP} {\bf 11} (2011) 012, [\href{http://xxx.lanl.gov/abs/1105.5135}{{\tt
  arXiv:1105.5135}}].

\bibitem{Fan:2012jf}
J.~Fan, M.~Reece, and J.~T. Ruderman, {\it {A Stealth Supersymmetry Sampler}},
  {\em JHEP} {\bf 07} (2012) 196,
  [\href{http://xxx.lanl.gov/abs/1201.4875}{{\tt arXiv:1201.4875}}].

\bibitem{Barbier:2004ez}
R.~Barbier {\em et.~al.}, {\it {R-parity violating supersymmetry}},  {\em Phys.
  Rept.} {\bf 420} (2005) 1--202,
  [\href{http://xxx.lanl.gov/abs/hep-ph/0406039}{{\tt hep-ph/0406039}}].

\bibitem{Csaki:2011ge}
C.~Csaki, Y.~Grossman, and B.~Heidenreich, {\it {MFV SUSY: A Natural Theory for
  R-Parity Violation}},  {\em Phys. Rev. D} {\bf 85} (2012) 095009,
  [\href{http://xxx.lanl.gov/abs/1111.1239}{{\tt arXiv:1111.1239}}].

\bibitem{Papucci:2011wy}
M.~Papucci, J.~T. Ruderman, and A.~Weiler, {\it {Natural SUSY Endures}},  {\em
  JHEP} {\bf 09} (2012) 035, [\href{http://xxx.lanl.gov/abs/1110.6926}{{\tt
  arXiv:1110.6926}}].

\bibitem{Dimopoulos:1995mi}
S.~Dimopoulos and G.~F. Giudice, {\it {Naturalness constraints in
  supersymmetric theories with nonuniversal soft terms}},  {\em Phys. Lett. B}
  {\bf 357} (1995) 573--578,
  [\href{http://xxx.lanl.gov/abs/hep-ph/9507282}{{\tt hep-ph/9507282}}].

\bibitem{Cohen:1996vb}
A.~G. Cohen, D.~B. Kaplan, and A.~E. Nelson, {\it {The More minimal
  supersymmetric standard model}},  {\em Phys. Lett. B} {\bf 388} (1996)
  588--598, [\href{http://xxx.lanl.gov/abs/hep-ph/9607394}{{\tt
  hep-ph/9607394}}].

\bibitem{Fox:2002bu}
P.~J. Fox, A.~E. Nelson, and N.~Weiner, {\it {Dirac gaugino masses and
  supersoft supersymmetry breaking}},  {\em JHEP} {\bf 08} (2002) 035,
  [\href{http://xxx.lanl.gov/abs/hep-ph/0206096}{{\tt hep-ph/0206096}}].

\bibitem{Kribs:2012gx}
G.~D. Kribs and A.~Martin, {\it {Supersoft Supersymmetry is Super-Safe}},  {\em
  Phys. Rev. D} {\bf 85} (2012) 115014,
  [\href{http://xxx.lanl.gov/abs/1203.4821}{{\tt arXiv:1203.4821}}].

\end{thebibliography}\endgroup

\end{document}